\documentclass[
 nofootinbib,
 superscriptaddress,
 twocolumn,
 amsmath,amssymb,
 prd,
]{revtex4-2}

\usepackage{graphicx}
\usepackage{dcolumn}
\usepackage{bm}
\usepackage{lineno}
\usepackage{fix-cm}
\usepackage{amsmath}
\usepackage{soul}
\usepackage{hyperref}
\usepackage{CJK}
\usepackage{appendix}
\usepackage{xcolor}
\usepackage{tikzducks}
\usepackage{orcidlink}
\usepackage{ulem}
\usepackage{aas_macros}
\usepackage{gensymb}
\usepackage{svg}

\graphicspath{{./}{figures/}{emojis/}{wind/}{1d-figures/}{stress/}{amt/}{amt-sum/}}

\begin{document}
\preprint{APS/123-QED}
\onecolumngrid           
\begin{center}

\textbf{\normalsize BMAD -- Circumbinary Magnetically Arrested Disks around Stellar or Black Hole Binaries:}\\
\textbf{\normalsize Hot Accretion Flows, Disk Properties, and Angular Momentum Transfer}\\[0.6cm]

{\normalsize
{Hai‑Yang Wang \orcidlink{0000-0001-7167-6110}~}$^{\includegraphics[height=1.4ex]{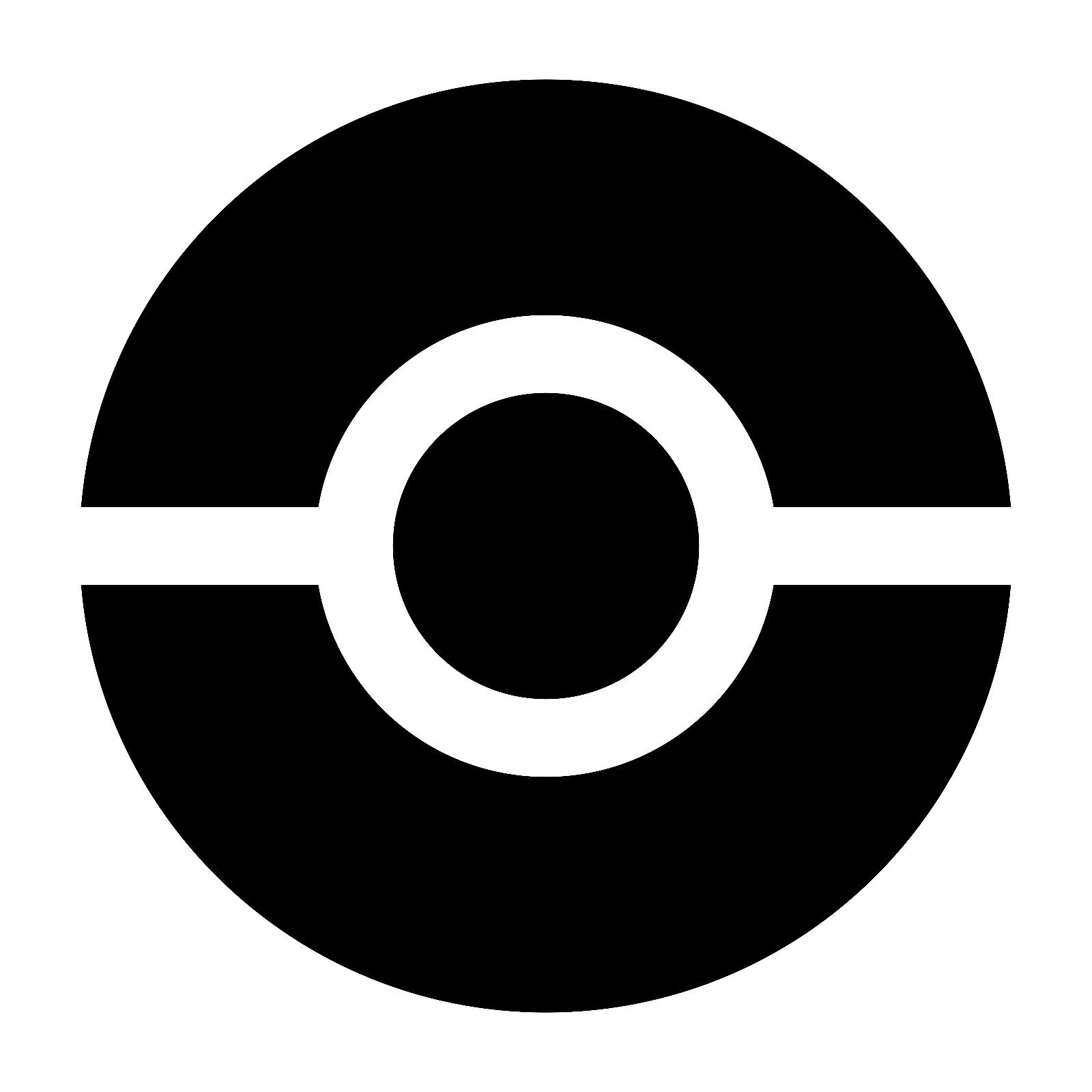},\,\includegraphics[height=1.4ex]{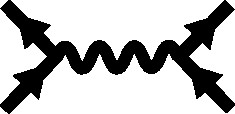}}$,
{Elias R.~Most \orcidlink{0000-0002-0491-1210}~}$^{\includegraphics[height=1.4ex]{emojis/pb.jpg},\,\includegraphics[height=1.4ex]{emojis/fd.jpg}}$, and
{Philip F.~Hopkins \orcidlink{0000-0003-3729-1684}~}$^{\includegraphics[height=1.4ex]{emojis/pb.jpg},\,\includegraphics[height=1.4ex]{emojis/fd.jpg}}$
}\\[0.4cm]

\textit{\footnotesize
$^{\includegraphics[height=1.4ex]{emojis/pb.jpg}}$\,TAPIR, Mailcode 350-17, California Institute of Technology, Pasadena, CA 91125, USA\\
$^{\includegraphics[height=1.4ex]{emojis/fd.jpg}}$\,Walter Burke Institute for Theoretical Physics, California Institute of Technology, Pasadena, CA 91125, USA}

\end{center}
\twocolumngrid         

\begin{abstract}
Binary systems surrounded by a circumbinary accretion flow can be subject to strong magnetic fields, potentially altering the character of the accretion flow itself, the evolution of the orbital dynamics, and outflow properties from the system. 
Here we focus on a regime where magnetic fields become so strong that the outer circumbinary flow becomes magnetically arrested, establishing a (circum)binary magnetically arrested disk (\textit{BMAD}) state. Such flows feature quasi-periodic magnetic flux eruptions, power jet-like magnetic tower outflows, and consequently alter the predominant contribution to angular momentum transfer inside the circumbinary disk.
In this work, we provide a comprehensive analysis of the properties of these flows around equal-mass binary systems on circular orbits ultilizing massively parallel three-dimensional Newtonian magnetohydrodynamics simulations. 
We investigate the impact of the equation of state and of dynamical cooling, as well as that of the (large-scale) magnetic field topology. 
Our findings are as follows: 
(1) A magnetically arrested accretion flow through the cavity can generally be achieved, so long as the initial seed field is
strong enough.
(2) The cavity, and magnetic flux tube properties and their subsequent propagation are subject to the choice of equation of state/cooling physics. 
(3) We find tentative evidence that in some regimes the BMAD state, {particularly during a flux eruption cycle}, can aid shrinking of the binary's orbit.
The regimes we explore have implications for multi-messenger transients to stars, supermassive and stellar black hole binaries and their orbital evolution in gaseous environments.

\end{abstract}

\maketitle

\section{Introduction}

Binary systems~\cite{Lai:2022ylu} have attracted considerable interested in recent years, in particular with regard to  protoplanetary disks
\cite{2011ARA&A..49...67W}, stellar binary systems
\cite{2008ApJ...681..375K,2024ApJ...964..133M}, stellar mass black hole binaries embedded in active galactic nuclei (AGN) disks \cite{2021ApJ...911..124L,2022ApJ...928L..19L}
and supermassive black hole binaries (SMBHBs) (e.g., \cite{Mayer:2007vk}).

The recent tentative detection of a stochastic (low frequency) gravitational wave background by an international team of pulsar timing networks, including the NANOGrav
collaboration \cite{Nanograv2023a,Nanograv2023b,InternationalPulsarTimingArray:2023mzf,NANOGrav:2023pdq,NANOGrav:2023vfo,NANOGrav:2023hde,NANOGrav:2023ctt,NANOGrav:2023tcn,Xu:2023wog},
has also highlighted the importance of understanding the evolution of
SMBHBs in gaseous environments, which
might be the leading source of this background.
Looking ahead, future space-based gravitational wave missions, such as the 
Laser Interferometer Space Antenna (LISA) \cite{lisa2023} and TianQin
\cite{tianqin} will be able to detect the merging population of SMBHBs having masses around $10^6 M_\odot$.

Despite this pressing need to predict the evolution of SMBHBs, several aspects still remain uncertain.
Efficient gas fueling into the center of the gas-rich galaxy merger remnant
\cite{Barnes:1991zz,Barnes:1992rm,Mihos:1995ri,Barnes:2002sh,DiMatteo:2005ttp,Hopkins:2005fb,Hopkins:2012fd,Mayer:2007vk,Cox:2007mn,Johansson:2008ib,Capelo:2014gqa,Capelo:2016vif}
has been shown to aid the formation of a circumbinary disk (CBD) around the central SMBHB
\cite{Mayer:2007vk,Dunhill:2014oka,Goicovic:2015kda,Goicovic:2016dul,Goicovic:2018xxi,Wang:2025mit}.
Gas-assisted orbital evolution of the binary is generally regarded as one of the most promising
bridge connecting large scales and small scales, in which case the stellar
dynamical friction
\cite{Quinlan:1996vp,Milosavljevic:2001vi,2003AIPC..686..201M,Berczik:2006tz,Sesana:2006xw,Rantala:2016rng,Chen:2022MNRAS.510..531C,Chen:2022MNRAS.514.2220C}
and gravitational radiation \cite{Peters:1964zz} separately extracts angular momentum from the SMBHB. As a result, one of the main motivations of CBD research is to understand the long-term angular
momentum transfer between the binary and the disk, possibly settling the so-called `final-parsec problem' \cite{Begelman1980,2003AIPC..686..201M} (when an SMBHB has a separation around $10^{-3}$ to $10^{-1}$ parsec).

The theory of binary-disk interaction has undergone several major changes over the past decade (for a detailed review, see e.g., \cite{Lai:2022ylu}). 
First, the shape and size of the cavity are found to deviate from the
secular (analytic) theory of binary-disk interactions \cite{Artymowicz:1994bw,Artymowicz1996}. 
It has also been realized that disk microphysics (such as magnetic fields, thermodynamics, chemical
composition, radiation), see, e.g. Refs. \cite{Most:2021uck,Bai:2013pi,Jiang:2014tpa}, play an important role in describing the evolution of the disk.  
The gravitational potential of the binary exerts a tidal torque on the
circumbinary disk, leading to an opened low-density cavity and a truncated
inner disk edge/cavity wall (e.g., \cite{Munoz:2018tnj}). 

At the cavity wall, the viscous torque trying
to close the cavity and the gravitational torque trying to widen it balance each other. The estimated cavity truncation radius around
the inner Lindblad resonances given by pioneering analytical works
\citep{Artymowicz:1994bw,2015MNRAS.452.2396M} matches the results of both
two-dimensional \cite{Miranda2017,Munoz:2018tnj} and three-dimensional \cite{Moody2019} hydrodynamical simulations {within the same order of magnitude}.
The density truncation on the inner and outer edges of the CBD can act as a
potential well, effectively trapping eccentric modes/waves
\cite{Teyssandier:2016MNRAS.458.3221T,Munoz:2020azx} for a timescale having
orders of magnitude similar to the disk lifetime. 
Furthermore, as shown in
a recent work \citep{paper1}, when the CBD is strongly magnetized, the
cavity shrinks by a significant fraction, and the cavity edge is constantly
disrupted and reforms.
Weakly magnetized accretion flows have also been studied in this context \cite{Shi:2011us,Noble:2012xz,Noble:2021vfg,Avara:2023ztw}.

Crucially, the understanding of angular momentum exchange between the binary
and the disk has also evolved significantly.
Early analytical works typically suggest that specific angular momentum is transferring from
the binary to the outer disk or ring, through extending the theory of
planet-disk (extreme mass-ratio) interaction \cite{lin1979,Artymowicz:1994bw}.
It is generally believed that the coupling of the binary and the disk is through deposition of angular momentum at the location of Lindblad resonances and corotation resonances, following which the angular momentum
is re-distributed in the disk through propagating spiral density
waves \cite{Goldreich:1979zz,Goldreich:1980wa,Lin:1984ApJ...285..818P}. 
The binary is rotating faster than the gas parcels inside the
CBD, leading to the formation of trailing spirals, lagging
the binary's rotation. 
{The nonlinear nature of the binary-disk coupling, however, makes this aforementioned perturbative approach give qualitative incorrect results.
By carrying out long-term two-dimensional hydrodynamical simulations, \cite{Miranda2017, Munoz:2018tnj} found that the binary can extract angular momentum from the disk and harden (shrink)
its orbit. 
This finding challenges the picture of binary SMBH mergers
within a Hubble time since the inward migration of the binary is stopped
indefinitely. Further investigations on this problem found that the binary's
orbital evolution can be sensitive to binary parameters, including binary mass ratio \cite{Duffell2020,Derdzinski:2020wlw,Siwek2023b,Siwek2023a}, eccentricity
\cite{Zrake:2020zkw,DOrazio2021,Siwek2023b,Siwek2023a}; disk aspect ratio \cite{Tiede2020,Dittmann2023b,Tiede:2024mwn}, viscosity \cite{Dittmann2023,Dittmann2023b,Penzlin:2024MNRAS.532.3166P,Penzlin:2025MNRAS.537.2422P}, equation of state
\cite{Sudarshan2022,Wang2022,Wang2023,Pieren2023}, and the relative mass between the disk and
the binary when self-gravity of the gas is included \cite{Franchini2021,Bourne2023}. 
While most of these hydrodynamical simulations are carried out by
integrating out the vertical direction (two-dimensional), there are a few
studies investigating the torque evolution in full three dimensions
\citep{Moody2019,Bourne2023}. 

In reality, an effective disk viscosity can be provided by magnetohydrodynamical turbulence, among many other mechanisms. Starting with Ref. \cite{Shi:2011us}, magnetized
CBDs have been investigated in global three-dimensional
numerical simulations, in Newtonian gravity \cite{Shi:2011us,2024ApJ...964..133M} 
and both numerical \cite{2012PhRvL.109v1102F,Gold:2013zma,Gold:2014dta,Paschalidis:2021ntt,Ennoggi:2025nht} and post-Newtonian relativity \cite{Noble:2012xz,Noble:2021vfg,Avara:2023ztw}. 
The inner cavity in these works is either excised with a dipole
boundary condition \cite{Noble:2012xz,Shi:2011us} or evolved similarly as the hydrodynamical simulations, because of the short evolution
time with the central region being included and the weak magnetic field
strength to start with. 

Recently, motivated by progress with large-scale SMBH accretion simulations
\cite{Ressler:2018yhi,2023ApJ...946...26G,Guo2024,Hopkins:2023ipv,Hopkins2023,Cho:2023wqr,Cho:2024wsp,Shi:2024wgh,Shi2024}, it is
gradually becoming clear that disks around single black holes can be much more strongly magnetized than previous estimation. 
{This is consistent with} observations of {Faraday rotation by the Event Horzion Telescope Collaboration} {likely originating from} near the event horizon, {preferentially favoring a magnetically arrested accretion state around the central black hole in M87} \cite{EventHorizonTelescope:2021srq,Yuan:2022mkw}).
In \cite{Wang:2025mit}, which assumed ISM cooling rate proportional to the
square of the local density field \cite{Guo2024}, it is found that the CBD formed from a
galaxy background can be hyper-magnetized and multi-phased. But in that
simulation the binary accretion is similar to the standard and normal
evolution (SANE) \cite{Davis:2020wea} instead of magnetically arrested (MAD)
\cite{Narayan2003,Igumenshchev:2007bh,Tchekhovskoy:2011zx}.  It has also
been shown that under a locally isothermal equation of state (strong
cooling), the accretion flow around a binary system can become magnetically arrested
\cite{paper1}. 
In this regime, the cavity can
rapidly be filled with vertical net magnetic flux establishing a regime where accretion into the cavity is regulated by the magnetic field. The two mini-disks around each gravitating
components are fully disrupted and the Rayleigh-Taylor fingers are
identified inside the cavity \cite{paper1}, similar to the scenario of accretion onto magnetospheres of the neutron star (e.g., \cite{Parfrey:2017nby}), protostar (e.g., \cite{Zhu2024MNRAS.528.2883Z,Zhu:2025uhl}), or single black hole surrounded by a truncated disk (e.g., \cite{Liska:2022jdy}).

In this work, building upon the recent demonstration of circumbinary magnetically
arrested accretion flows (BMAD) in Ref. \cite{paper1}, we adopt a
more realistic initial magnetic field topology and equations of state,
systematically exploring the subsequent evolution of magnetically arrested
cavities and their interaction with the outer CBD.
To bridge the two thermodynamic extremes (isothermal and adiabatic
conditions), we adopt a simplified orbital cooling scheme following
Ref. \cite{Gammie:2001bw}.  
Our results demonstrate that the BMAD
state-characterized by (1) strong, vertically aligned magnetic flux within
the cavity region, (2) periodic eruptions of magnetic flux into the bulk of
the circumbinary disk, and (3) magnetic tower-like outflows launched above
the cavity—can be universally attained, provided the initial seed magnetic
field is sufficiently strong.

This work is structured as follows:
We begin with an executive summary in Sec. \ref{sec:MAD}. 
We present the details of our method and diagnostics used in Section
\ref{sec:methods}. The results focusing on disk properties and the MAD accretion state in Sections \ref{sec:thermo-limit} and \ref{sec:cavity}, as well as outflows and associated angular momentum transfer are shown in Sections \ref{sec:outflows} and \ref{sec:amt}, respectively. We finally conclude in Section \ref{sec:conclusion}.

\begin{figure*}
    \centering
    \includegraphics[width=0.9\linewidth]{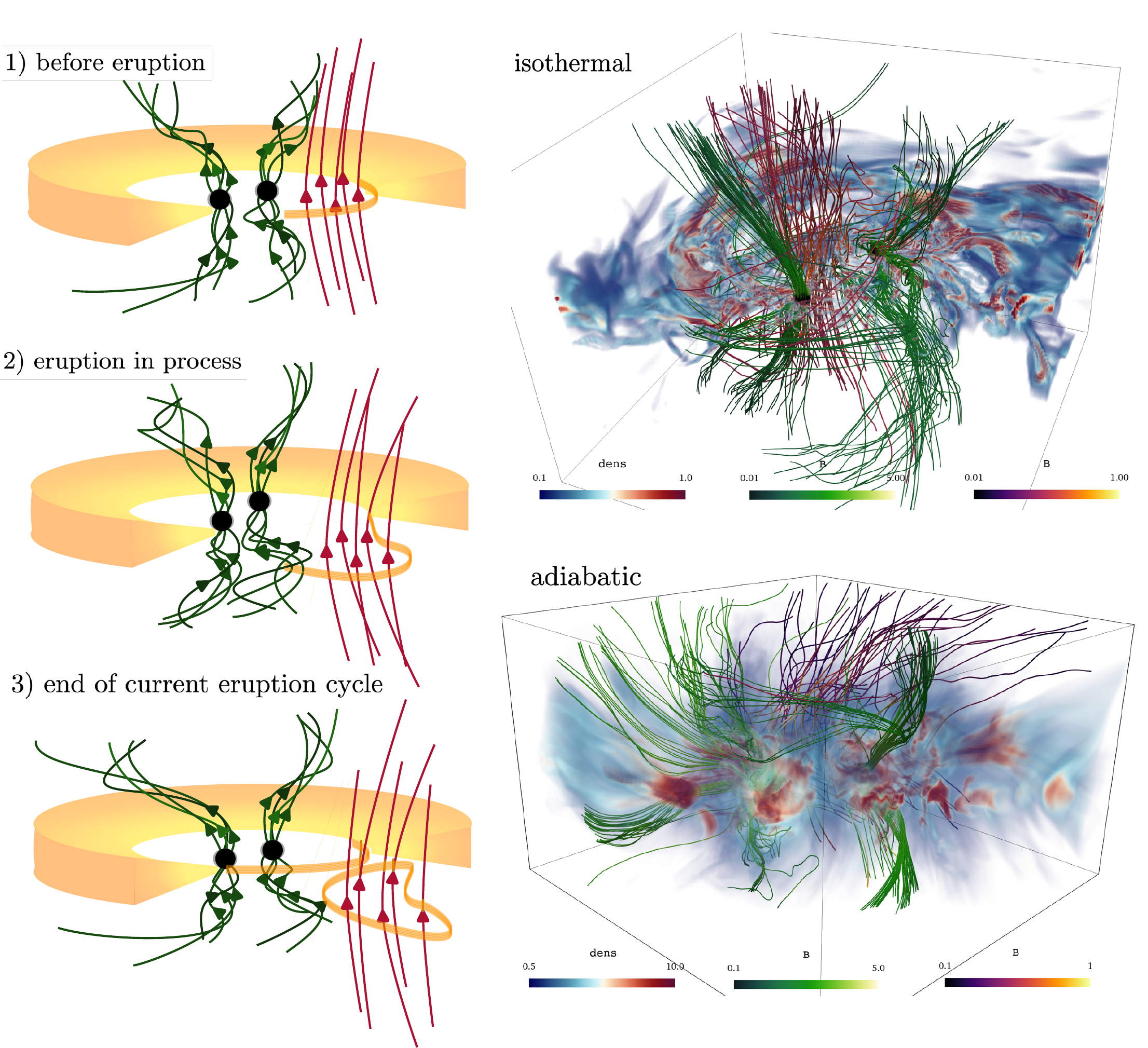}
    \caption{
    (\textit{Left}) Schematic illustration of the magnetic flux ejection process in a circumbinary magnetically arrested disk. (\textit{Top left}) Pre-eruption stage when strong vertical magnetic flux is gradually building-up at the disk cavity wall. (\textit{Center left}) Interchange instabilities at the cavity wall allow the flux to escape into the disk, (\textit{Bottom left}) the ejected magnetic flux tube is mixed into the disk. 
    (\textit{Right}) Three-dimensional volume rendering of a circumbinary magnetically arrested accretion disk simulation in a non-erupting state. 
    Shown are two different cooling states: isothermal (instant cooling; \textit{top right}) and adiabatic (no cooling; \textit{bottom right}).
    Red isosurfaces indicate concentrated mass densities in spiral accretion streams in the disk and the mini-spheres around each gravitational component. 
    Green magnetic field lines indicate magnetic tower structures directly connected to the sinks (black spheres), which drive collimated outflows. 
    Red magnetic field lines indicate large pockets of vertical net flux that regulate accretion onto the cavity and will be ejected during magnetic flux eruption events. 
    } 
    \label{fig:cartoon}
\end{figure*}

\section{BMAD Overview} 
\label{sec:MAD}
While the concept of magnetically arrested disk (MAD) accretion states has been well-established over the past two decades \cite{Bisnovatyi-Kogan:1974Ap&SS..28...45B,Bisnovatyi-Kogan:1976Ap&SS..42..401B,Narayan2003, Igumenshchev:2003rt, Yuan:2014gma, Davis:2020wea}, only recently has this framework been extended to circumbinary accretion flows \cite{paper1,Most:2024onq}. 
In this section, we provide a high-level overview of MAD and BMAD states, summarizing the underlying physical mechanisms responsible for their formation, as well as discussing their implications for the subsequent disk evolution.

\textbf{MAD state around a single black hole.} 
The MAD state has been extensively studied in the context of single accreting black holes \cite{Tchekhovskoy:2011zx, Ripperda:2021zpn}. In this scenario, the truncation of the accretion disk at the innermost stable circular orbit (ISCO) naturally leads to the formation of a central cavity. 
{In this regime, } accretion onto the black hole leads not just to mass growth, {but also to a strong} accumulation of magnetic flux at the horizon. {This leads to a magnetic pressure driven enlargement of the funnel region}.
This can temporarily halt or reduce accretion, effectively pushing the inner disk away from the black hole. 
At the same time, reconnection of magnetic field lines of opposite polarity near the equatorial plane of the black hole leads to an ejection of magnetic flux from the black hole \cite{Ripperda:2021zpn}.
These ejected magnetic flux tubes, consisting of non-anchored vertical magnetic flux, will then leave the cavity via an interchange instability  at the cavity wall \cite{Begelman:2021ufo}, regulating the accumulation of magnetic flux and periodically allowing accretion to resume \cite{Tchekhovskoy:2011zx,Porth:2020txf,Ripperda:2021zpn}. 
In this scenario, accretion will proceed through Rayleigh-Taylor fingers -- accretion streams penetrating the cavity, while simultaneously causing magnetic flux to be ejected \cite{Tchekhovskoy:2011zx,McKinney:2012vh}. 
These flux eruptions drive angular momentum transport in the disk \cite{Chatterjee2022} and can be a source of non-thermal radiation \cite{Davelaar:2019jxr,Dexter:2020cuv,Porth:2020txf,Hakobyan:2022alv,Zhdankin:2023wch}.

\textbf{BMAD state around binaries.}
Once sufficient magnetic flux accretes onto the binary, circumbinary accretion flows can behave in a similar way. 

We review the traditional hydrodynamical accretion paradigm onto the binary, which is also applicable to the magnetized standard and normal accretion (SANE) state.

A central cavity is cleared up around the binaries of comparable masses by the balance of tidal and viscous torques \cite{Artymowicz:1994bw,Artymowicz1996,2015MNRAS.452.2396M}, see, e.g., Refs. \cite{DOrazio:2012vqt,DOrazio:2015shf}. 
Two mini-disks are formed and truncated at the outer edges around each black hole by the same mechanism \cite{Artymowicz:1994bw,Artymowicz1996,2015MNRAS.452.2396M}. 
Accretion onto the BHs proceeds in tidal accretion streams penetrating the cavity, which circularizes and forms the mini-disks (see Ref. \cite{Lai:2022ylu} and references therein). 
In these hydrodynamical studies, accretion can proceed driven by the viscous stress of the disk, e.g., via turbulence triggered by the magnetorotational instability (MRI) \cite{Balbus:1991ay}. 
This picture remains qualitatively true in MHD simulations when the CBD is thermal-pressure dominated (i.e., high plasma $\beta$ case; see \cite{Avara:2023ztw}). 

\begin{figure*}
    \centering
    \includegraphics[width=0.8\linewidth]{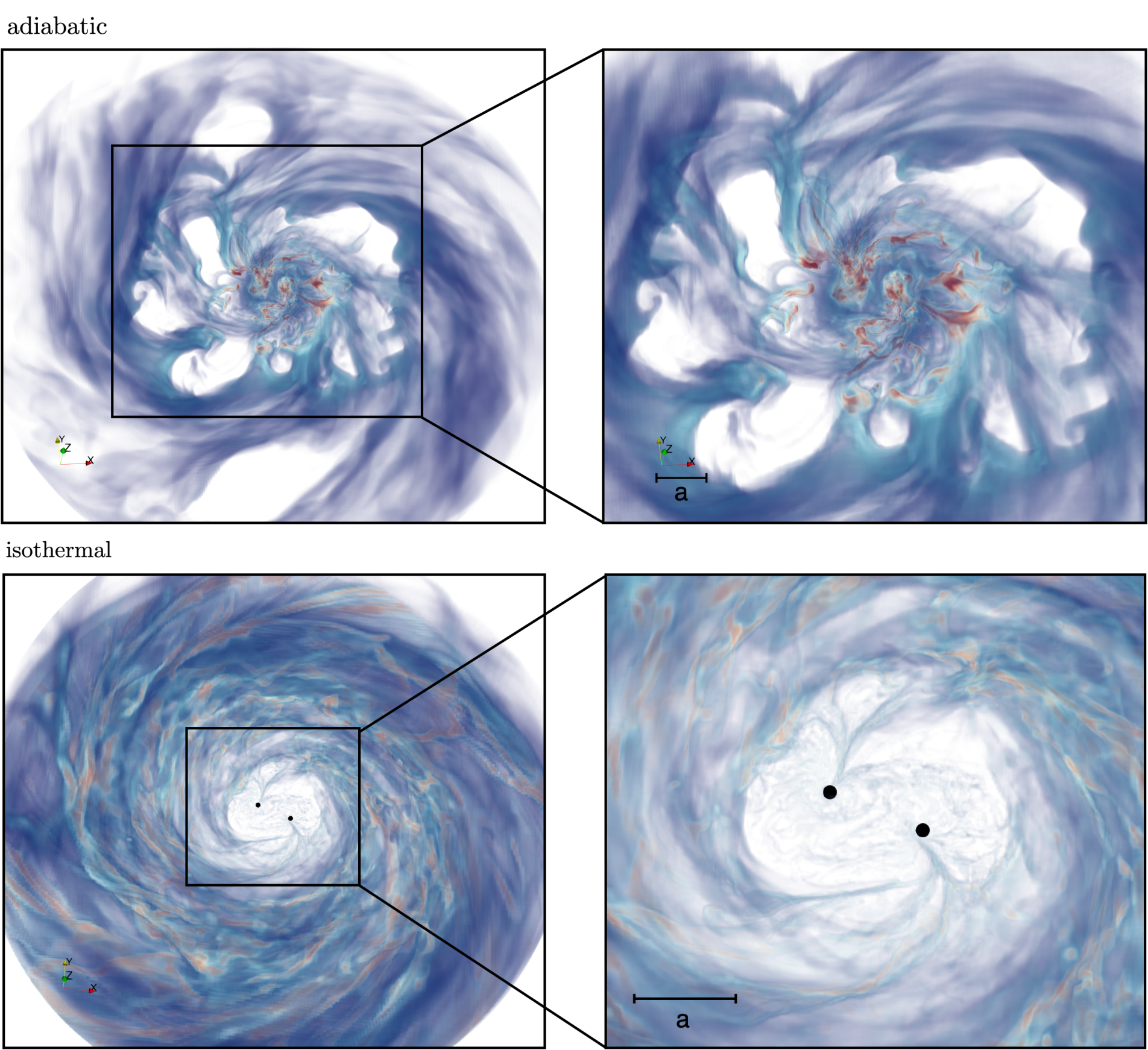 }
    \caption{Three-dimensional rendering of the quasi-steady magnetically arrested circumbinary disk shown in terms of the mass density $\rho$ around the central binary (shown with black spheres). The same colormap is used as in Fig. \ref{fig:cartoon}. 
    {\it (Top)} Instant cooling (locally isothermal). Spiral accretion streams feed the accretion sinks (black spheres) through the central cavity. In the strongly magnetically arrested regime, mini-disk formation is suppressed (disrupted by flux eruption at each sink). Density fluctuations near the cavity wall result from a consequence of magnetic flux eruptions, governing the exchange between the net vertical magnetic field between the cavity and the disk. But instant cooling in the disk causes the collapse of the flux tubes once they leave the cavity and maintains a quasi-stable cavity size. 
    {\it (Bottom)}  
    {Adiabatic system. Instead of being cleared by tidal forces and forming an eccentric edge, the cavity becomes a combination of multiple large magnetic flux bundles.
    The resulting disk edge location is substantially larger than the inner Lindblad radius, and large errupted flux tubes can easily travel outward.}
    The cavity is filled with hot gas, surrounding each sink. 
    More specifically, each sink is contained inside a almost-spherical hot gas sphere,which due to the lack of cooling cannot thermally collapse into a mini-disk.}
    \label{fig:overview-3D}
\end{figure*}

It has recently been demonstrated that this paradigm changes qualitatively when the CBD is initially threaded with a sufficiently strong magnetic field, leading to the establishment of a magnetically arrested state of the CBD \cite{paper1}.
We show an example of a BMAD rendering in Fig. \ref{fig:cartoon} as in Ref. \cite{paper1}.
In the MAD state, the strong magnetic field regulates the accretion dynamics. 
Here, we further elaborate and refine the interpretation of the MAD and BMAD states.
In particular, for the BMAD scenario, we emphasize the  hierarchical nature of magnetic flux eruptions, occurring both individually around each black hole and collectively at the CBD cavity.

\subsection{Flux Eruption Cycles}
\textit{Flux eruption on `horizon/ISCO' scales.} 
In the single BH MAD state, flux eruptions happen near the innermost stable circular orbit (ISCO). 
Assuming realistically the individual BHs in a SMBHB system would be in a MAD state, ejected magnetic flux tubes could propagate out to the Bondi scale \cite{Cho:2024wsp,Guo:2025sjb,Lalakos:2025msz}, and in the case of the SMBHB accumulate inside the cavity filling it with net vertical flux (see also Ref. \cite{Ennoggi:2025nht}).
This picture naturally carries over to our Newtonian simulations as well. 
Over time each sink becomes magnetically saturated and threaded by vertically aligned magnetic fields.
Reconnection near the sinks leads to an ejection of net vertical flux that then accumulates inside the cavity, over time driving it to a saturated state of the cavity itself.

\textit{Flux eruptions at the cavity scale.} 
A crucial feature of the magnetically arrested cavity is that it is only intermittently stable. 
Though being perturbed by the time-varying tidal potential of the binary, the quasi-periodicity switching from the quiescent state and the eruption state is determined by the timescale of magnetic saturation of the cavity (see Figs. \ref{fig:cartoon} and \ref{fig:overview-3D}).
When accreting from the disk onto the cavity, the cavity wall is subject to a magnetic interchange instability \cite{Spruit:1995fr}. 
This instability, also present in single accreting BH in a MAD state, allows for the accretion into the cavity via Rayleigh-Taylor fingers \cite{Zhdankin:2023wch, Kulkarni:2008vk,Zhu2024MNRAS.528.2883Z,Zhu:2025uhl}, also observed in accreting young stellar objects (e.g., T Tauri stars \cite{Takasao:2022glf}), or in low mass x-ray binaries with a neutron star accretor \cite{Parfrey:2023swe}. 
At the same time, as matter is being accreted into the cavity, a magnetic flux bundle (flux tube) is ejected into the disk (see cartoon in Fig. \ref{fig:cartoon}). 
This ejection has two important implications:
First, the flux tube is filled with hot, reconnected plasma \cite{Ripperda:2021zpn}. 
In the context of Sgr A* observations of the GRAVITY probe \cite{Gravity2018,Gravity2021}, it has been suggested that the these flux tubes may be ideal sites for particle acceleration and non-thermal emission \cite{Dexter:2020cuv,Porth:2020txf,Ripperda:2020bpz,Ripperda:2021zpn}, e.g., through a Rayleigh-Taylor instability of the flux tube itself \cite{Zhdankin:2023wch}, potentially power TeV flares \cite{Hakobyan:2022alv}.  {Though it is worth noticing that the magnetic field environment is substantially different from the well-studied split monopole geometry in the near-event horizon regime \cite{Tchekhovskoy:2009ba,Ripperda:2021zpn}.
}

Another important aspect of magnetic flux eruptions is their ability to regulate angular momentum transport in the wind and the disk \cite{Chatterjee2022,Manikantan:2023vcw}, as the vertical field in the magnetic field bundle can much more efficiently transport angular momentum vertically comparing to in-plane viscous transport \cite{Blandford:1982xxl,Bai:2013pi}. It is worth noticing that this picture is fundamentally distinct from naively tuning the viscosity in the hydrodynamical simulations \cite{Duffell:2019uuk,Dittmann:2022obl,Dittmann:2023ztg,Penzlin:2024MNRAS.532.3166P,Penzlin:2025MNRAS.537.2422P}, as the MAD magnetically field dynamics are fundamentally a non-perturbative far-from-equilibrium effect, that cannot be captured with a viscous prescription.

\subsection{Magnetic Field Structure Inside the Cavity}
We provide a detailed description of the properties of the magnetic field inside the cavity of CBD in the quasi-steady state.
As we will demonstrate in this work, the cavity dynamics will depend strongly on the equation of state (which in turn reflects the efficiency of radiative cooling). 
This can quantitatively change the physical systems the BMAD state is applicable to. 
For clarity, we illustrate here two ends of this thermodynamics spectrum, strong/instant cooling (i.e., locally isothermal equation of state), and no cooling (i.e., adiabatic equation of state), which we also show in Fig. \ref{fig:overview-3D}.

\textit{Instant cooling.}
After accreting a sufficient amount of vertical magnetic flux, the cavity settles into a quasi-steady state configuration shown in Fig. \ref{fig:cartoon}. 
Focusing first on the magnetic field configuration, we can identify two regions:
Field lines threading the sinks (shown in green), and those constituting non-anchored vertical net flux inside the cavity (shown in yellow).
Those field lines conceptually behave very differently. 
The field lines directly connected to the sinks form a magnetic tower \cite{Lynden-Bell:1996MNRAS.279..389L,Lynden-Bell:2003MNRAS.341.1360L}, with rotation at the base creating a helically wound magnetic field structure \cite{Blandford:1977ds,Blandford:1982xxl}. The effective magnetic pressure gradient in the tower then drives a net outflow \cite{Blandford:1982xxl,Heyvaerts:1989ApJ...347.1055H,Bromberg:2016MNRAS.456.1739B,Lynden-Bell:1996MNRAS.279..389L,Lynden-Bell:2003MNRAS.341.1360L}.
Different from, e.g., a Blandford-Znajek jet \cite{Blandford:1977ds}, the outflow is not driven by the sink itself, but by the plasma supplied from the accretion flow, making it intermittent and strongly dependent on the feeding of the sink, see also Ref. \cite{paper1}.

Free net vertical magnetic flux in the cavity has a very different dynamics. It accumulates inside the cavity up to the point where the pressure balance at the cavity wall will become unstable to an interchange instability \cite{Spruit:1995fr}. This will lead to an intermittent flux ejection (and subsequent mixing dependent on the thermodynamics as we will discuss later, see also Fig. \ref{fig:overview-3D}).

The origin of this flux is likely associated with reconnection near the sink (in reality with eruptions from the BH which can travel out to the Bondi radius \citep{Lalakos:2025msz}.), or at transiently formed mini-disks.
The overall cavity structure is closely resembling magnetically truncated disks \cite{Liska:2022jdy}, or magnetically mediated accretion in the ergomagnetosphere scenario in single BH accretion \cite{Blandford:2022Galax..10...89B,Blandford:2022MNRAS.514.5141B}. 
There, the jet region is separated from a truncated accretion disk through vertical net flux extending beyond the ISCO.
However, in contrast to the ergomagnetosphere proposal, the presence of tidal accretion streams (red regions in Fig. \ref{fig:cartoon}) in the circumbinary make the cavity highly asymmetric and transient in nature, as we will describe in the following.

\textit{Adiabatic equation of state (no cooling).}
The overall cavity dynamics in the adiabatic/no-cooling case is very similar, with differences concerning primarily the plasma filling of the cavity, and additional thermal support of ejected flux tubes.
As we can see in Fig. \ref{fig:cartoon}, the individual sinks continue to launch tower-like outflows. However, since the sinks are always surrounded by gaseous halo-like mini disk (which cannot thermally collapse due to lack of cooling), the tower outflows will feature less intermittency. 
The cavity itself is now threaded with hot gas, and never fully clears out as in the strongly cooled scenario.
The cavity truncation radius is now also further out compared to the strongly cooled case, and the cavity wall features strong eruption dynamics. Ejected magnetic flux tubes now have a strong thermal support, allowing them to travel large distances inside the accretion disk after being ejected (Fig. \ref{fig:overview-3D}).

\section{Methods} 
\label{sec:methods}
In this work, we perform ideal Newtonian magnetohydrodynamic simulations \citep{Stone:2008mh} of circumbinary accretion flows around an equal-mass binary on a circular orbit. 
To this end, we model the accretion flow as a fluid specified by mass density $\rho$, gaseous pressure $P$, velocity $\boldsymbol{v}$, internal energy density $e$, and magnetic field $\boldsymbol{B}$,
\begin{align}
 \frac{\partial \rho}{\partial t}+\nabla \cdot(\rho \boldsymbol{v}) &=s_{\rho}\,\label{governeq-1}, \\
 \frac{\partial(\rho \boldsymbol{v})}{\partial t}+\nabla \cdot(\rho \boldsymbol{v} \boldsymbol{v} + \bar{P} \mathbb{I} - \boldsymbol{B}\boldsymbol{B})&=\boldsymbol{s_p}-\rho \nabla \Phi\,\label{governeq-2}, \\
 \frac{\partial E}{\partial t}+\nabla \cdot[(E+\bar{P}) \boldsymbol{v} - \left(\boldsymbol{B}\cdot \boldsymbol{v}\right) \boldsymbol{B}]&=s_E-\rho \boldsymbol{v} \cdot \nabla \Phi + \Lambda\,\label{governeq-3},\\
 \frac{\partial \boldsymbol{B}}{\partial t}-\nabla \times ( \boldsymbol{v} \times \boldsymbol{B} )&=0\,\label{governeq-4},
\end{align}
where we have used the conserved energy $E$, and effective pressure $\bar{P}$,
\begin{align}
    E = e + \frac{1}{2} \rho v^2 + \frac{1}{2}B^2\,,\\
    \bar{P} = P + \frac{1}{2}B^2\,.
\end{align}
We have also introduced effective sink, $s_i$, and cooling terms, $\Lambda$,  which are going to discuss in the following section.
The gravitational potential of the binary is modeled as softened point masses
\begin{align}
  \Phi = 
  - \frac{GM_1}{(\left|\boldsymbol{r}-\boldsymbol{r}_1\right|^2 + \epsilon^2)^{1/2}}
  - \frac{GM_2}{(\left|\boldsymbol{r}-\boldsymbol{r}_2\right|^2 + \epsilon^2)^{1/2}}\,,
\end{align}
where $M_i$ and $\boldsymbol{r}_i$ are the location and mass of the binary
constituents respectively, $G$ is the gravitational constant, $\boldsymbol{r}$
is the radial coordinate to the center of mass of the binary, and $\epsilon=0.05a$ is the plummer gravitational softening length \cite{paper1}. 
We consider an equal mass binary, $M_1=M_2=M/2$ where $M$ is the total mass.
A unit convention of $G=M=a=1$ is adopted in our simulations. 

\begin{table}
\centering
\begin{tabular}{ c c c c c c  }
\hline
label & $B$ \ topology & $\beta$  & EoS & $\beta _{\rm cooling}$ & $N_{\rm orbit}$\\
\hline
\texttt{iso-p} & poloidal & 20 & isothermal & 0  & 300\\
\texttt{iso-t} & toroidal & 20 & isothermal & 0 &   300\\
\texttt{adi-p} & poloidal & 20 & adiabatic  & 0 &   300\\
\texttt{adi-t} & toloidal & 20 & adiabatic  & 0  &   300\\
\hline
\texttt{adi-p-c0.6} & poloidal & 20 & adiabatic & 3/5    & 300+100\\
\texttt{adi-p-c10 } & poloidal & 20 & adiabatic & 10     & 300+100\\
\hline
\end{tabular}
\caption{Simulations presented in this work. These are characterized by their magnetic field ($B$) topology, the initial plasma $\beta$-parameter, the equation of state (EoS), the cooling parameter, $\beta_{\rm cool}$, and the number of orbits, $N_{\rm orbit}$. Note that the simulations with $\beta-$cooling where started based the isothermal configuration, \texttt{adi-p}, after 300 binary orbits.  All simulations reach a circumbinary magnetically arrested disk (BMAD) state.}
\label{tab:runs}
\end{table}

\subsection{Simulation setup}

In this Section, we briefly describe the numerical setup used in our simulations, which are summarized in Table \ref{tab:runs}. Since we are interested in difference cooling regimes, we adopt three fundamental prescriptions to capture disk thermodynamics. 

The choice of equation of state (EoS) closes the governing equations \eqref{governeq-1}-\eqref{governeq-4}.
The isothermal or adiabatic equation of states, as two ideal thermodynamic limits, are equivalent to either instant cooling or no cooling in the accretion disks, respectively. 
While a full radiative MHD treatment would be desirable \cite{Jiang:2014tpa,2023ApJ...949..103W,Zhang:2025uug}, the sometimes prohibitive computational cost of such simulations forces one to either adopt one of the two limiting cases (e.g., see \cite{Lai:2022ylu} and references therein), or choose a manually prescribed cooling time. 
The approach we use is similar to that in Ref. \cite{Wang2023}. 

For the strongly cooled case, we use a locally isothermal prescription commonly employed in circumbinary accretion \citep{Munoz:2016ApJ...827...43M,Munoz:2018tnj,Miranda2017}, which is valid for optically thin disks and when the cooling timescale of the disk is short. 
This amounts to setting $P=T_{\rm iso} \left(R\right) \rho$, the temperature, $T_{\rm iso}$, is forced to be only a function the fluid location projected onto the disk midplane in the corotating frame of the binary $T_{\rm iso}= c_s^2 = \left|\Phi_{\rm mid}\right|/\mathcal{M}^2$, where the sound speed is $c_s$ and $\mathcal{M}=10$ is the sonic Mach number (equivalent to disk aspect ratio $h=1/\mathcal{M}=0.1$). \\

In line with two-dimensional studies, we adopt an equatorial density profile (see Ref. \cite{Duffell:2024fwy} for a recent comparison),
\begin{equation}
    \Sigma_{\mathrm{iso}}\left(R\right) = \Sigma_0 \exp{\left[ -\left( \frac{R}{R_{\mathrm{edge}}} \right)^{-6} \right]}\,,
\end{equation}
where $R_{\rm edge}=2.5a$ is the effective disk truncation radius, and $\Sigma_0=1.0$. The vertical scale height of the disk is determined by imposing vertical hydrostatic equilibrium at constant temperature $T(R,z)= T_{\rm iso} = c_s^2 = \rm const$ at fixed $R$,
\begin{align}\label{eqn:iso_disk}
\rho_{\rm iso} \left(R,z\right) =  \Sigma_{\mathrm{iso}}\left(R\right) \exp\left[ -\frac{1}{c_s^2} \left( \Phi(R, z) - \Phi(R,0) \right) \right]
\end{align}
We then assign an energy density by adopting an ideal fluid prescription, $P = e \left(\gamma-1\right)$, where $\gamma=4/3$ is chosen assuming the disk is supported by radiation pressure.

For the other cases, we adopt an adiabatic (uncooled) prescription with an optional cooling function added. For the purely adiabatic case only, we set up an initial isentropic profile via a polytropic equation of state $P = \kappa \rho ^ \gamma $ for the evolution, and a distinct initial (vertical) density profile from hydrostatic equilibrium as follows
\begin{align}
    \rho(R, z) 
    = &\Sigma_{\mathrm{iso}}(R)\times \\
    &\left[ 1 + \frac{\left(\gamma-1\right)}{c_{s,0}^2} \left( \Phi\left(R,z\right) -  \Phi\left(R,0\right)\right)\right]^{1/(\gamma-1)}\,,
\end{align}
where $c_{s,0}^2 = \gamma \kappa \Sigma_{\rm iso}^{\gamma-1}$, and $\kappa = c_{s,{\,\rm edge}}^2 \rho(R_{\mathrm{edge}},0)^{1-\gamma}$ can be fixed via the sound speed, $c_{s,{\,\rm edge}}^2$, at the cavity wall, $R=R_{\rm edge}$.
The polytropic constant, $\kappa$, is fixed by adjusting the sound speed at the edge of the cavity, $R=R_{\mathrm{edge}}$.
A detailed derivation is provided in Appendix \ref{app:id}.
For all cases, the initial radial and vertical velocities are set to zero, and angular velocity slightly modified from the local Keplerian angular velocity (see e.g., \cite{Duffell:2024fwy}). 

For some of the adiabatic models, we systematically relax the instant cooling assumption of the isothermal case by adopting a $\beta$-cooling prescription \cite{Gammie:2001bw},
\begin{align}
    \Lambda = - \frac{\Sigma}{\gamma -1} \left(T-T_{\rm iso}\right)\frac{\tilde{\Omega}_K}{\beta}\,,
\end{align}
where $\beta$ is the cooling coefficient, and $\tilde{\Omega}_K=[(GM_1/|\boldsymbol{r}-\boldsymbol{r}_1|^3)^2+(GM_2/|\boldsymbol{r}-\boldsymbol{r}_2|^3)^2]^{1/4}$ is the effective orbital rotation rate \cite{Wang2023}\footnote{Ref. \cite{Wang2023} contains a typographical error in their Eq. (18). The effective orbital rotation rate is the same the one used here.}. The cooling coefficient controls how quickly we approach the isothermal case. When adopting a dynamical cooling prescription, we initialize our simulations using the `adiabatic' simulation after $300$ binary orbits.

\subsubsection{Magnetic field topology} \label{sec:ini-magtop}
{Since the BMAD regime depends critically on the establishment of strong net vertical magnetic flux {inside the cavity, it is important to understand the impact of magnetic flux supply from large scales or small-scale dynamo/conversion processes.}.} 
{{Studies of single BH} MAD states found that net vertical magnetic flux supply to the black hole can guarantee reaching a MAD state \cite{Tchekhovskoy:2011zx,Tchekhovskoy:2012MNRAS.423L..55T,McKinney:2012vh}. 
Later it was also found that starting from purely toroidal flux, the disk can generate large-scale poloidal magnetic flux in situ {due to disk dynamo processes} and reach a MAD accretion state \cite{Liska:2018btr,Jacquemin-Ide:2023qrj}.
Here, to achieve a BMAD state characterized by sufficiently supply of net vertical magnetic flux within the disk cavity, we begin our simulations with disk threaded by poloidally dominated flux.}
Additionally, we vary the initial magnetic field topology in the circumbinary disk between purely poloidal and the mixture of toroidal and poloidal fields. The vector potential we use is given by 
\begin{equation}
    \boldsymbol{A} =(1-\eta) A_\phi \hat{\phi} + \eta A_z \hat{z}\,,
\end{equation}
where  $A_\phi = A_z= R^2 A_0 {\max} \left(\rho - 0.04\,\rho_{\max}, 0\right)^2$, and $\eta=0$ corresponds to a purely poloidal field, and $\eta=1$ to a purely toroidal field. 
All our simulations retain an initially poloidal contribution, $\eta < 1$, in particular, we pick $\eta=0.9$ in toroidal field dominated simulations. Here $A_0$ is chosen to ensure that {the initial disk magnetization parameter $\beta_{\rm ini}=2P_{\rm max}/B_{\rm max}^2 \simeq 20$}, consistent with our previous study on BMAD systems \cite{paper1}{\footnote{In \cite{paper1}, the actual initial disk magnetization is $\beta=24$. But the results obtained are consistent after reaching a quasi-steady state.}}.

 \begin{figure}
    \centering
    \includegraphics[width=0.85\linewidth]{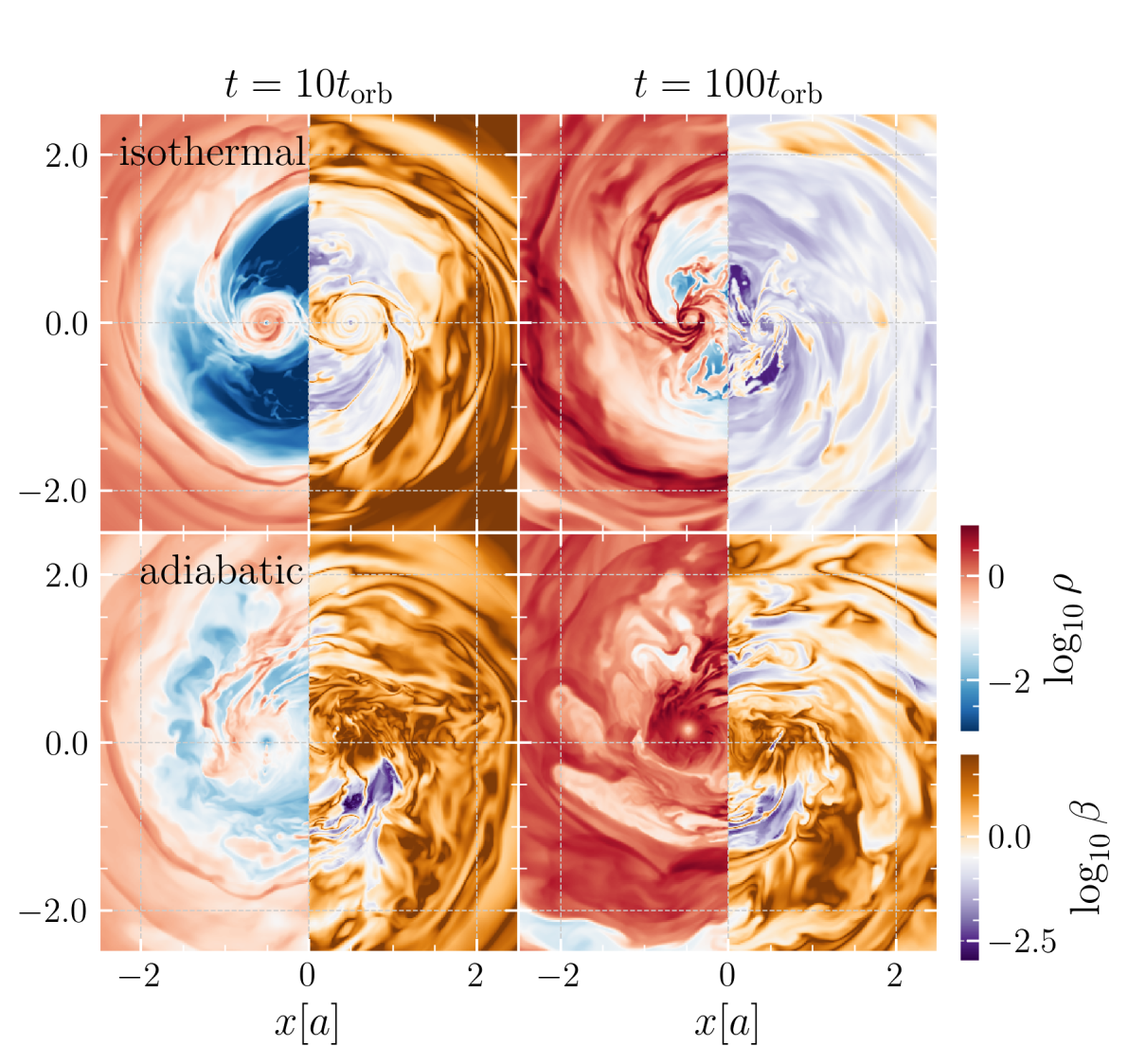}
    \caption{Initial circumbinary disk evolution towards a magnetically arrested state, represented by the mass density, $\rho$ ({\it upper half panel}) and plasma parameter, $\beta$ ({\it lower half panel}) in the disk midplane at 10 and 100 binary orbits, respectively. We separately show models for purely poloidal initial magnetic fields, but varying the equation of state to be isothermal (\texttt{iso-p}), and adiabatic (\texttt{adi-p}). Evolution time $t$ and distance $r$ are stated relative to the binary orbital period $t_{\mathrm{orb}}$ and binary separation $a$. 
    } 
    \label{fig:wtm}
\end{figure}

\begin{figure}
    \centering
    \includegraphics[width=0.99\linewidth]{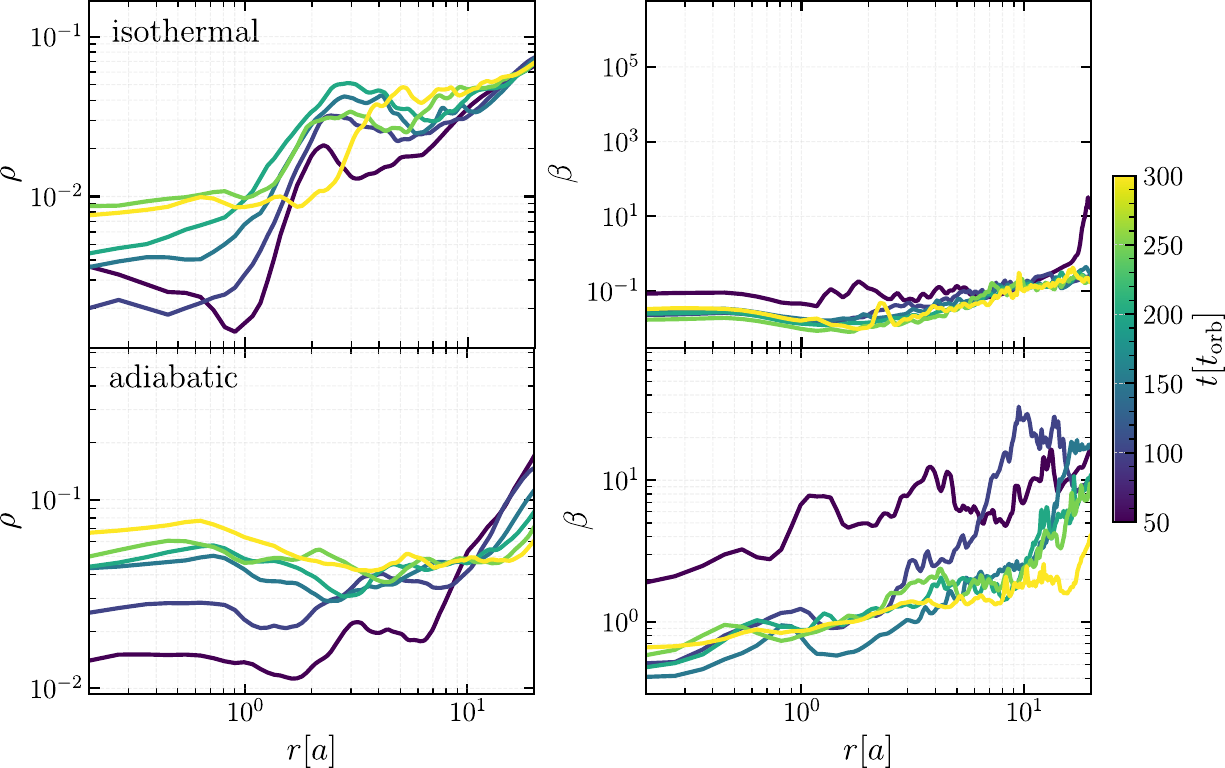}
    \caption{
    Radial profiles of mass density, $\rho$, and plasma parameter, $\beta$, throughout the evolution of the binary for models \texttt{iso-p} (isothermal) and \texttt{adi-p} (adiabatic). 
    Shown quantities are averages over azimuthal and vertical direction in cylindrical coordinates.
    Times (in color) are stated relative to the binary orbital period, $t_{\rm orb}$.
    } 
    \label{fig:wtm-1d}
\end{figure}

\subsubsection{Sink Prescription}\label{sec:sink}
Since our binary gravitational sources accrete matter, we need to provide an effective mass sink prescription. 
However, the orbital separation is much larger than the size of the event horizons of the supermassive black hole binaries, so that an effective sink prescription is neither trivial or unique (see, e.g., Refs. \cite{2020ApJ...892L..29D,Dittmann:2021wzj} for a detailed discussion in the vertically integrated setup).
In practice, we choose a simple mass removal prescription draining mass proportional to the local Keplerian rate,
$$
\begin{aligned}
s_{\rho} & =s_{\rho, 1}+s_{\rho, 2} \\
& =-({\rho}\Omega_{K,1}+{\rho}\Omega_{K,2}) \,,\\
\boldsymbol s_p & =\boldsymbol s_{p, 1}+\boldsymbol s_{p, 2}\\
& =-({\rho}\boldsymbol{v}_1^{\mathrm{res}}\Omega_{K,1}+{\rho}\boldsymbol{v}_2^{\mathrm{res}}\Omega_{K,2})\,, \\
s_E & =s_{E, 1}+s_{E, 2} \\
& =-({\rho}E_1^{\mathrm{res}}\Omega_{K,1}+{\rho}E_2^{\mathrm{res}}\Omega_{K,2})\,, \\
\boldsymbol{v}_i^{\mathrm{res}} & =\left(\boldsymbol{v}-\boldsymbol{v}_i\right) \,, \\
E_i^{\mathrm{res}} & =\frac{1}{2} \boldsymbol v_i^{\mathrm{res}} \cdot \boldsymbol v_i^{\mathrm{res}}+\frac{P}{(\gamma-1)\rho}\,,
\end{aligned}
$$
where the sink radius is $r_s=0.07a$. 
The removal rate is chosen to be the local Keplerian frequency around each sink $\Omega_{K,i} = (GM_i)^{1/2}/(\left|\boldsymbol{r}-\boldsymbol{r}_i\right|^2 +\epsilon^2)^{3/2}$. 
To avoid material accumulating inside the sink region, we directly set the density, momentum, and energy {to the floor value} within a depletion radius $r_{\mathrm{dep}}=0.5r_{\mathrm{s}}$.

\subsection{Numerical setup}
In this work, we use the new performance portable version, \texttt{AthenaK}\footnote{\url{https://github.com/IAS-Astrophysics/athenak}} \cite{athenak}, of the \texttt{Athena++} code \citep{Stone2020}. \texttt{AthenaK} is based on the \texttt{Kokkos} library \citep{Trott2021} and allows us to run our simulations on hundreds to thousands of GPUs. We use piecewise parabolic reconstruction \citep{Colella1984}, an HLLD Riemann solver \citep{Miyoshi2005}, and a constraint transport algorithm \citep{Gardiner2008} for the divergence-free magnetic field evolution.
Similar to our previous study \cite{paper1}, we adopt a Cartesian grid with 6 levels of mesh refinement, an outer boundary of the domain of $\pm 80 a$, and a finest resolution of $\Delta x \simeq 0.0098 a$ covering the innermost domain of $[-1.25a, 1.25a]\times[-1.25a, 1.25a]\times[-1.25a, 1.25a]$ . 
We then numerically integrate the system on 600 V100 GPUs on the DOE OLCF Summit system for 300 binary orbits. Overall, the cost of the simulation presented here is about 130k V100 GPU-hours.

\begin{figure*}
    \centering
    \includegraphics[width=0.95\linewidth]{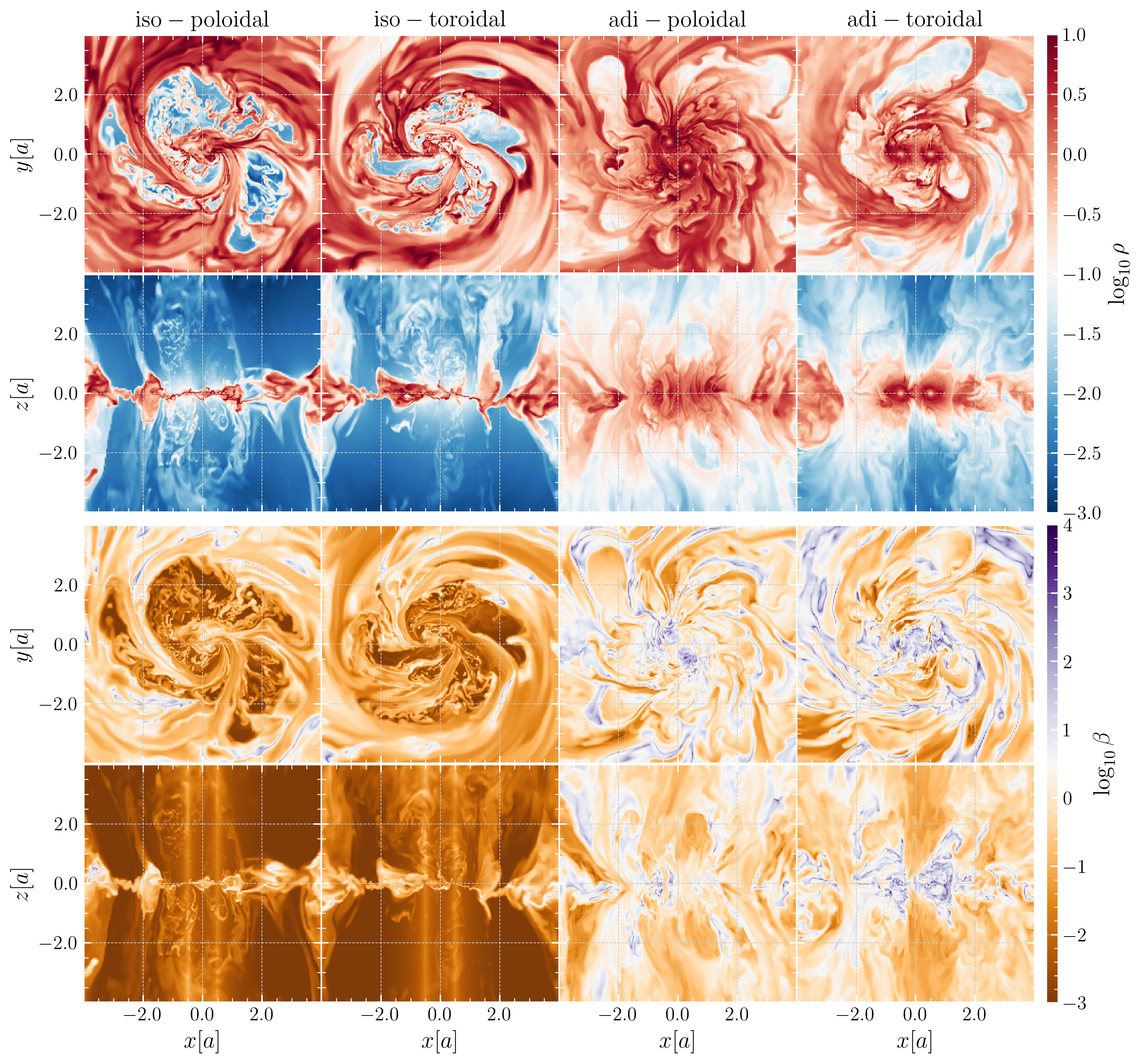}
    \caption{
    Magnetically arrested cavities across different thermodynamics (isothermal (iso), and adiabatic (adi)) and initial magnetic field topology regimes (poloidal and toroidal). 
    {All cases have reached a quasi-steady state and the cavity edges are unstable to magnetic flux eruptions. 
    In the isothermal cases, the central cavity is clearly distinct from the disk and strongly magnetized with plasma-$\beta\sim10^{-3}$. A thin current sheet extends through the whole cavity.
    In contrast, in the adiabatic cases the cavity is composed of a combination of flux bundles where the magnetic and thermal pressure roughly reach equipartition. The cavity is also more extended.}
    }
    \label{fig:all-summary}
\end{figure*}

\begin{figure}
   \centering
   \includegraphics[width=1.0\linewidth]{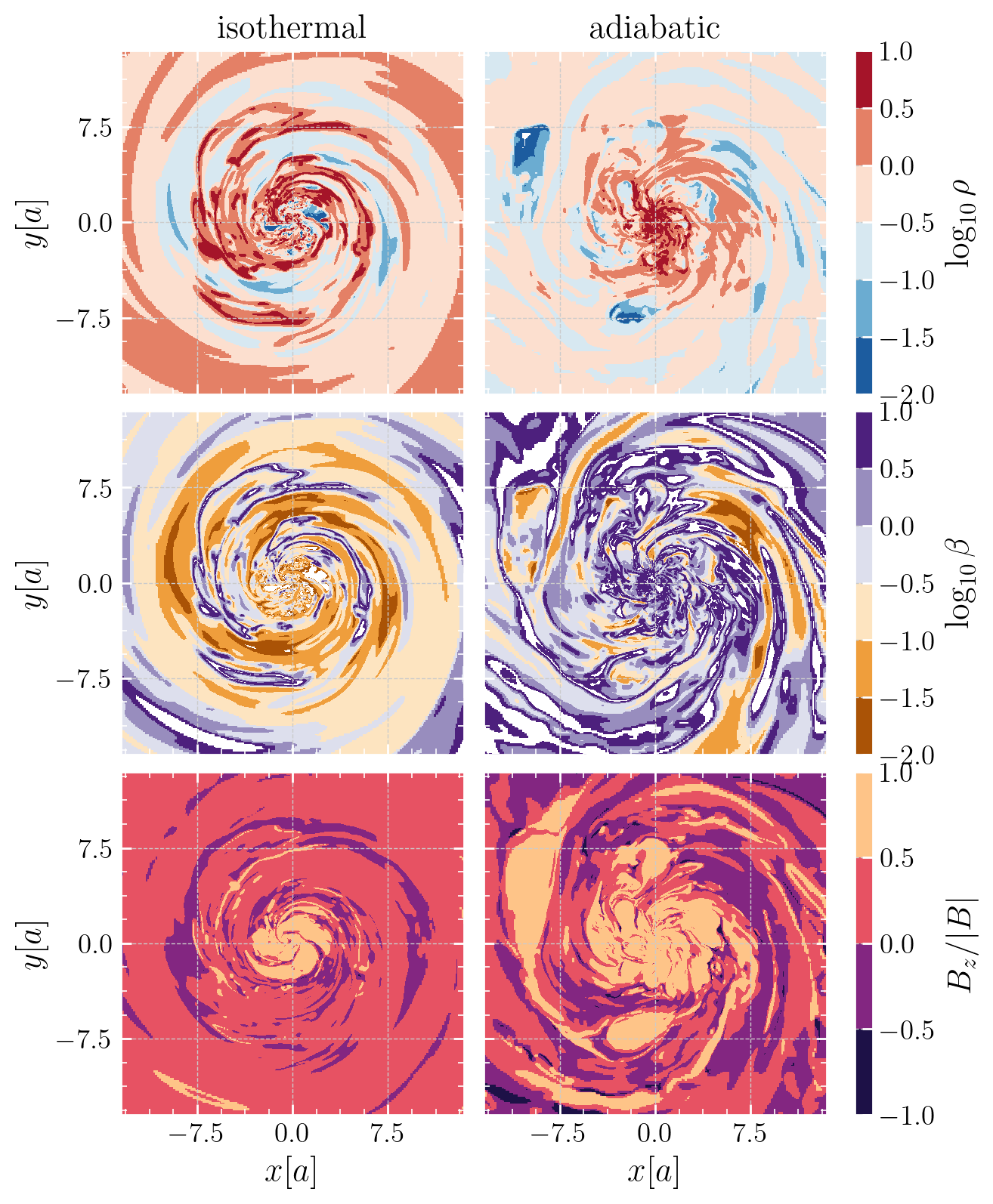}
   \caption{Flux eruption properties for two thermodynamic limits. Discrete colors are used to highlight the cavity and coherent flux tube locations. From top to bottom are the density $\rho$, plasma parameter $\beta$, and the normalized vertical magnetic field component $B_z/|B|$. In the isothermal scenario, vertical magnetic flux remains confined within a compact central cavity, approximately 2-3 times the binary separation in radius, and no outward-traveling flux tubes are observed. In contrast, the adiabatic case exhibits a significantly larger and more irregular cavity shape, with a radius extending to roughly 5-6 binary separations. Multiple outward-propagating flux tubes are visible, including prominent structures located toward the right side and at the bottom center of the panel. The distorted cavity in the adiabatic case actively generates these flux tubes, highlighting the dynamic nature of the flux eruption process under adiabatic conditions.} 
    \label{fig:fe-cycle}
\end{figure}

\begin{figure*}
    \centering
    \includegraphics[width=\linewidth]{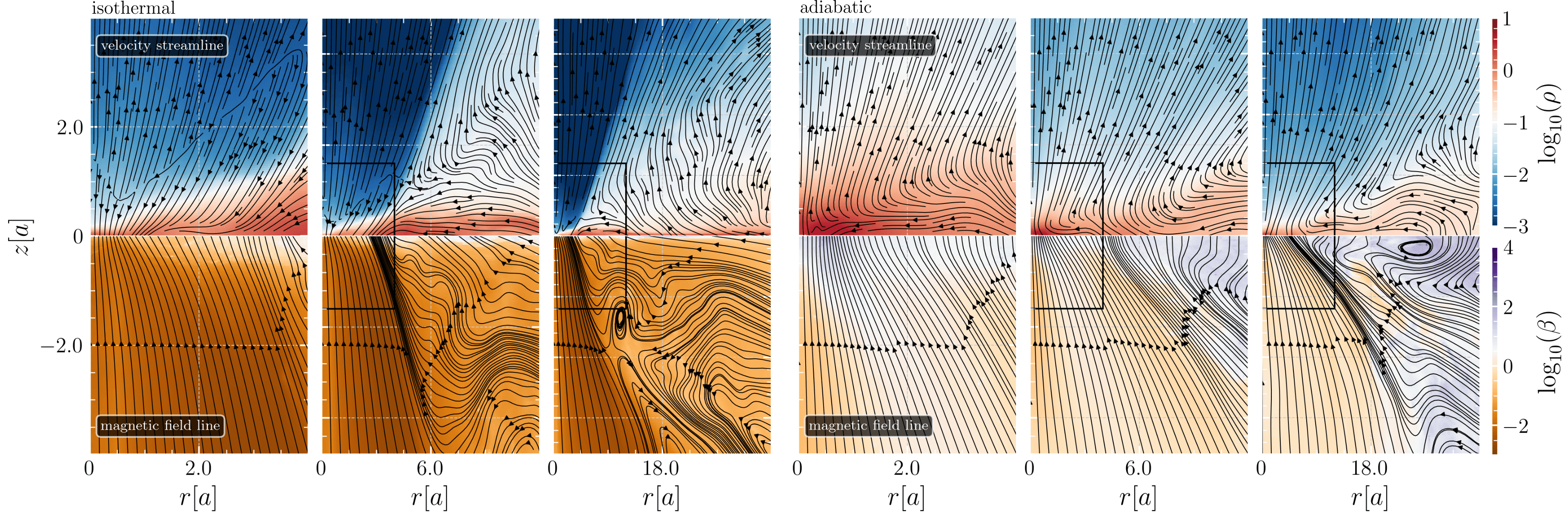}
    \caption{
    Time and azimuthally averaged circumbinary disk profiles in the magnetically arrested regime for an isothermal (\texttt{iso-p}, instant cooling, left) 
    and adiabatic (\texttt{adi-p}, no cooling, right) magnetically arrested accretion flow. 
    Shown quantities are the mass density $\rho$ (\textit{Upper half}) and plasma-$\beta$ (\textit{Lower half}) on increasing radial scales (\textit{Left to Right}). Velocity and magnetic field lines are overlaid on top of the upper and lower plane, respectively.
    For the isothermal case ({\it left}) We can clearly identify two regions, with a transition at $3a$: the cavity (inner region) features a large coherent amount of net vertical magnetic flux; the disk region consists of a thermally collapsed midplane with a vertically thick outflowing magnetized atmosphere,
   with an inclination angle between $40-42\degree$ be clearly identified.
   For the adiabatic case ({\it right}), the cavity is twice the size, at $6 a$, as shown by the plasma-$\beta$ transition. 
   The density transition between the funnel region and the magnetized disk atmosphere becomes smooth.  
  The disk scale height is also extended compared to the isothermal case.
    }
    \label{fig:iso-p-wind}
\end{figure*}

\begin{figure*}
    \centering
    \includegraphics[width=0.8\linewidth]{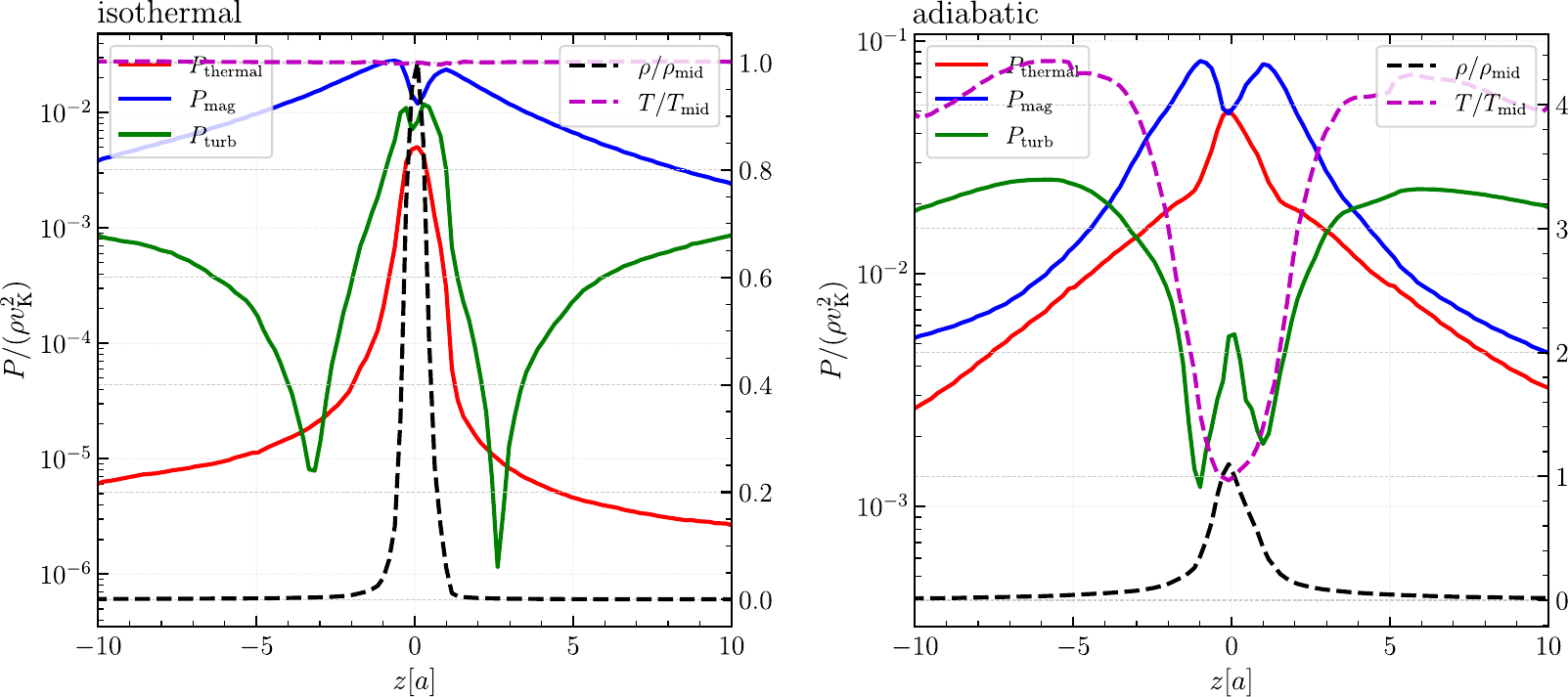}
    \caption{
    Time and azimuthal averaged vertical profile of mass density, $\rho$, temperature, $T$, thermal pressure, $P_{\rm thermal}$, magnetic pressure, $P_{\rm mag}$, and turbulent pressure, $P_{\rm turb}$. $\rho_{\rm mid}$ and $T_{\rm mid}$ are the relevant quantites on the disk midplane ($z=0$).  All quantities are measured near the cavity wall at $r=3a$ for models \texttt{iso-p} (isothermal) and \texttt{adi-p} (adiabatic).
    } 
    \label{fig:disk_scale_height}
\end{figure*}

\begin{figure*}
    \centering
    \includegraphics[width=.9\linewidth]{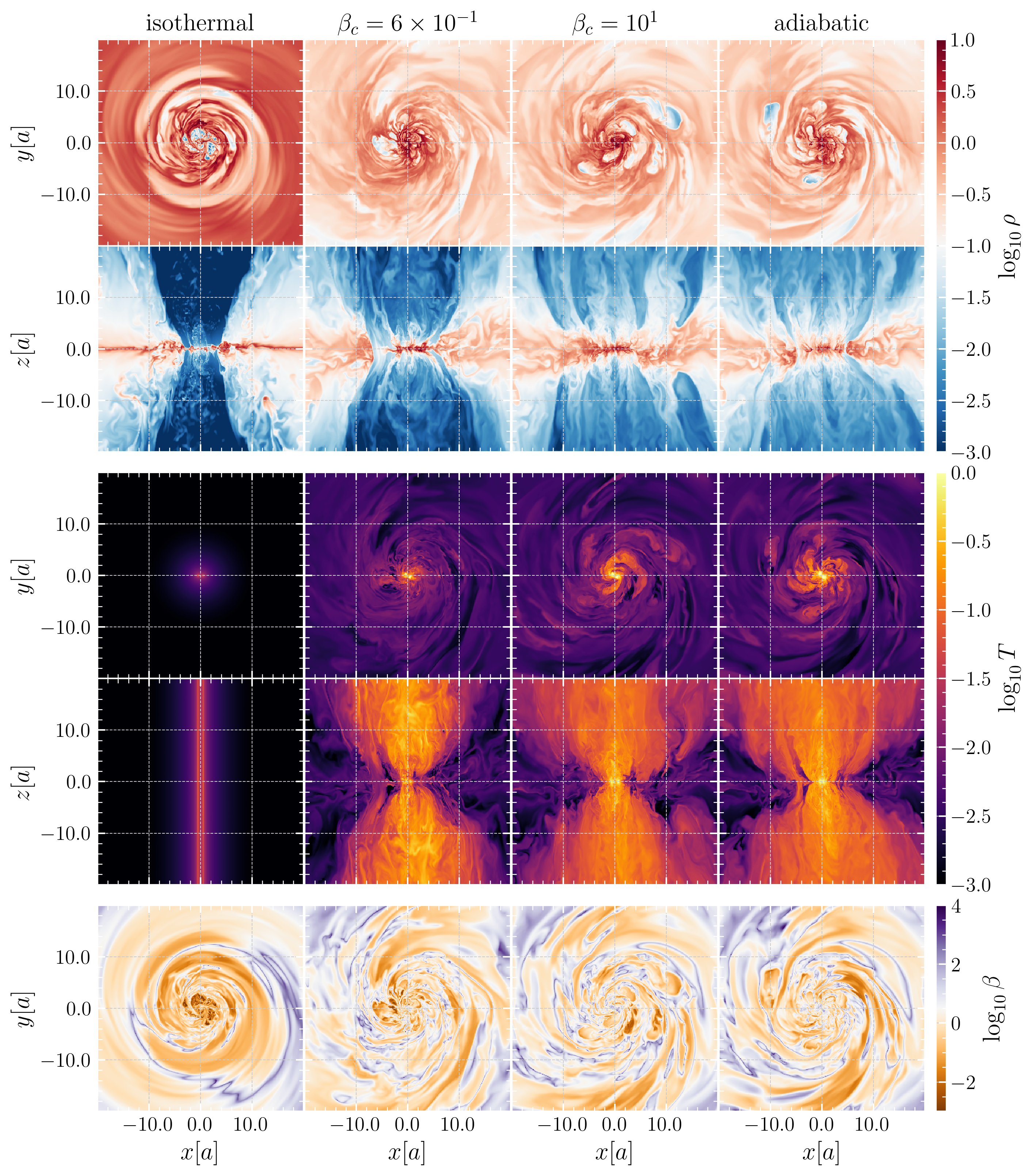}
    \caption{Comparison of different cooling regimes. ({\it Left to right}) Instantaneous cooling (isothermal), intermediate ($\beta-$)cooling, and no cooling (adiabatic). Shown are the mass density $\rho$, temperature $T$, and plasma parameter $\beta$ on the equatorial and meridional planes.
    } 
    \label{fig:cooling-summary-out}
\end{figure*}

\begin{figure*}[t]
    \centering
\includegraphics[width=0.9\linewidth]{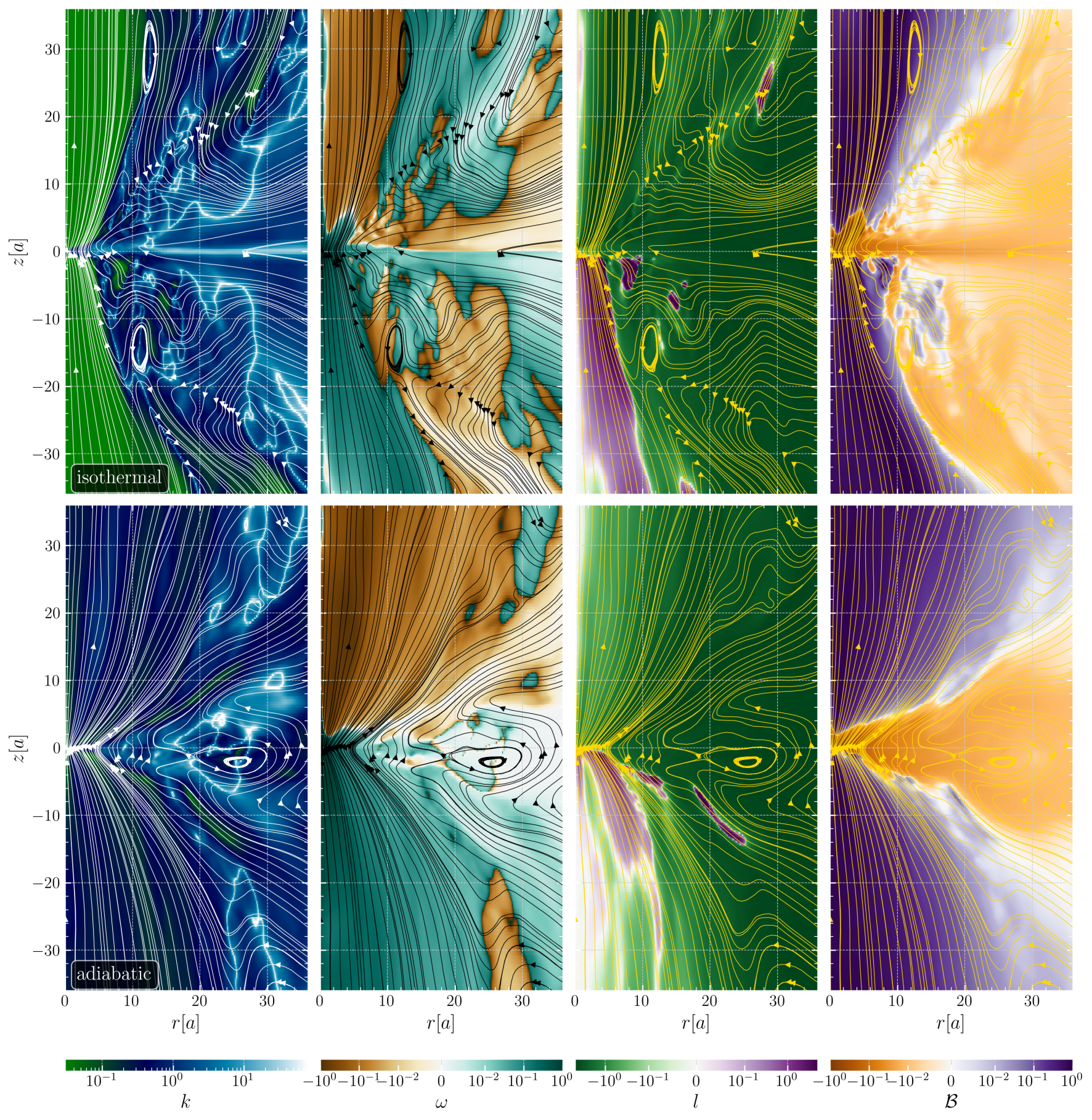}
    \caption{Blandford-Payne wind analysis for the isothermal (\texttt{iso-p}, {\it top}), and adiabatic (\texttt{adi-p}, {\it bottom}) models, averaged both in the azimuthal direction and in time between $t=300-305\, t_{\mathrm{orb}}$. Streamlines indicate the magnetic field. Shown in color are the mass loading parameter $k$, the magnetically modified angular velocity, $\omega$, and the specific angular momentum $l$ of the wind, and the Bernoulli parameter $\mathcal{B}$. We can clearly see a magnetically driven wind component in both systems, with the isothermal wind being confined to a narrow funnel, whereas the adiabatic disk has a strong wind component.}
    \label{fig:iso-p-bp}
\end{figure*}

\section{Results}
In the following, we provide a detailed account of various aspects of circumbinary disk dynamics in the BMAD state. 
We focus on cavity, disk, and outflow properties and angular momentum transport, {both between the disk and the binary and within the disk}.
To simplify the comparison of different regimes, we primarily focus on the two thermodynamic limits: locally isothermal (instant cooling) and adiabatic (no cooling).
First, we examine the evolution toward the MAD state, focusing on magnetic flux accumulation within the central cavity and the subsequent launching of magnetic outflows around each BH.
By introducing an effective cooling function, we also clarify and explore the intermediate regime between the two thermodynamic extreme regimes we primarily consider.
We then analyze the cyclic nature of accretion, magnetic flux eruptions, the properties of these magnetic outflows, and angular momentum transfer between the disk and the binary. 
Additionally, we investigate the role played by the initial magnetic field topology, in particular of toroidal magnetic fields.

\subsection{Overview}
\label{sec:thermo-limit}

We begin by describing the process of flux accumulation and the onset of flux eruption in this section (also see Section 3.1 in Ref. \cite{paper1}). Fig. \ref{fig:wtm} shows the mid-plane density $\rho$ and plasma parameter $\beta$ at 10 and 100 binary orbits for simulations {with purely poloidal magnetic field, but different thermodynamic limits,} \texttt{iso-p} and \texttt{adi-p}, representing locally isothermal and adiabatic conditions, respectively.
In both cases, at 10 binary orbits, the accretion sets in through spiral accretion streams.
The streams circularize around each of the gravitational components and form mini-disks in the isothermal case and mini-spheres in the adiabatic case. 
This early evolution phase resembles {commonly considered} two and three-dimensional hydrodynamical simulations (see, e.g., \cite{Lai:2022ylu}). 
{In this regime}, the bulk of the disk has $\beta \gtrsim 10$, the binary is exciting $m=2$ spiral density waves in the disk, {and accretion onto the binary is largely driven by effective turbulent viscosity in the disk} indicative of predominantly hydrodynamical behavior, {akin to the Standard and Normal (SANE) accretion regime in BH accretion}.
{Under these conditions, the} cavity wall {is clearly delineated and remains stable over many orbital periods}. 
It is worth noticing that the run \texttt{adi-p} already exhibits Rayleigh-Taylor instabilities at the boundary of the cavity. {Due to the largely high $\beta-$regime in which this happens, the instability is purely hydrodynamical.}
The cavity {is, however, starting to accumulate} $\beta$ ranging from $10^{-2}$ to $1$ {locally} in both cases, which {indicates the onset of} magnetic flux accumulation.

After 100 binary orbits, {the bulk of the disk is also growing to be more magnetized with $\beta$ between 0.1 and 10.} The morphologies of the mini-disks (mini-spheres), the cavities, and the disks now all drastically differ. {This is a direct result of the cavity accumulating so much vertical magnetic flux, $\beta\sim 10^{-3}$, that the dynamics are largely magnetospheric (see Sec. \ref{sec:MAD}).}
{As a result, the morphology of the cavity is fundamentally altered.}
In the isothermal case, the mini-disks are disrupted, {and the sinks now} directly connect to the current sheet spanning through the entire cavity region and the spiral accretion streams. 
In the adiabatic case, the mini-spheres remain largely intact.
The degree of magnetization of the cavity, spiral accretion streams, and the cavity is similar to the initial stage, while evidently we can spot the narrow-shaped region having $\beta \sim 0.1-10^{-3}$ escaping from the sink region to the cavity (see the narrow purple region at the center of the right panel {in Fig. \ref{fig:wtm}}), and escaping from the cavity to the outer region of the disk (upper right region of the right panel {in Fig. \ref{fig:wtm}}).

We can further assess how relaxed the disk configuration is at this point, by comparing averaged radial profiles of various quantities (Fig. \ref{fig:wtm-1d}). We find that the density in the disk reaches a quasi-steady state after about 200-250 orbits for both fiducial systems. However, the magnetization parameter, $\beta$, in the adiabatic case drops much slower than in the isothermal case. This is likely because magnetic flux eruptions lead to a constant ejection of magnetic flux from the cavity, which can propagate outwards, whereas in the isothermal limit flux rapidly becomes trapped inside the cavity. Overall, we take the dynamics we report after 300 orbits to be representative of the BMAD state.

\subsection{Magnetically arrested cavity dynamics}\label{sec:cavity}
\textit{Quasi-steady state. } {Once magnetically saturated, the cavity wall will become unstable to interchange instabilities \cite{Spruit:1995fr,Begelman:2021ufo}, leading to the ejection (eruption) of magnetic flux (see Sec. \ref{sec:MAD}).}  {We demonstrate the background configuration of a magnetically arrested cavity in} Fig. \ref{fig:all-summary} {for four different configurations}, varying the equation of state and the initial magnetic field topology (models \texttt{iso-p}, \texttt{iso-t}, \texttt{adi-p}, and \texttt{adi-t}). Morphologically, the choice of the equation of state can alter the results most significantly. We will briefly describe the main features below.

In the isothermal case, the cavity region (within $\sim 2a$) consists of low mass density ({$\rho \ll 10^{-2}\rho_{\rm disk}$ with $\rho_{\rm disk}$) being the average disk density near the cavity wall,} and high magnetization ($\beta \sim 10^{-3}$). Spiral accretion streams are still present,{ but smaller Rayleigh-Taylor fingers emerge, which are characteristic of magnetospheric accretion \cite{Kulkarni:2008vk,Parfrey:2023swe}.}
{In addition, the} two spiral accretion streams episodically merge into one large spiral arm and split into two. The disk surrounding the cavity is less magnetized, but still having $\beta \sim 0.1-10^2$. 
Inside the cavity, the accretion proceeds {primarily through magnetospheric streams, in addition to accretion through the} midplane.

\begin{figure*}
    \centering
    \includegraphics[width=0.7\linewidth]{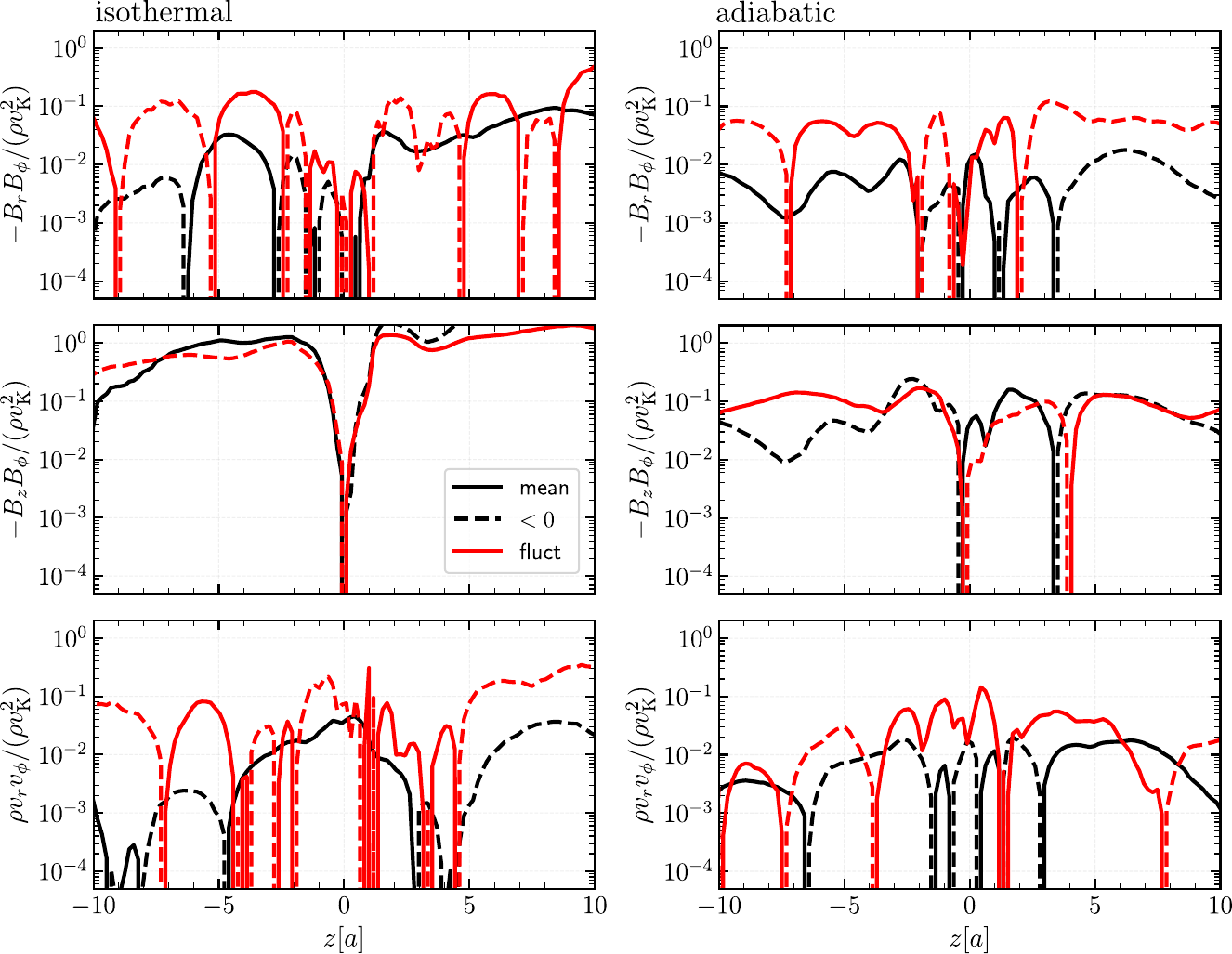}
    \caption{
    Stresses inside the accretion disk. Shown are time and azimuthal averaged vertical profile of the $R\phi$- and $z\phi$-components of Maxwell stress, $B_i B_j$, and the Reynolds stress, $\rho v_i v_j$, where $\rho$ is the mass density, and $v_K$ denotes the Keplerian velocity. Mean and fluctuating components are denoted separately with solid and dashed lines, respectively. All quantities are measured at $r=3a$ for adiabatic (\texttt{adi-p}) and isothermal (\texttt{iso-p}) models.
    } 
    \label{fig:disk_stress}
\end{figure*}

\begin{figure*}[t]
    \centering
    \includegraphics[width=0.95\linewidth]{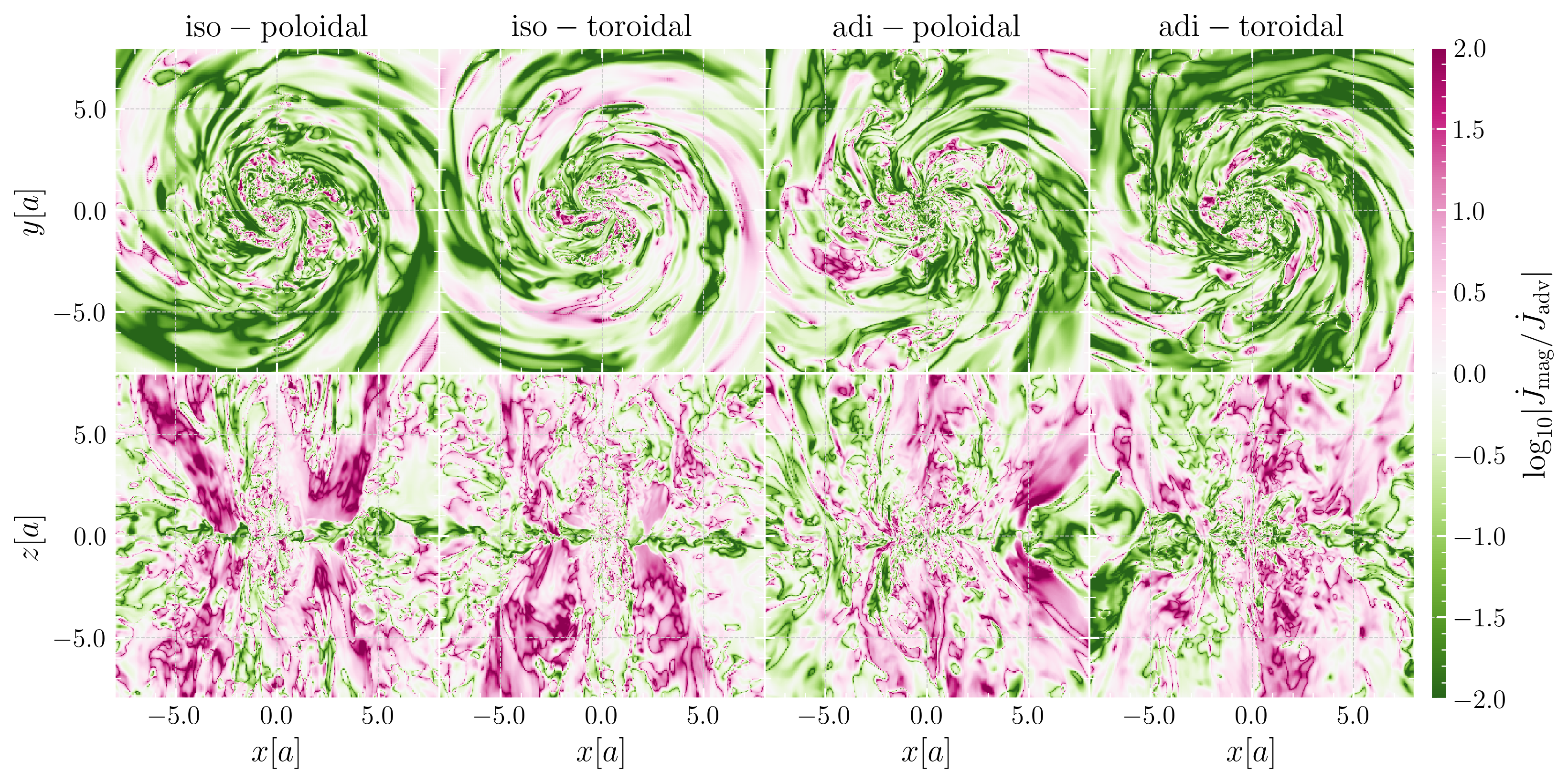}
    \caption{
    Magnetically driven angular momentum transport in the circumbinary magnetically arrested (BMAD) state for different models. 
    Shown is the ratio between the magnetic contribution, $\dot J_{\rm mag}$, and the advective contribution, $\dot J_{\rm adv}$, to the angular momentum current. 
    The equatorial and meridional planes are shown separately in the top and bottom rows. 
    Different columns denote runs with no cooling (adiabatic), instantaneous cooling (isothermal); and different initial magnetic field topologies (poloidal and toroidal), see Sec. \ref{sec:methods} for details.
    In the BMAD state, angular momentum transfer through the cavity is largely dominated by magnetic contributions in the highly magnetized region (i.e., the funnel region and the flux tubes (pink) ); 
    while the advection component (green) dominates in the bulk of the disk.
    } 
    \label{fig:amt-summary}
\end{figure*}

In the adiabatic case, both the funnel region and the disk are less magnetized compared to the isothermal scenario, having {equipartition values}, $\beta \simeq 1$. The distinction between the disk and the cavity also becomes less clear.  
The to-be ejected flux tubes have plasma parameter $\beta$ ranging from $10^{-1}$ to $1$, {not unlike flux tubes ejected in single BH MAD accreting systems \cite{Ripperda:2021zpn}}.
{Accretion in this region proceeds almost exclusively through} large Rayleigh-Taylor fingers penetrating the cavity \cite{Chatterjee2022}.

\textit{Flux eruption cycles.}
A distinctive feature of the MAD state is the quasi-periodic switching between two accretion phases—the quiescent and eruption states—commonly referred to as flux eruption cycles \cite{Tchekhovskoy:2011zx,Ripperda:2021zpn}. 
A similar mechanism operates in the BMAD state, {though it} differs {in some of its details} between the isothermal and adiabatic limits.
We first summarize the general characteristics of these two accretion states (Fig. \ref{fig:fe-cycle}). 

The key difference distinguishing the BMAD state from the conventional MAD state lies in the natural truncation radius of the outer disk, which shifts from the ISCO (single black hole scenario) to the tidal truncation radius of the cavity \cite{paper1}, {overall more similar to magnetically truncated disk models \cite{Liska:2022jdy}.}
Inside this truncation radius, gas parcels follow geodesic trajectories and freely fall toward the central gravitational objects. 

In the quiescent state, force equilibrium is established at the cavity boundary, with magnetic pressure within the cavity balancing the combined thermal and magnetic pressure from the circumbinary disk. 
As mass accretion onto the two sinks proceeds, magnetic flux gradually accumulates inside the cavity, causing it to expand adiabatically. 
The quiescent state persists until the cavity radius surpasses the tidal truncation radius of the CBD, {making the cavity wall unstable by triggering} an interchange instability \cite{Spruit:1995fr,Begelman:2021ufo}.
Accretion, distinct from the spiral accretion streams, then occurs via Rayleigh-Taylor fingers, ejecting coherent magnetic flux tubes into the main body of the disk (Fig. \ref{fig:fe-cycle}).  During this eruption state, mass accretion is enhanced, the cavity size shrinks, and the net vertical magnetic flux within the cavity decreases. Subsequently, the system returns to a quiescent state, restoring stability at the cavity boundary.

{The dynamics of flux tube propagation in this background also proceeds slightly differently.}
{In the isothermal limit,} although the cavity undergoes interchange instability and forms Rayleigh-Taylor fingers, the flux tubes {do not efficiently} propagate outward, {and only partly mix into the disk} (Fig. \ref{fig:fe-cycle}). This limitation results from the instantaneous cooling scheme, which imposes a temperature profile proportional to the gravitational potential in the disk midplane. Consequently, flux tubes that move outward from the gravitational sources experience rapid thermal pressure reduction and collapse, causing the trapped magnetic flux to be promptly re-accreted into the cavity on a local dynamical timescale. Similar dynamics have been observed in full radiative simulations of magnetically truncated accretion disk \cite{Liska:2022jdy}. In the adiabatic limit cooling of the flux tubes is not present, and they are able to propagate to large distances inside the disk (Fig. \ref{fig:fe-cycle}). 

{In order to better compare cavity morphologies, we provide azimuthally averaged disk and cavity profiles in Fig. \ref{fig:iso-p-wind}.
In the isothermal limit, the boundary between the cavity and the disk is sharp and clearly delineated. The cavity is threaded with a strongly magnetized net vertical flux component ($\beta\sim10^{-3}$), whereas the disk is thermally collapsed, has an extended corona, and features wind-like outflows.}
The transition between the funnel region and magnetized disk atmosphere is abrupt. This could result from the fact that we are adopting a temperature profile only as a function of cylindrical radius $r$. 
The disk truncation radius is around $3a$.\\
In the uncooled (adiabatic) limit, the cavity is filled with hot gas, making the transition between cavity region and circumbinary disk gradual (Fig. \ref{fig:iso-p-wind}). We can define the radial extend of the CBD cavity using the transition in the plasma parameter $\beta$ profile (see also Refs. \cite{Dittmann2023,Most:2024onq} for alternative definitions using the density).
The truncation radius is around $6a$. The opening angle of the funnel region is also wider than the isothermal case. The disk features a strong magnetocentrifugal wind component.\\

In addition to the cavity properties, we also briefly comment on the vertical structure of the disk, which has gained recent interest in terms of the possibility of supporting the disk magnetically with strong toroidal magnetic fields \cite{Gaburov:2012jd,Hopkins2023,Hopkins:2023lgk}, and a potential dependence on cooling physics \cite{Guo:2025glc}. While none of our models enter such a hypermagnetized state, Ref. \cite{Wang:2025mit} has recently demonstrated that parsec-scale circumbinary disks formed from intermediate galactic scale accretion may be in such a regime.
Following Ref. \cite{Guo:2025glc}, we here briefly comment on the vertical disk structure by separately computing the magnetic, thermal, and turbulent contribution for the two thermodynamic limiting cases (Fig. \ref{fig:disk_scale_height}).
As expected, we find that for the strong cooling (isothermal) limit, the disk plane is fully collapse, having small verical extent in terms of thermal pressure. We can see that outside the midplane (indicated by the density contour), the disk has a strong magnetic pressure contribution indicating an extended (puffy) coronal region above the disk, which we will comment on more in Section \ref{sec:outflows}. Since we fix the temperature relative to the gravitational potential on the midplane, the disk temperature is constant along the vertical scale axis.\\
The uncooled (adiabatic) disk on the other hand is not collapsed, featuring an extended thermally supported disk region, with a heated corona. This disk drives strong outflows from a region with roughly equal magnetic and thermal pressure. Different from the isothermal case, the turbulent pressure in the disk mid plane is low. Though, we caution that the especially the magnetic field in terms of $\beta$ is not fully converged in time, yet, in this part of the disk (Fig. \ref{fig:wtm-1d}).

\begin{figure*}
    \centering
    \includegraphics[width=.45\linewidth]{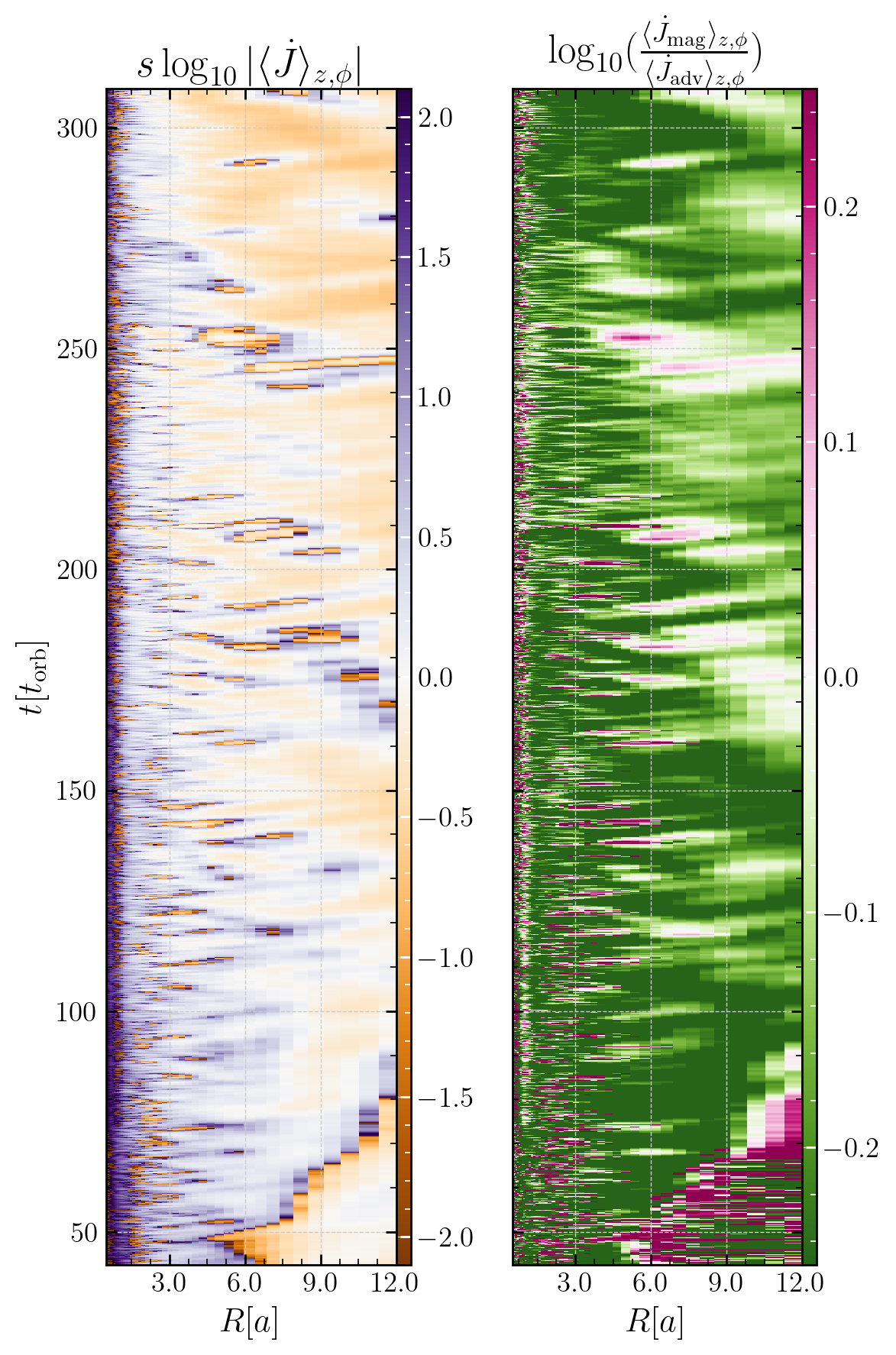}
    \includegraphics[width=.45\linewidth]{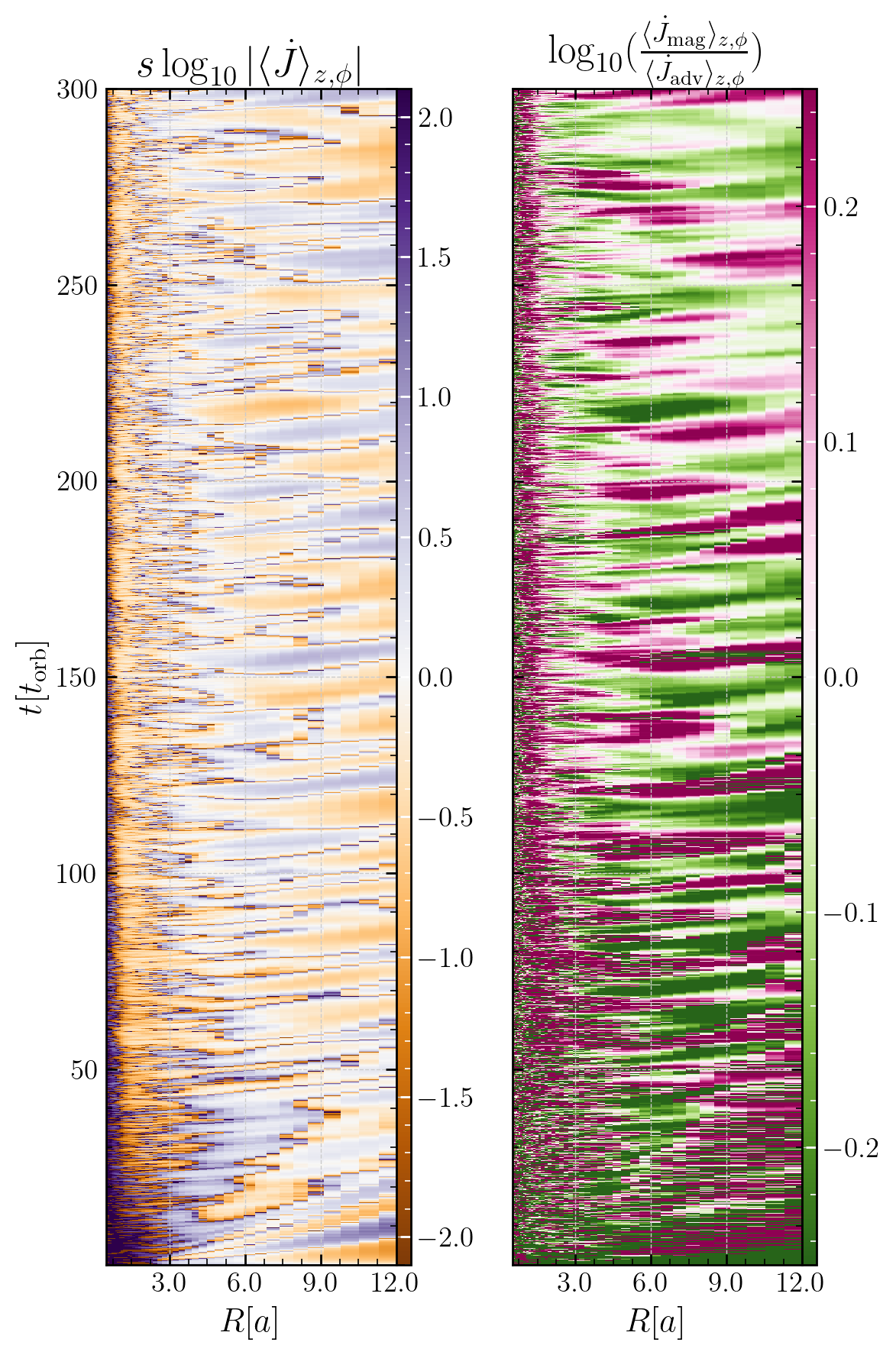}
    \caption{
    Azimuthally and vertically integrated radial angular momentum fluxes shown as a spacetime diagram, for the isothermal, \texttt{iso-p} (\textit{Left}), and adiabatic, \texttt{adi-p} (\texttt{Right}), models. For each simulation, the total radial angular momentum flux, $\left<\dot{J}\right>_{z,\phi}$, is shown on the left-hand side and the ratio of magnetic, $\left<\dot{J}_{\rm mag}\right>_{z,\phi}$, and advective, $\left<\dot{J}_{\rm adv}\right>_{z,\phi}$, flux is shown on the right-hand side.
    Each purple region in the spacetime diagram corresponds to an ejected magnetic flux tube, which for isothermal runs quickly mix into the disk, or get ejected to large distances in the adiabatic model. Both cases reach a quasi-steady state between $200-250$ orbits. Compared to the isothermal case, angular momentum transfer in the adiabatic scenario proceeds primarily through magnetic transport rather than advective transport.
    } 
    \label{fig:flux-trans-iso-p-adi-p}
\end{figure*}

\subsubsection{Impact of dynamical cooling}
\label{sec:cooling}
{Having described the cavity dynamics in the magnetically arrested regime for two thermodynamic limiting cases (instant cooling and no cooling), we now want to briefly illustrate the transition between these two limits.}
{Within the framework of $\beta-$cooling, the} isothermal case can be regarded as instantaneous cooling (cooling time $t_{\rm cool}=\beta_c t_{\rm orb}/2\pi\to0$, where $t_{\rm orb}$ is the binary orbital period) and adiabatic case as zero cooling ($t_{\rm cool}\to\infty$).
Instead of starting with the same initial condition as run \texttt{adi-p} (see Appendix \ref{app:id}), we turn on cooling only after the quasi-steady state has bee reach for run \texttt{adi-p} after $300$ orbits. {This allows us to more efficiently survey the different cooling regimes.}

We present a comparison of circumbinary disk morphology spanning from the isothermal to the adiabatic limits in Fig.~\ref{fig:cooling-summary-out}. The cavity shape and properties of magnetic flux tubes undergo substantial changes with varying cooling timescales. Specifically, as the cooling time increases, the magnetic flux tubes grow larger and propagate radially outward to greater distances:
\begin{itemize}
    \item In the \textit{isothermal} case, flux tubes are unable to escape from the cavity, as they instantly cool, collapse, and get sheared into the disk.
    \item In the case with cooling parameter $\beta_c=0.6$, larger flux tubes form within the disk midplane and escape the cavity; however, these tubes do not remain coherent at larger radii due to continued cooling, and begin to deflate.
    \item In the case with cooling parameter $\beta_c=10$, flux tubes propagate to significantly larger radii and rotate coherently along with the disk.
    \item The fully adiabatic limit is essentially identical to the $\beta_c=10$ case.
\end{itemize}
The third row of Fig.~\ref{fig:cooling-summary-out} provides clearer evidence of the above-described transition. 
In the adiabatic limit, the flux tubes maintain significantly higher internal temperatures compared to the surrounding disk material, enabling coherent propagation. 
In the intermediate cooling case ($\beta_c=0.6$), the partially cooled flux tubes exhibit irregular shapes, reflecting the competition between cooling and flux eruption dynamics. 
Conversely, in the isothermal scenario, the imposed vertically isothermal temperature profile prevents successful flux tube ejection altogether.

\begin{figure*}
    \centering
    \includegraphics[width=.45\linewidth]{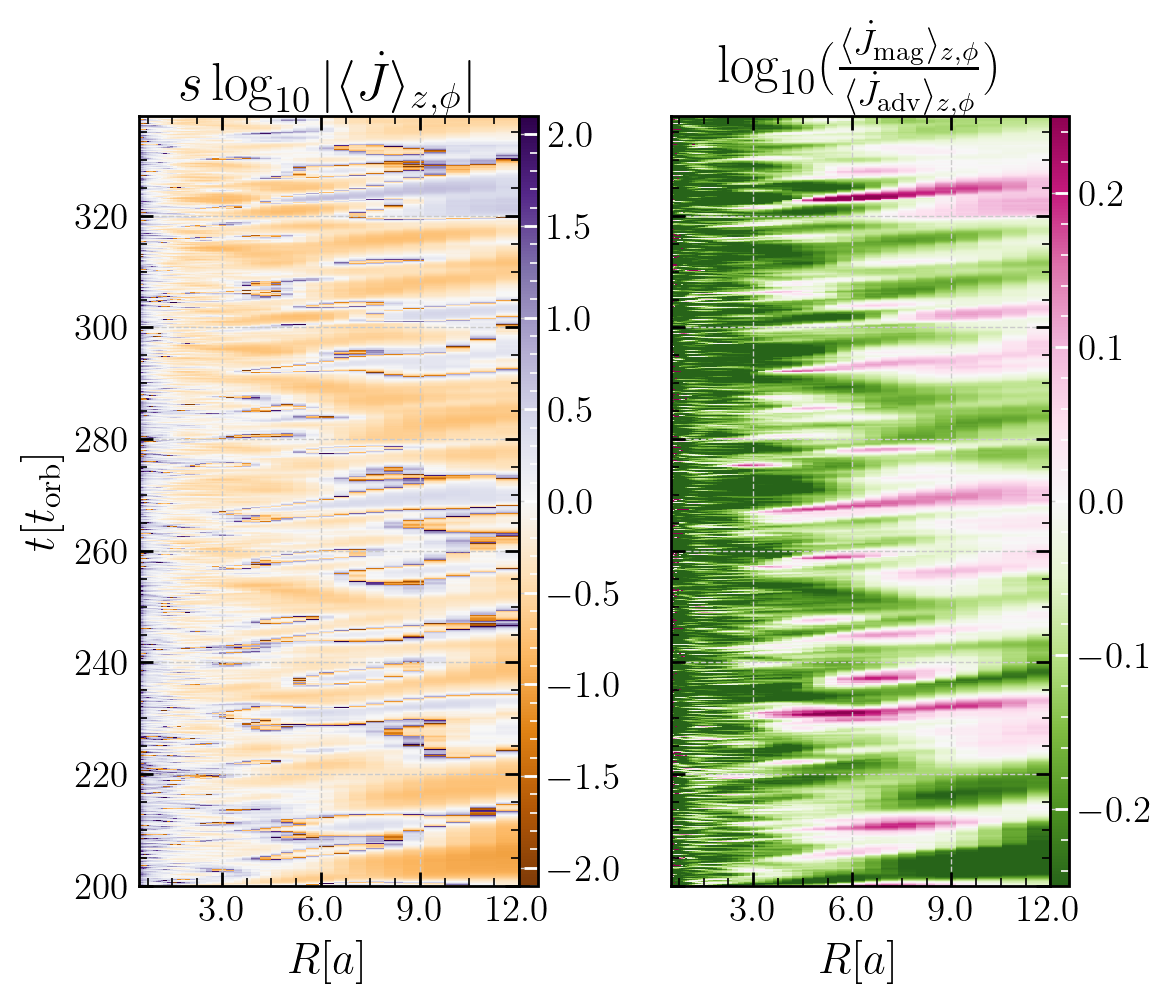}
    \includegraphics[width=.45\linewidth]{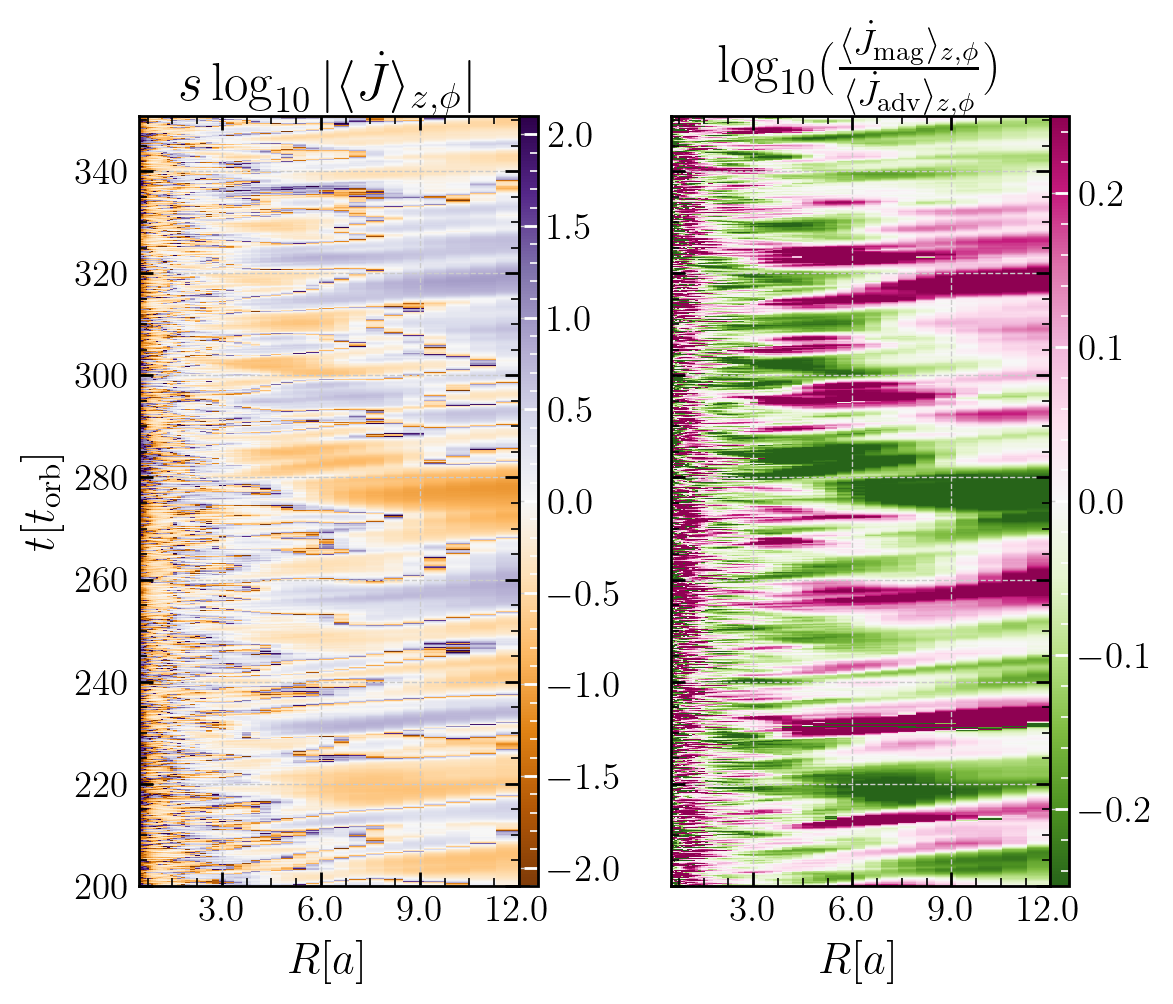}
    \caption{Same as Fig. \ref{fig:flux-trans-iso-p-adi-p}, but for $\beta-$cooling models \texttt{adi-p-c0.6} and \texttt{adi-p-c10}.
    } 
    \label{fig:flux-trans-cooling}
\end{figure*}

\subsection{Outflow Properties}\label{sec:outflows}

{One important aspect of simulating fully magnetized accretion onto a binary is the potential of magnetically driven outflows. These can either be launched from the sink itself, akin to dual jet outflows \cite{Palenzuela:2010nf,Ressler:2024tan}, or from the disk \cite{2018ApJ...857...34Z}. The presence of these outflows can have important implications for electromagnetic signatures \cite{Yuan:2021jjt}, and angular momentum transport \cite{Bai:2013pi,2022ApJ...925..161S,2024ApJ...964..133M}, as we will discuss in the next Section.}

{We had already observed the presence of strong disk outflows when discussing the azimuthally average disk profiles of the accretion disk (see Fig. \ref{fig:iso-p-wind}).}
{We now provide a detailed analysis of these outflows, for which we}
average over the final 5 orbits with a sampling frequency of $1/20 t_{\rm{orb}}$, where $t_{\rm{orb}}$ is the binary orbital period.

Though our wind is turbulent compared to the steady axisymmetric outflows, the time and azimuthal averaged profiles still exhibit aligned magnetic field lines in the cavity and above the disk magnetic scale height/magnetic corona. 
We measure the four quantities that are conserved in magnetocentrifugal wind theory \citep{Blandford:1982xxl}, along the azimuthally averaged poloidal fields. {To this end, we s}eparately decompose velocity and magnetic field $\boldsymbol{v}$ and $\boldsymbol{B}$ into poloidal and toroidal components ($\boldsymbol{v} = \boldsymbol{v}_p + \Omega R \hat \phi$ and $\boldsymbol{B} = \boldsymbol{B}_p + \boldsymbol{B}_\phi$). 
The wind is then characterized using
the mass load parameter $k$, magnetically modified angular velocity $\omega$, Bernoulli constant $\mathcal{B}$, and specific angular momentum of the wind $l$,
\begin{align}
k &= \frac{\rho \boldsymbol{v}_p}{\boldsymbol{B}_p}\,, \\
\omega &= \Omega-\frac{k B_\phi}{\rho R}\,, \\
\mathcal{B} &= \frac{1}{2} v^2+\Phi+h+\frac{B^2}{\rho}-\frac{\boldsymbol{B} \cdot v}{k}\,,\\
l &= R\left(v_\phi-\frac{B_\phi}{k}\right)\,, 
\end{align}
where we will primarily focus on the first three (Fig. \ref{fig:iso-p-bp}).
{We can see that in the isothermal case mass loading is strongest in the disk wind region, whereas in the adiabatic case it is gradual, with a substantial contribution coming from the central cavity region. By comparing the magnetic field line structure with the $\omega$-profile, we find that the these winds are largely magneto-centrifugally driven, with a large extended funnel like region being present in the adiabatic case. In particular, by considering unbound material $(\mathcal{B}>0)$, we can spot an extended wind structure in the adiabatic case. On the other hand, in the isothermal case this wind region is narrow, with a turbulent disk coronal region indicating a more complex wind launching mechanism, including magnetic tower outflows from the near sink region \cite{paper1}. }\\

In order to get more quantitative insight into the stresses operating inside the disk and corona, we now focus on vertical profiles of Maxwell and Reynolds stresses (Fig. \ref{fig:disk_stress}). By comparing the net stresses, we find that for the isothermal model turbulent Reynolds stresses dominate in the disk midplane. In coronal regions ($\left| z \right| \gtrsim a$), however, the magnetic stresses become equally important, in particular in terms of vertical transport consistent with our previous wind analysis discussion. In the adiabatic case, Maxwell and Reynolds stresses are roughly equal.

\subsection{Angular Momentum Transport}\label{sec:amt}

{One of the most important aspects of circumbinary accretion research is the determination of the angular momentum transport rate. In the context of the final parsec problem \cite{Begelman:1980vb}, understanding under which conditions the binary shrinks, or expands is crucial towards determining the role gaseous accretion plays in the overall evolution of the system. While detailed studies have been undertaken to characterize angular momentum transport in hydrodynamical regimes \cite{Munoz:2018tnj,DOrazio:2021kob,Duffell:2019uuk,Dittmann:2023ztg,Siwek:2023rlk}, these do not necessarily reflect findings in MHD simulations \cite{Noble:2012xz,Noble:2021vfg}. Especially the BMAD regime may qualitatively operate differently due to the presence of interchange instabilities at the cavity wall \cite{paper1}. This resembles similar findings in MAD single BH accretion disk \cite{Chatterjee2022}. Here, we extend our previous investigation \cite{paper1}, by providing a detailed analysis angular momentum transport in the BMAD regime.}

We calculate the angular momentum flux {as a function of radius inside the cavity and within the circumbinary disk}. Compared to the angular momentum transfer equation in hydrodynamical simulations (e.g., \cite{Miranda2017,Moody2019,Munoz:2018tnj}), in our case magnetocentrifugal/magnetic-turbulent wind can also carry angular momentum flux \cite{2024ApJ...964..133M}, these terms appear as the boundary terms in a cylindrical coordinate representation (see Appendix \ref{app:amt} for a detailed derivation). {In summary,} the MHD angular momentum transfer equation in the disk can be written as:
\begin{align}
 \left<\dot J_{B}\right>_t=\left<\dot{J}_{\phantom{B}}\right>_t= &\left< \dot{J}_{\mathrm{adv} ,R}\right> _{t} +\left< \dot{J}_{\mathrm{mag} ,R}\right> _{t} \nonumber \\
 &+\left< \frac{dT_{\mathrm{adv} ,z}}{dR}\right> _{R,t} +\left< \frac{dT_{mag,z}}{dR}\right> _{R,t} \nonumber\\ 
 &+\left< \frac{dT_{\mathrm{grav}}}{dR}\right> _{R,t}\,,
\end{align}
with individual components,
\begin{equation*}
\begin{aligned}
\frac{dJ}{dR} & = \left< j\right> _{\phi ,z}\,,\\
\dot{J}_{\mathrm{adv} ,R} & = \left< jv_{R}\right> _{\phi ,z}\,,\\
\left< \frac{dT_{\mathrm{adv} ,z}}{dR}\right> _{R} & = \int dR^\prime \left< jv_{z}\right> _{\phi } |_{z_{\mathrm{lower}}}^{z_{\mathrm{upper}}}\,,\\
\dot{J}_{\mathrm{mag} ,R} & = -R^{2}\left< B_{R} B_{\phi }\right> _{\phi ,z}\,,\\
\left< \frac{dT_{mag,z}}{dR}\right> _{R} & = -\int dR^\prime R^{\prime 2}\left< B_{z} B_{\phi}\right> _{\phi } |_{z_{\mathrm{lower}}}^{z_{\mathrm{upper}}}\,,\\
\left< \frac{dT_{\mathrm{grav}}}{dR}\right> _{R} & = \int dR^\prime R^\prime \left< \rho \partial _{\phi } \Phi _{g}\right> _{\phi ,z}\,,
\end{aligned}
\end{equation*}
where $\dot J_{B}$ is the angular momentum flux added to the binary, $\dot J$ is the total angular momentum flux in the disk, $\dot{J}_{\mathrm{adv} ,R}$ is the advective angular momentum flux, $\left< {dT_{\mathrm{adv} ,z}}/{dR}\right> _{R}$ is the magnetic torque from vertical transport, $\dot{J}_{\mathrm{mag} ,R}$ is the magnetic angular momentum flux, $\left< {dT_{{\rm mag},z}}/{dR}\right> _{R} $ is the magnetic torque, and $\left<{dT_{\mathrm{grav}}}/{dR}\right> _{R}$ is the gravitational torque. 
{The coordinates $z_{\rm lower/upper}$ denote the upper and lower height of the cylinders over which the angular momentum transfer is computed.} Following the convention in (GR)MHD simulations \citep{Chatterjee:2020wef}, we define, $\dot J > 0$, corresponding to an outward angular momentum flux, converse to the sign used in the hydrodynamical simulations \citep{Miranda2017, Munoz:2018tnj, Moody2019}.

\begin{figure*}
    \centering
    \includegraphics[width=0.99\linewidth]{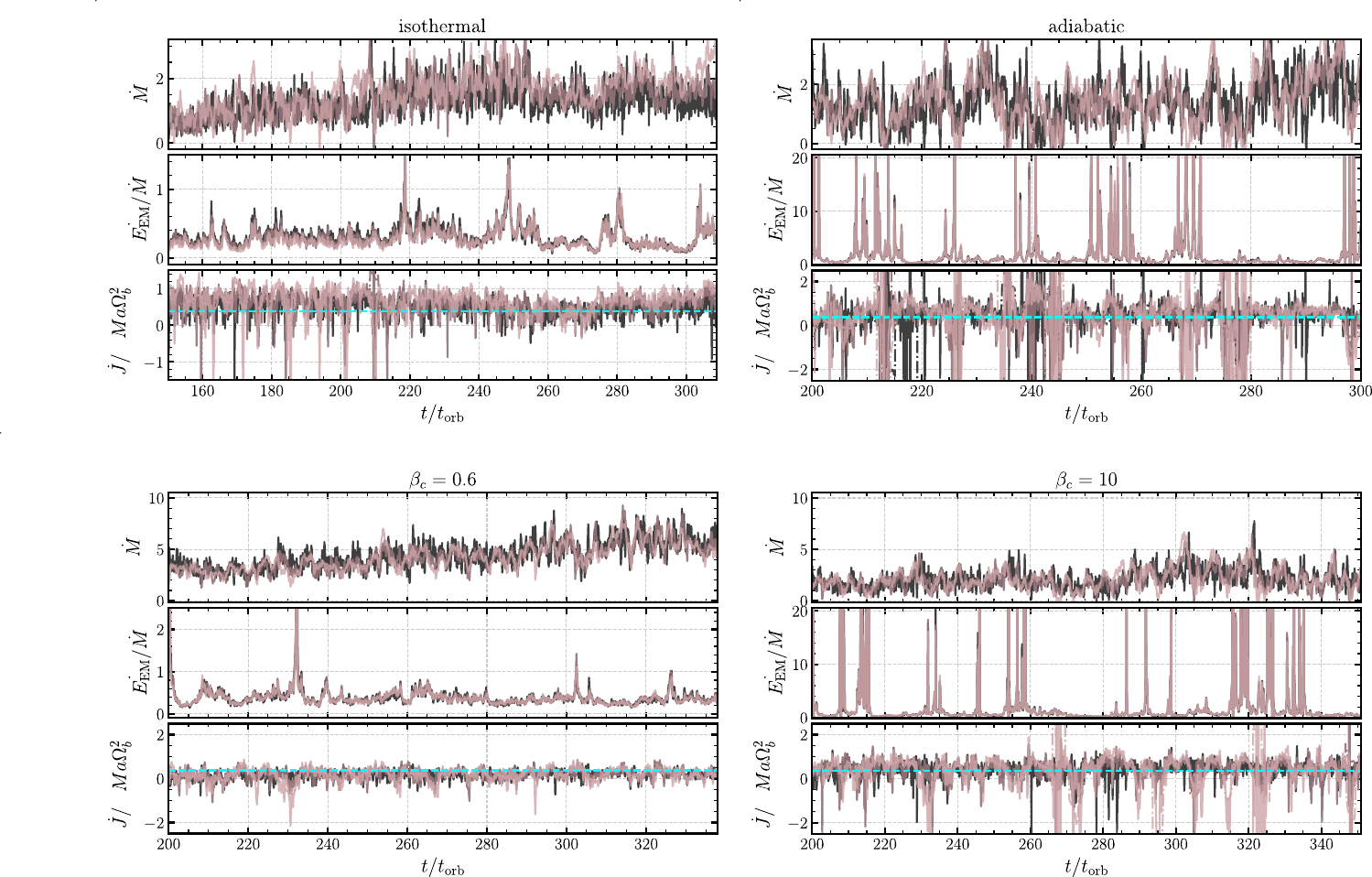}
    \caption{
    Angular momentum transport onto the binary, and outflow properties of the cavity for different thermodynamic models with initially poloidal magnetic field: {\it (Top Left)} isothermal (\texttt{iso-p}),  (\textit{Top Right}) adiabatic (\texttt{adi-p}), and (\textit{Bottom Left and Right}) $\beta_c-$cooling (\texttt{adi-p-c0.6} and \texttt{adi-p-c10}). In each figure from top to bottom we show the mass accretion rate $\dot M$ (\textit{Top}), electromagnetic energy flux $\dot E_{\rm{EM}}$ (\textit{Middle}), and the angular momentum flux $\dot J$ (\textit{Bottom}). The over plotted black, dark pink, and light pink lines are the values measured at radius $2, 3,$ and $4 a$, where $a$ is the binary separation. All extraction radii are in good agreement, indicating an overall convergent inflow equilibrium near the cavity. In the bottom panels, 
    the cyan dashed line shows the threshold for binary inward or outward migration. Times are stated relative to the binary orbital time $t_{\rm orb}$.
    } 
    \label{fig:amt-all-ideal}
\end{figure*}

{In order to illustrate the importance of the various terms, we here compare the magnetic and advective contributions (Fig. \ref{fig:amt-summary}). We can see that irrespective of the cooling prescription or initial magnetic field topology, the disk is always largely dominated by hydrodynamical (advective) transport. However, as discussed previously, the adiabatic (no cooling) and weakly cooling regimes support the propagation of ejected magnetic flux tubes through the disk. Works in GRMHD have suggested that this transport of flux tubes can dominate the angular momentum transport \cite{Chatterjee2022}. Consequently, we clearly identify these flux tubes as small magnetically dominated regions inside the adiabatic disks (coherent pink regions in Fig. \ref{fig:amt-summary}). In general, we can also see that angular momentum transport inside the funnel/cavity region is always magnetically dominated, in line the fundamental nature of the BMAD regime \cite{paper1}. This demonstrate that the BMAD regime clearly does have non-hydrodynamical angular momentrum transport channels.}\\
{In order to better quantify the impact of ejected flux tubes on the angular momentum transport budget, we now consider radial spacetime diagrams showing the evolution of angular momentum transport inside the disk (Fig. \ref{fig:flux-trans-iso-p-adi-p}). In the adiabatic case, we can clearly spot (pink) regions of magnetically dominated outwards directed transport. These are the flux tubes that propagate from the cavity wall outwards and transport angular momentum magnetically. On the other hand, in non-erupting intervals, transport is predominantly inwards and operates hydrodynamically (green regions). In the case of the isothermal disk, we can also spot small flux eruptions (pink) in inner regions of the disk. However, since the flux tubes are strongly cooled and deflated, they are unable to transport angular momentum to large distances. As a result, angular momentum transport inside the disk remains largely advective (hydrodynamical).} {Since the cooling regime seems to have a very strong impact on the overall angular momentum transport, we confirm this trend with two $\beta-$cooling simulations (Fig. \ref{fig:flux-trans-cooling}). We can clearly see that with increasing cooling the transport due to flux eruptions (pink regions) weakens strongly, and that overall less total angular momentum is transported outwards. In summary, this confirms that only weakly cooled disks evolve in analogy with (uncooled) GRMHD disks. Furthermore, we caution that in the limit of strong radiative cooling in single black hole accretion \cite{Liska:2022jdy}, similar transport states can be achieved.}\\

{We now compute the net angular momentum transport onto the binary. This is important if we want to understand whether the binary shrinks or expands \cite{Miranda2017,Moody:2019nes,Duffell:2019uuk,Dittmann2022,Siwek:2023rlk}.}
For equal-mass circular binaries, the secular orbital evolution rate is given by \citep{Lai:2022ylu},
\begin{align}
\frac{\dot{a}}{a_{\mathrm{B}}}=8\left(\frac{l_0}{\Omega_{\mathrm{B}} a_{\mathrm{B}}^2}-\frac{3}{8}\right) \frac{\dot{M}}{M_{\mathrm{B}}},
\end{align}
where the total torque per unit of accreted mass is
\begin{align}
l_0=\left\langle\dot{J}_{\text {tot }}\right\rangle /\langle\dot{M}\rangle \text {. }
\end{align}
Therefore, whether binary shrink or expand depends on whether $l_0$ is greater or smaller than $3 \Omega_{\mathrm{B}} a_{\mathrm{B}}^2 / 8$.

We show the mass accretion rate $\dot M$, Poynting flux $\dot E_{\rm EM}$, and total specific angular momentum transfer rate $\dot J/(\dot M a \Omega_b^2)$ for the isothermal, adiabatic and $\beta-$cooling models in Fig. \ref{fig:amt-all-ideal}. 
{Interestingly, all cases show the rate of specific angular momentum transfer between the disk and the binary around the threshold of binary orbital inspiral. However, the isothermal case shows strongest variations, consistent with the fact that flux tube mediated angular momentum transport is not as efficient as in the weakly cooled or uncooled cases. In fact, it appears that over the 150 orbits considered in this analysis the angular momentum transport rate oscillates around the threshold value. In the case of the weakly or uncooled cases, the binary appears to be shrinking, highlighting the importance of the BMAD regime for the binary's evolution. We can confirm the eruption picture by comparing the net electromagnetic outflow rate, $\dot{E}_{\rm EM}$, with the specific angular momentum transfer rate. Indeed, we find that strongly negative trends in the outflow rate approximately correlate with shrinking events in the angular momentum transport evolution, lending additional credibility to the magnetic flux tube transport picture described above, with more longer-lived flux tubes and more violent magnetic flux eruption events in the case of weakly cooled systems.}

\section{Conclusions}
\label{sec:conclusion}

In this work, we have a performed detailed numerical investigation of a new accretion regime onto binary systems with strong magnetic fields: the magnetic arrested (circum)binary accretion disk state (BMAD), which had only been proposed recently \cite{paper1}.
By performing a parameter survey over different cooling regimes and initial magnetic field topologies, we have provided a first systematic exploration of the impact this accretion regime can have, e.g., on the evolution of supermassive black hole binaries in gaseous environments. We provide a detailed summary of our results in Section \ref{sec:MAD}.\\

Similar to the MAD state in single black hole accretion, the BMAD state features strong jet-like magnetic tower outflows, and has an intrinsic intermittent magnetic flux eruption cycle, leading to an ejection of magnetic flux tubes into the disk. While the subsequent propagation of flux tubes and the strength of magnetically driven winds depend sensitively on the cooling regime of the disk, the overall MAD features are present in all models we consider. In particular, the cavity -- opened by gravitational torques -- always fills and saturates with magnetic flux for the low $\beta$-disks we use.
However, strongly cooled BMAD and uncooled BMAD states differ greatly in flux eruption shape, cavity size, and angular momentum transfer, with the cavity radius becoming as large as $6a$ ($a$: binary separation), which is significantly larger than that from the hydrodynamic simulations. In terms of angular momentum transfer, we find that in line with previous work on MAD systems \cite{Chatterjee2022}, that especially weakly cooled systems transport angular momentum primarily through magnetic flux eruptions, since flux tubes remain coherent as they propagate through the disk for these systems. We show that in this limit, the binary is around or below the hardening threshold, unlike hydrodynamical systems \citep{Munoz:2018tnj}.

Our work highlights the importance of the BMAD regime, and its robustness and applicability for a range of parameters. 
This connects well with the recent applications of hot accretion flows to binary systems~\cite{Tiede:2025llq,Wen:2025xpa}.
While our study has considered a relatively large number of configurations given the numerical cost of our study, several areas remain to be explored. 
These include the importance of mass ratio studies, including the limit to gap-like system \cite{DOrazio:2015shf}, effects of general-relativity in particular with regards to realistic sink dynamics \cite{Ressler:2024tan}, and the calculation of near horizon radiative signatures in the BMAD regime \cite{Hakobyan:2022alv,Hakobyan:2025ywq}.
Recent simulations of common envelope interaction also indicate the onset of a potential MAD regime in this case \cite{Vetter:2024loo,Vetter:2025fby}. 
BMAD-like systems could also potentially arise around intermediate-mass black holes in dense star clusters, which have been proposed to have strongly-magnetized circumbinary disks \citep{Shi2024}, and even potentially some large-separation stellar binaries if these can form from envelopes with large accreted magnetic fluxes \citep{luo:2024.zoomin.bh.1star}. 
It may be interesting to more directly explore such connections in future work.

\begin{acknowledgements}
The authors are grateful for discussions with Xue-Ning Bai, Christopher Bambic, Luciano Combi, Alexander Dittmann, Minghao Guo, Yoonsoo Kim, Julian Krolik, Yuri Levin, Aretaios Lalakos, Amir Levinson, Douglas N. C. Lin, Matthew Liska, Vikram Manikantan, Kohta Murase, Ramesh Narayan, Scott Noble, E. Sterl Phinney, Alexander Philippov, Sean Ressler, Bart Ripperda, Re'em Sari, James M. Stone, Alexander Tchekhovskoy, and Zhaohuan Zhu. 
HYW acknowledges partial support from the B. Thomas Soifer Chair's Fellowship. 
ERM and HYW acknowledge support from the National Science Foundation through award NSF-AST2508940.
PFH was supported by a Simons Investigator Grant.
The simulations were performed on DOE OLCF Summit under allocations AST198 and AST226.
This research used resources of the Oak Ridge Leadership Computing Facility at the Oak Ridge National Laboratory, which is supported by the Office of Science of the U.S. Department of Energy. 
This research used resources of the National Energy Research Scientific Computing Center (NERSC), a Department of Energy User Facility using NERSC awards NP-ERCAP0028480 and ERCAP0034803.
\end{acknowledgements}

\begin{widetext}
\appendix

\section{Vertical Structure of Accretion Disks Varying Equation of States}\label{app:id}

When neglecting the contributions of the magnetic fields, the force balance in the vertical direction can be written as,
\begin{equation}
    - \rho \frac{\partial \Phi}{\partial z} - \frac{\partial P}{\partial z} = 0\,.
\end{equation}
Taylor expanding the gravitational potential $\Phi$ (or simply considering the projection of gravitational force along the vertical $z$-direction) gives,
\begin{equation}
    g_z = -\frac{\partial \Phi}{\partial z} = - \Omega_z^2 z = - \frac{GM}{(R^2+z^2)^{3/2}}z\,.
\end{equation}
where we have used the vertical frequency of the disk $\Omega_z=\Omega$ in the Keplerian case, and we are assuming a single gravitational source residing at the center. 

\subsection{Locally Isothermal}

When adopting a locally isothermal equation of state, $P=\rho T = \rho c_s^2$, the vertical profile of the disk in the cylindrical coordinates is,
\begin{align}
    \rho(R, z) &= \rho(R, 0) \exp\left[ -\frac{1}{c_s^2} \left( \Phi(R, z) - \Phi(R,0) \right) \right]\,, \\
    &= \rho(R, 0) \exp\left[ \frac{GM}{c_s^2} \left( \frac{1}{\sqrt{R^2+z^2}} - \frac{1}{R} \right) \right]\,.
\end{align}

\subsection{Adiabatic/Polytropic Relation}
When adopting a polytropic equation of state, $P=K \rho ^{\gamma}$, the vertical profile of the disk is (see, e.g., \cite{Kato2004}),
\begin{align}
    \rho(R, z) &= \rho(R, 0) \left[ 1 - \frac{1}{c_{s,0}^2} \frac{\gamma-1}{\gamma}\left( \Phi(R, z) - \Phi(R,0)\right)\right]^{1/(\gamma-1)}\,, \\
    &= \rho_0 \exp{\left[ -\left( \frac{R}{R_{\mathrm{edge}}} \right)^{-6} \right]} \left[ 1 + \frac{GM}{c_{s,0}^2} \frac{\gamma-1}{\gamma}\left( \frac{1}{\sqrt{R^2+z^2}} - \frac{1}{R} \right)\right]^{1/(\gamma-1)}\,,
\end{align}
where the constant $K$ in the equation of state is determined at a reference radius $r_0=a$: 
$K = c_{s,0}^2 \rho(R_{\mathrm{edge}},0)^{1-\gamma}$. The resulting pressure is given by,
\begin{equation}
    P(R,z) = c_{s,0}^2 \rho(R_{\mathrm{edge}},0)^{1-\gamma} \times \rho(R,z)^\gamma\,.
\end{equation}
The sound speed $c_{s,0}$ is chosen to be value at the edge of the cavity $R=R_{\mathrm{edge}}$.

\section{Angular momentum transport in cylindrical coordinates}
\label{app:amt}

In this section we want to briefly review the equations governing the transport of angular momentum in the circumbinary system. We do so by starting out with the ideal MHD equations in cylindrical coordinates. Denoting %
$R$ as the cylindrical radius and $z$ as the height coordinate, the conserved angular momentum density $j$
is then given by
\begin{align}
    {j = R^2 \rho v_\phi}\,.
\end{align}
{Using the momentum equation Eq. \eqref{governeq-2}, we then obtain},
\begin{align}
{
    \partial_t j + \partial_R \left( j v_R + R^2 B_R B_\phi \right) + \partial_\phi \left( {j} + R B_\phi B_\phi + R \bar{P}\right) + \partial_z \left(j v_z  + R^2 B_z B_\phi\right) - R \rho \partial_\phi \Phi_g = 0\,.}
\end{align}
In practice, we are mainly interested in the transport of angular momentum in radius $R$ and height $z$, so we will average over the $\phi$ direction. This is appropriate for both the outer and the individual minidisks.
Defining the azimuthally averaged angular momentum density,
\begin{align}
    \left<j\right>_\phi = \int_0^{2\pi}{\rm d}\, \phi j\,,
\end{align}
{we obtain}
\begin{align}
\partial_t \left<j\right>_\phi + \partial_R \left( \left<j v_R \right>_\phi + R^2 \left<B_R B_\phi\right>_\phi \right) + \partial_z \left(\left< j v_z \right>_\phi + R^2 \left<B_z B_\phi \right>_\phi \right) - R \left<\rho \partial_\phi \Phi_g\right>_\phi = 0\,,
\end{align}
where the azimuthal fluxes cancel out in the averaging operation.

Angular momentum flux advected along the disk can be calculated by doing the integration in the $\phi$ and z directions. For convenience, in the following we will use the notation:
\begin{align*}
\displaystyle \left< Q \right>_{\phi ,z}( R) &=\int _{0}^{2\pi } d\phi^\prime \int _{z_{\mathrm{lower}}}^{z_{\mathrm{upper}}} dz^\prime\cdot Q( R,\phi^\prime ,z^\prime) \,,\\
\left< Q \right>_{R} &=\int_{0}^{R} dR^\prime\cdot Q(R^\prime) \,,\\
\left< Q \right>_{t} &=\int dt^\prime\cdot Q(t^\prime)\,.
\end{align*}
The momentum equation can be further modified as,
\begin{align*}
\int dz[ \partial _{t}\left< j\right> _{\phi }]  = &-\int dz\partial _{R} \left< jv_{R}\right> _{\phi } -\int dz\partial _{z}\left< jv_{z}\right> _{\phi } \\
&-\int dz\partial _{R}\left( R^{2} \left< B_{R} B_{\phi }\right> _{\phi }\right) -\int dz\partial _{z}\left( R^{2} \left< B_{z} B_{\phi}\right> _{\phi }\right) -\int dzR \left< \rho \partial _{\phi } \Phi _{g}\right> _{\phi }\,,\\
\partial _{t} \left< j\right> _{\phi ,z}  = &-\partial _{R} \left< jv_{R} \right> _{\phi ,z} - \left< jv_{z} \right> _{\phi } |_{z_{\mathrm{lower}}}^{z_{\mathrm{upper}}} \\
&-\partial _{R}\left( R^{2} \left< B_{R} B_{\phi }\right> _{\phi ,z}\right) -R^{2} \left< B_{z} B_{\phi }\right> _{\phi } |_{z_{\mathrm{lower}}}^{z_{\mathrm{upper}}} -R \left< \rho \partial _{\phi } \Phi _{g} \right> _{\phi ,z}\,.
\end{align*}
We can reformulate the radial balance into a more compact form: 
\begin{equation*}
\partial _{t}\left(\frac{dJ}{dR}\right) =\frac{\partial \dot{J}_{\mathrm{adv} ,R}}{\partial R} +\frac{dT_{\mathrm{adv} ,z}}{dR} +\frac{\partial \dot{J}_{\mathrm{mag} ,R}}{\partial R} +\frac{dT_{mag,z}}{dR} +\frac{dT_{\mathrm{grav}}}{dR}\,,
\end{equation*}
where we have defined,

\begin{equation*}
\begin{aligned}
 & ( 1) \ \mathrm{radial\ dependence\ of\ total\ angular\ mometum} & \frac{dJ}{dR} & =- \left< j\right> _{\phi ,z}\,,\\
 & ( 2) \ \mathrm{inward\ angular\ momentum\ flux\ from\ advection} & \dot{J}_{\mathrm{adv} ,R} & =- \left< jv_{R} \right> _{\phi ,z}\,,\\
 & ( 3) \ \mathrm{torque\ from\ vertical\ transport( wind)} & \left< \frac{dT_{\mathrm{adv} ,z}}{dR}\right> _{R} & =-\int dR^\prime \left< jv_{z}\right> _{\phi } |_{z_{\mathrm{lower}}}^{z_{\mathrm{upper}}}\,,\\
 & ( 4) \ \mathrm{inward\ angular\ momentum\ flux\ from\ magnetic\ transport\ in\ the\ radial} & \dot{J}_{\mathrm{mag} ,R} & =-R^{2} \left< B_{R} B_{\phi }\right> _{\phi ,z}\,,\\
 & ( 5) \ \mathrm{magnetic\ torque\ per\ unit\ radius} & \left< \frac{dT_{mag,z}}{dR}\right> _{R} & =-\int dR^\prime R^{\prime 2} \left< B_{z} B_{\phi }\right> _{\phi } |_{z_{\mathrm{lower}}}^{z_{\mathrm{upper}}}\,,\\
 & ( 6) \ \mathrm{gravitational\ torque\ per\ unit\ radius} & \left< \frac{dT_{\mathrm{grav}}}{dR}\right> _{R} & =-\int dR^\prime R^\prime \left< \rho \partial _{\phi } \Phi _{g}\right> _{\phi ,z}\,.
\end{aligned}
\end{equation*}
After time averaging of the above equation and assuming a steady state of accretion $\displaystyle ( \partial _{t} =0)$, the accretion disk reaches a steady state, 
\begin{equation*}
\frac{d\left<\dot{J}\right>_t}{dR} =\frac{\partial }{\partial R}(\left< \dot{J}_{\mathrm{adv} ,R}\right> _{t} +\left< \dot{J}_{\mathrm{mag} ,R}\right> _{t}) +\left< \frac{dT_{\mathrm{adv} ,z}}{dR}\right> _{t} +\left< \frac{dT_{mag,z}}{dR}\right> _{t} +\left< \frac{dT_{\mathrm{grav}}}{dR}\right> _{t}\,.
\end{equation*}
Integrating along the radial direction from the coordinate pole (which is also the location where different torque components vanishes due to $\displaystyle R\equiv 0$), we have,
\begin{equation*}
 \left<\dot J_{B}\right>_t=\dot{\left<J\right>_t}=\left< \dot{J}_{\mathrm{adv} ,R}\right> _{t} +\left< \dot{J}_{\mathrm{mag} ,R}\right> _{t} +\left< \frac{dT_{\mathrm{adv} ,z}}{dR}\right> _{R,t} +\left< \frac{dT_{mag,z}}{dR}\right> _{R,t} +\left< \frac{dT_{\mathrm{grav}}}{dR}\right> _{R,t}\,,
\end{equation*}
where $\dot J_{B}$ is the total torque exerted directly onto the binary.

\end{widetext}
\bibliography{cbd_ref}

\begin{thebibliography}{183}%
\makeatletter
\providecommand \@ifxundefined [1]{%
 \@ifx{#1\undefined}
}%
\providecommand \@ifnum [1]{%
 \ifnum #1\expandafter \@firstoftwo
 \else \expandafter \@secondoftwo
 \fi
}%
\providecommand \@ifx [1]{%
 \ifx #1\expandafter \@firstoftwo
 \else \expandafter \@secondoftwo
 \fi
}%
\providecommand \natexlab [1]{#1}%
\providecommand \enquote  [1]{``#1''}%
\providecommand \bibnamefont  [1]{#1}%
\providecommand \bibfnamefont [1]{#1}%
\providecommand \citenamefont [1]{#1}%
\providecommand \href@noop [0]{\@secondoftwo}%
\providecommand \href [0]{\begingroup \@sanitize@url \@href}%
\providecommand \@href[1]{\@@startlink{#1}\@@href}%
\providecommand \@@href[1]{\endgroup#1\@@endlink}%
\providecommand \@sanitize@url [0]{\catcode `\\12\catcode `\$12\catcode `\&12\catcode `\#12\catcode `\^12\catcode `\_12\catcode `\%12\relax}%
\providecommand \@@startlink[1]{}%
\providecommand \@@endlink[0]{}%
\providecommand \url  [0]{\begingroup\@sanitize@url \@url }%
\providecommand \@url [1]{\endgroup\@href {#1}{\urlprefix }}%
\providecommand \urlprefix  [0]{URL }%
\providecommand \Eprint [0]{\href }%
\providecommand \doibase [0]{https://doi.org/}%
\providecommand \selectlanguage [0]{\@gobble}%
\providecommand \bibinfo  [0]{\@secondoftwo}%
\providecommand \bibfield  [0]{\@secondoftwo}%
\providecommand \translation [1]{[#1]}%
\providecommand \BibitemOpen [0]{}%
\providecommand \bibitemStop [0]{}%
\providecommand \bibitemNoStop [0]{.\EOS\space}%
\providecommand \EOS [0]{\spacefactor3000\relax}%
\providecommand \BibitemShut  [1]{\csname bibitem#1\endcsname}%
\let\auto@bib@innerbib\@empty
\bibitem [{\citenamefont {Lai}\ and\ \citenamefont {Mu{\~n}oz}(2023)}]{Lai:2022ylu}%
  \BibitemOpen
  \bibfield  {author} {\bibinfo {author} {\bibfnamefont {D.}~\bibnamefont {Lai}}\ and\ \bibinfo {author} {\bibfnamefont {D.~J.}\ \bibnamefont {Mu{\~n}oz}},\ }\bibfield  {title} {\bibinfo {title} {{Circumbinary Accretion: From Binary Stars to Massive Binary Black Holes}},\ }\href {https://doi.org/10.1146/annurev-astro-052622-022933} {\bibfield  {journal} {\bibinfo  {journal} {Ann. Rev. Astron. Astrophys.}\ }\textbf {\bibinfo {volume} {61}},\ \bibinfo {pages} {517} (\bibinfo {year} {2023})},\ \Eprint {https://arxiv.org/abs/2211.00028} {arXiv:2211.00028 [astro-ph.HE]} \BibitemShut {NoStop}%
\bibitem [{\citenamefont {{Williams}}\ and\ \citenamefont {{Cieza}}(2011)}]{2011ARA&A..49...67W}%
  \BibitemOpen
  \bibfield  {author} {\bibinfo {author} {\bibfnamefont {J.~P.}\ \bibnamefont {{Williams}}}\ and\ \bibinfo {author} {\bibfnamefont {L.~A.}\ \bibnamefont {{Cieza}}},\ }\bibfield  {title} {\bibinfo {title} {{Protoplanetary Disks and Their Evolution}},\ }\href {https://doi.org/10.1146/annurev-astro-081710-102548} {\bibfield  {journal} {\bibinfo  {journal} {Ann. Rev. Astron. Astrophys.}\ }\textbf {\bibinfo {volume} {49}},\ \bibinfo {pages} {67} (\bibinfo {year} {2011})},\ \Eprint {https://arxiv.org/abs/1103.0556} {arXiv:1103.0556 [astro-ph.GA]} \BibitemShut {NoStop}%
\bibitem [{\citenamefont {{Kratter}}\ \emph {et~al.}(2008)\citenamefont {{Kratter}}, \citenamefont {{Matzner}},\ and\ \citenamefont {{Krumholz}}}]{2008ApJ...681..375K}%
  \BibitemOpen
  \bibfield  {author} {\bibinfo {author} {\bibfnamefont {K.~M.}\ \bibnamefont {{Kratter}}}, \bibinfo {author} {\bibfnamefont {C.~D.}\ \bibnamefont {{Matzner}}},\ and\ \bibinfo {author} {\bibfnamefont {M.~R.}\ \bibnamefont {{Krumholz}}},\ }\bibfield  {title} {\bibinfo {title} {{Global Models for the Evolution of Embedded, Accreting Protostellar Disks}},\ }\href {https://doi.org/10.1086/587543} {\bibfield  {journal} {\bibinfo  {journal} {Astrophys. J.}\ }\textbf {\bibinfo {volume} {681}},\ \bibinfo {pages} {375} (\bibinfo {year} {2008})},\ \Eprint {https://arxiv.org/abs/0709.4252} {arXiv:0709.4252 [astro-ph]} \BibitemShut {NoStop}%
\bibitem [{\citenamefont {{Matsumoto}}(2024)}]{2024ApJ...964..133M}%
  \BibitemOpen
  \bibfield  {author} {\bibinfo {author} {\bibfnamefont {T.}~\bibnamefont {{Matsumoto}}},\ }\bibfield  {title} {\bibinfo {title} {{Angular Momentum Transport in Binary Star Formation: The Enhancement of Magnetorotational Instability and Role of Outflows}},\ }\href {https://doi.org/10.3847/1538-4357/ad25ee} {\bibfield  {journal} {\bibinfo  {journal} {Astrophys. J.}\ }\textbf {\bibinfo {volume} {964}},\ \bibinfo {eid} {133} (\bibinfo {year} {2024})},\ \Eprint {https://arxiv.org/abs/2402.03212} {arXiv:2402.03212 [astro-ph.SR]} \BibitemShut {NoStop}%
\bibitem [{\citenamefont {{Li}}\ \emph {et~al.}(2021)\citenamefont {{Li}}, \citenamefont {{Dempsey}}, \citenamefont {{Li}}, \citenamefont {{Li}},\ and\ \citenamefont {{Li}}}]{2021ApJ...911..124L}%
  \BibitemOpen
  \bibfield  {author} {\bibinfo {author} {\bibfnamefont {Y.-P.}\ \bibnamefont {{Li}}}, \bibinfo {author} {\bibfnamefont {A.~M.}\ \bibnamefont {{Dempsey}}}, \bibinfo {author} {\bibfnamefont {S.}~\bibnamefont {{Li}}}, \bibinfo {author} {\bibfnamefont {H.}~\bibnamefont {{Li}}},\ and\ \bibinfo {author} {\bibfnamefont {J.}~\bibnamefont {{Li}}},\ }\bibfield  {title} {\bibinfo {title} {{Orbital Evolution of Binary Black Holes in Active Galactic Nucleus Disks: A Disk Channel for Binary Black Hole Mergers?}},\ }\href {https://doi.org/10.3847/1538-4357/abed48} {\bibfield  {journal} {\bibinfo  {journal} {Astrophys. J.}\ }\textbf {\bibinfo {volume} {911}},\ \bibinfo {eid} {124} (\bibinfo {year} {2021})},\ \Eprint {https://arxiv.org/abs/2101.09406} {arXiv:2101.09406 [astro-ph.HE]} \BibitemShut {NoStop}%
\bibitem [{\citenamefont {{Li}}\ \emph {et~al.}(2022)\citenamefont {{Li}}, \citenamefont {{Dempsey}}, \citenamefont {{Li}}, \citenamefont {{Li}},\ and\ \citenamefont {{Li}}}]{2022ApJ...928L..19L}%
  \BibitemOpen
  \bibfield  {author} {\bibinfo {author} {\bibfnamefont {Y.-P.}\ \bibnamefont {{Li}}}, \bibinfo {author} {\bibfnamefont {A.~M.}\ \bibnamefont {{Dempsey}}}, \bibinfo {author} {\bibfnamefont {H.}~\bibnamefont {{Li}}}, \bibinfo {author} {\bibfnamefont {S.}~\bibnamefont {{Li}}},\ and\ \bibinfo {author} {\bibfnamefont {J.}~\bibnamefont {{Li}}},\ }\bibfield  {title} {\bibinfo {title} {{Hot Circumsingle Disks Drive Binary Black Hole Mergers in Active Galactic Nucleus Disks}},\ }\href {https://doi.org/10.3847/2041-8213/ac60fd} {\bibfield  {journal} {\bibinfo  {journal} {Astrophys. J. Lett.}\ }\textbf {\bibinfo {volume} {928}},\ \bibinfo {eid} {L19} (\bibinfo {year} {2022})},\ \Eprint {https://arxiv.org/abs/2112.11057} {arXiv:2112.11057 [astro-ph.HE]} \BibitemShut {NoStop}%
\bibitem [{\citenamefont {Mayer}\ \emph {et~al.}(2007)\citenamefont {Mayer}, \citenamefont {Kazantzidis}, \citenamefont {Madau}, \citenamefont {Colpi}, \citenamefont {Quinn},\ and\ \citenamefont {Wadsley}}]{Mayer:2007vk}%
  \BibitemOpen
  \bibfield  {author} {\bibinfo {author} {\bibfnamefont {L.}~\bibnamefont {Mayer}}, \bibinfo {author} {\bibfnamefont {S.}~\bibnamefont {Kazantzidis}}, \bibinfo {author} {\bibfnamefont {P.}~\bibnamefont {Madau}}, \bibinfo {author} {\bibfnamefont {M.}~\bibnamefont {Colpi}}, \bibinfo {author} {\bibfnamefont {T.~R.}\ \bibnamefont {Quinn}},\ and\ \bibinfo {author} {\bibfnamefont {J.}~\bibnamefont {Wadsley}},\ }\bibfield  {title} {\bibinfo {title} {{Rapid Formation of Supermassive Black Hole Binaries in Galaxy Mergers with Gas}},\ }\href {https://doi.org/10.1126/science.1141858} {\bibfield  {journal} {\bibinfo  {journal} {Science}\ }\textbf {\bibinfo {volume} {316}},\ \bibinfo {pages} {1874} (\bibinfo {year} {2007})},\ \Eprint {https://arxiv.org/abs/0706.1562} {arXiv:0706.1562 [astro-ph]} \BibitemShut {NoStop}%
\bibitem [{\citenamefont {Agazie}\ \emph {et~al.}(2023{\natexlab{a}})\citenamefont {Agazie} \emph {et~al.}}]{Nanograv2023a}%
  \BibitemOpen
  \bibfield  {author} {\bibinfo {author} {\bibfnamefont {G.}~\bibnamefont {Agazie}} \emph {et~al.} (\bibinfo {collaboration} {NANOGrav}),\ }\bibfield  {title} {\bibinfo {title} {{The NANOGrav 15 yr Data Set: Evidence for a Gravitational-wave Background}},\ }\href {https://doi.org/10.3847/2041-8213/acdac6} {\bibfield  {journal} {\bibinfo  {journal} {Astrophys. J. Lett.}\ }\textbf {\bibinfo {volume} {951}},\ \bibinfo {pages} {L8} (\bibinfo {year} {2023}{\natexlab{a}})},\ \Eprint {https://arxiv.org/abs/2306.16213} {arXiv:2306.16213 [astro-ph.HE]} \BibitemShut {NoStop}%
\bibitem [{\citenamefont {Agazie}\ \emph {et~al.}(2023{\natexlab{b}})\citenamefont {Agazie} \emph {et~al.}}]{Nanograv2023b}%
  \BibitemOpen
  \bibfield  {author} {\bibinfo {author} {\bibfnamefont {G.}~\bibnamefont {Agazie}} \emph {et~al.} (\bibinfo {collaboration} {NANOGrav}),\ }\bibfield  {title} {\bibinfo {title} {{The NANOGrav 15 yr Data Set: Constraints on Supermassive Black Hole Binaries from the Gravitational-wave Background}},\ }\href {https://doi.org/10.3847/2041-8213/ace18b} {\bibfield  {journal} {\bibinfo  {journal} {Astrophys. J. Lett.}\ }\textbf {\bibinfo {volume} {952}},\ \bibinfo {pages} {L37} (\bibinfo {year} {2023}{\natexlab{b}})},\ \Eprint {https://arxiv.org/abs/2306.16220} {arXiv:2306.16220 [astro-ph.HE]} \BibitemShut {NoStop}%
\bibitem [{\citenamefont {Agazie}\ \emph {et~al.}(2024{\natexlab{a}})\citenamefont {Agazie} \emph {et~al.}}]{InternationalPulsarTimingArray:2023mzf}%
  \BibitemOpen
  \bibfield  {author} {\bibinfo {author} {\bibfnamefont {G.}~\bibnamefont {Agazie}} \emph {et~al.} (\bibinfo {collaboration} {International Pulsar Timing Array}),\ }\bibfield  {title} {\bibinfo {title} {{Comparing Recent Pulsar Timing Array Results on the Nanohertz Stochastic Gravitational-wave Background}},\ }\href {https://doi.org/10.3847/1538-4357/ad36be} {\bibfield  {journal} {\bibinfo  {journal} {Astrophys. J.}\ }\textbf {\bibinfo {volume} {966}},\ \bibinfo {pages} {105} (\bibinfo {year} {2024}{\natexlab{a}})},\ \Eprint {https://arxiv.org/abs/2309.00693} {arXiv:2309.00693 [astro-ph.HE]} \BibitemShut {NoStop}%
\bibitem [{\citenamefont {Agazie}\ \emph {et~al.}(2023{\natexlab{c}})\citenamefont {Agazie} \emph {et~al.}}]{NANOGrav:2023pdq}%
  \BibitemOpen
  \bibfield  {author} {\bibinfo {author} {\bibfnamefont {G.}~\bibnamefont {Agazie}} \emph {et~al.} (\bibinfo {collaboration} {NANOGrav}),\ }\bibfield  {title} {\bibinfo {title} {{The NANOGrav 15 yr Data Set: Bayesian Limits on Gravitational Waves from Individual Supermassive Black Hole Binaries}},\ }\href {https://doi.org/10.3847/2041-8213/ace18a} {\bibfield  {journal} {\bibinfo  {journal} {Astrophys. J. Lett.}\ }\textbf {\bibinfo {volume} {951}},\ \bibinfo {pages} {L50} (\bibinfo {year} {2023}{\natexlab{c}})},\ \Eprint {https://arxiv.org/abs/2306.16222} {arXiv:2306.16222 [astro-ph.HE]} \BibitemShut {NoStop}%
\bibitem [{\citenamefont {Agazie}\ \emph {et~al.}(2024{\natexlab{b}})\citenamefont {Agazie} \emph {et~al.}}]{NANOGrav:2023vfo}%
  \BibitemOpen
  \bibfield  {author} {\bibinfo {author} {\bibfnamefont {G.}~\bibnamefont {Agazie}} \emph {et~al.} (\bibinfo {collaboration} {NANOGrav}),\ }\bibfield  {title} {\bibinfo {title} {{The NANOGrav 12.5 yr Data Set: Search for Gravitational Wave Memory}},\ }\href {https://doi.org/10.3847/1538-4357/ad0726} {\bibfield  {journal} {\bibinfo  {journal} {Astrophys. J.}\ }\textbf {\bibinfo {volume} {963}},\ \bibinfo {pages} {61} (\bibinfo {year} {2024}{\natexlab{b}})},\ \Eprint {https://arxiv.org/abs/2307.13797} {arXiv:2307.13797 [gr-qc]} \BibitemShut {NoStop}%
\bibitem [{\citenamefont {Agazie}\ \emph {et~al.}(2023{\natexlab{d}})\citenamefont {Agazie} \emph {et~al.}}]{NANOGrav:2023hde}%
  \BibitemOpen
  \bibfield  {author} {\bibinfo {author} {\bibfnamefont {G.}~\bibnamefont {Agazie}} \emph {et~al.} (\bibinfo {collaboration} {NANOGrav}),\ }\bibfield  {title} {\bibinfo {title} {{The NANOGrav 15 yr Data Set: Observations and Timing of 68 Millisecond Pulsars}},\ }\href {https://doi.org/10.3847/2041-8213/acda9a} {\bibfield  {journal} {\bibinfo  {journal} {Astrophys. J. Lett.}\ }\textbf {\bibinfo {volume} {951}},\ \bibinfo {pages} {L9} (\bibinfo {year} {2023}{\natexlab{d}})},\ \Eprint {https://arxiv.org/abs/2306.16217} {arXiv:2306.16217 [astro-ph.HE]} \BibitemShut {NoStop}%
\bibitem [{\citenamefont {Agazie}\ \emph {et~al.}(2023{\natexlab{e}})\citenamefont {Agazie} \emph {et~al.}}]{NANOGrav:2023ctt}%
  \BibitemOpen
  \bibfield  {author} {\bibinfo {author} {\bibfnamefont {G.}~\bibnamefont {Agazie}} \emph {et~al.} (\bibinfo {collaboration} {NANOGrav}),\ }\bibfield  {title} {\bibinfo {title} {{The NANOGrav 15 yr Data Set: Detector Characterization and Noise Budget}},\ }\href {https://doi.org/10.3847/2041-8213/acda88} {\bibfield  {journal} {\bibinfo  {journal} {Astrophys. J. Lett.}\ }\textbf {\bibinfo {volume} {951}},\ \bibinfo {pages} {L10} (\bibinfo {year} {2023}{\natexlab{e}})},\ \Eprint {https://arxiv.org/abs/2306.16218} {arXiv:2306.16218 [astro-ph.HE]} \BibitemShut {NoStop}%
\bibitem [{\citenamefont {Agazie}\ \emph {et~al.}(2023{\natexlab{f}})\citenamefont {Agazie} \emph {et~al.}}]{NANOGrav:2023tcn}%
  \BibitemOpen
  \bibfield  {author} {\bibinfo {author} {\bibfnamefont {G.}~\bibnamefont {Agazie}} \emph {et~al.} (\bibinfo {collaboration} {NANOGrav}),\ }\bibfield  {title} {\bibinfo {title} {{The NANOGrav 15 yr Data Set: Search for Anisotropy in the Gravitational-wave Background}},\ }\href {https://doi.org/10.3847/2041-8213/acf4fd} {\bibfield  {journal} {\bibinfo  {journal} {Astrophys. J. Lett.}\ }\textbf {\bibinfo {volume} {956}},\ \bibinfo {pages} {L3} (\bibinfo {year} {2023}{\natexlab{f}})},\ \Eprint {https://arxiv.org/abs/2306.16221} {arXiv:2306.16221 [astro-ph.HE]} \BibitemShut {NoStop}%
\bibitem [{\citenamefont {Xu}\ \emph {et~al.}(2023)\citenamefont {Xu} \emph {et~al.}}]{Xu:2023wog}%
  \BibitemOpen
  \bibfield  {author} {\bibinfo {author} {\bibfnamefont {H.}~\bibnamefont {Xu}} \emph {et~al.},\ }\bibfield  {title} {\bibinfo {title} {{Searching for the Nano-Hertz Stochastic Gravitational Wave Background with the Chinese Pulsar Timing Array Data Release I}},\ }\href {https://doi.org/10.1088/1674-4527/acdfa5} {\bibfield  {journal} {\bibinfo  {journal} {Res. Astron. Astrophys.}\ }\textbf {\bibinfo {volume} {23}},\ \bibinfo {pages} {075024} (\bibinfo {year} {2023})},\ \Eprint {https://arxiv.org/abs/2306.16216} {arXiv:2306.16216 [astro-ph.HE]} \BibitemShut {NoStop}%
\bibitem [{\citenamefont {Seoane}\ \emph {et~al.}(2023)\citenamefont {Seoane} \emph {et~al.}}]{lisa2023}%
  \BibitemOpen
  \bibfield  {author} {\bibinfo {author} {\bibfnamefont {P.~A.}\ \bibnamefont {Seoane}} \emph {et~al.} (\bibinfo {collaboration} {LISA}),\ }\bibfield  {title} {\bibinfo {title} {{Astrophysics with the Laser Interferometer Space Antenna}},\ }\href {https://doi.org/10.1007/s41114-022-00041-y} {\bibfield  {journal} {\bibinfo  {journal} {Living Rev. Rel.}\ }\textbf {\bibinfo {volume} {26}},\ \bibinfo {pages} {2} (\bibinfo {year} {2023})},\ \Eprint {https://arxiv.org/abs/2203.06016} {arXiv:2203.06016 [gr-qc]} \BibitemShut {NoStop}%
\bibitem [{\citenamefont {Li}\ \emph {et~al.}(2025)\citenamefont {Li} \emph {et~al.}}]{tianqin}%
  \BibitemOpen
  \bibfield  {author} {\bibinfo {author} {\bibfnamefont {E.-K.}\ \bibnamefont {Li}} \emph {et~al.},\ }\bibfield  {title} {\bibinfo {title} {{Gravitational wave astronomy with TianQin}},\ }\href {https://doi.org/10.1088/1361-6633/adc9be} {\bibfield  {journal} {\bibinfo  {journal} {Rept. Prog. Phys.}\ }\textbf {\bibinfo {volume} {88}},\ \bibinfo {pages} {056901} (\bibinfo {year} {2025})},\ \Eprint {https://arxiv.org/abs/2409.19665} {arXiv:2409.19665 [astro-ph.GA]} \BibitemShut {NoStop}%
\bibitem [{\citenamefont {Barnes}\ and\ \citenamefont {Hernquist}(1991)}]{Barnes:1991zz}%
  \BibitemOpen
  \bibfield  {author} {\bibinfo {author} {\bibfnamefont {J.~E.}\ \bibnamefont {Barnes}}\ and\ \bibinfo {author} {\bibfnamefont {L.~E.}\ \bibnamefont {Hernquist}},\ }\bibfield  {title} {\bibinfo {title} {{Fueling starburst galaxies with gas-rich mergers}},\ }\href {https://doi.org/10.1086/185978} {\bibfield  {journal} {\bibinfo  {journal} {Astrophys. J. Lett.}\ }\textbf {\bibinfo {volume} {370}},\ \bibinfo {pages} {L65} (\bibinfo {year} {1991})}\BibitemShut {NoStop}%
\bibitem [{\citenamefont {Barnes}\ and\ \citenamefont {Hernquist}(1992)}]{Barnes:1992rm}%
  \BibitemOpen
  \bibfield  {author} {\bibinfo {author} {\bibfnamefont {J.~E.}\ \bibnamefont {Barnes}}\ and\ \bibinfo {author} {\bibfnamefont {L.~E.}\ \bibnamefont {Hernquist}},\ }\bibfield  {title} {\bibinfo {title} {{Dynamics of interacting galaxies}},\ }\href {https://doi.org/10.1146/annurev.aa.30.090192.003421} {\bibfield  {journal} {\bibinfo  {journal} {Ann. Rev. Astron. Astrophys.}\ }\textbf {\bibinfo {volume} {30}},\ \bibinfo {pages} {705} (\bibinfo {year} {1992})}\BibitemShut {NoStop}%
\bibitem [{\citenamefont {Mihos}\ and\ \citenamefont {Hernquist}(1996)}]{Mihos:1995ri}%
  \BibitemOpen
  \bibfield  {author} {\bibinfo {author} {\bibfnamefont {J.~C.}\ \bibnamefont {Mihos}}\ and\ \bibinfo {author} {\bibfnamefont {L.}~\bibnamefont {Hernquist}},\ }\bibfield  {title} {\bibinfo {title} {{Gasdynamics and starbursts in major mergers}},\ }\href {https://doi.org/10.1086/177353} {\bibfield  {journal} {\bibinfo  {journal} {Astrophys. J.}\ }\textbf {\bibinfo {volume} {464}},\ \bibinfo {pages} {641} (\bibinfo {year} {1996})},\ \Eprint {https://arxiv.org/abs/astro-ph/9512099} {arXiv:astro-ph/9512099} \BibitemShut {NoStop}%
\bibitem [{\citenamefont {Barnes}(2002)}]{Barnes:2002sh}%
  \BibitemOpen
  \bibfield  {author} {\bibinfo {author} {\bibfnamefont {J.~E.}\ \bibnamefont {Barnes}},\ }\bibfield  {title} {\bibinfo {title} {{Formation of gas disks in merging galaxies}},\ }\href {https://doi.org/10.1046/j.1365-8711.2002.05335.x} {\bibfield  {journal} {\bibinfo  {journal} {Mon. Not. Roy. Astron. Soc.}\ }\textbf {\bibinfo {volume} {333}},\ \bibinfo {pages} {481} (\bibinfo {year} {2002})},\ \Eprint {https://arxiv.org/abs/astro-ph/0201250} {arXiv:astro-ph/0201250} \BibitemShut {NoStop}%
\bibitem [{\citenamefont {Di~Matteo}\ \emph {et~al.}(2005)\citenamefont {Di~Matteo}, \citenamefont {Springel},\ and\ \citenamefont {Hernquist}}]{DiMatteo:2005ttp}%
  \BibitemOpen
  \bibfield  {author} {\bibinfo {author} {\bibfnamefont {T.}~\bibnamefont {Di~Matteo}}, \bibinfo {author} {\bibfnamefont {V.}~\bibnamefont {Springel}},\ and\ \bibinfo {author} {\bibfnamefont {L.}~\bibnamefont {Hernquist}},\ }\bibfield  {title} {\bibinfo {title} {{Energy input from quasars regulates the growth and activity of black holes and their host galaxies}},\ }\href {https://doi.org/10.1038/nature03335} {\bibfield  {journal} {\bibinfo  {journal} {Nature}\ }\textbf {\bibinfo {volume} {433}},\ \bibinfo {pages} {604} (\bibinfo {year} {2005})},\ \Eprint {https://arxiv.org/abs/astro-ph/0502199} {arXiv:astro-ph/0502199} \BibitemShut {NoStop}%
\bibitem [{\citenamefont {Hopkins}\ \emph {et~al.}(2006)\citenamefont {Hopkins}, \citenamefont {Hernquist}, \citenamefont {Cox}, \citenamefont {Di~Matteo}, \citenamefont {Robertson},\ and\ \citenamefont {Springel}}]{Hopkins:2005fb}%
  \BibitemOpen
  \bibfield  {author} {\bibinfo {author} {\bibfnamefont {P.~F.}\ \bibnamefont {Hopkins}}, \bibinfo {author} {\bibfnamefont {L.}~\bibnamefont {Hernquist}}, \bibinfo {author} {\bibfnamefont {T.~J.}\ \bibnamefont {Cox}}, \bibinfo {author} {\bibfnamefont {T.}~\bibnamefont {Di~Matteo}}, \bibinfo {author} {\bibfnamefont {B.}~\bibnamefont {Robertson}},\ and\ \bibinfo {author} {\bibfnamefont {V.}~\bibnamefont {Springel}},\ }\bibfield  {title} {\bibinfo {title} {{A Unified, merger-driven model for the origin of starbursts, quasars, the cosmic x-ray background, supermassive black holes and galaxy spheroids}},\ }\href {https://doi.org/10.1086/499298} {\bibfield  {journal} {\bibinfo  {journal} {Astrophys. J. Suppl.}\ }\textbf {\bibinfo {volume} {163}},\ \bibinfo {pages} {1} (\bibinfo {year} {2006})},\ \Eprint {https://arxiv.org/abs/astro-ph/0506398} {arXiv:astro-ph/0506398} \BibitemShut {NoStop}%
\bibitem [{\citenamefont {Hopkins}\ \emph {et~al.}(2013)\citenamefont {Hopkins}, \citenamefont {Cox}, \citenamefont {Hernquist}, \citenamefont {Narayanan}, \citenamefont {Hayward},\ and\ \citenamefont {Murray}}]{Hopkins:2012fd}%
  \BibitemOpen
  \bibfield  {author} {\bibinfo {author} {\bibfnamefont {P.~F.}\ \bibnamefont {Hopkins}}, \bibinfo {author} {\bibfnamefont {T.~J.}\ \bibnamefont {Cox}}, \bibinfo {author} {\bibfnamefont {L.}~\bibnamefont {Hernquist}}, \bibinfo {author} {\bibfnamefont {D.}~\bibnamefont {Narayanan}}, \bibinfo {author} {\bibfnamefont {C.~C.}\ \bibnamefont {Hayward}},\ and\ \bibinfo {author} {\bibfnamefont {N.}~\bibnamefont {Murray}},\ }\bibfield  {title} {\bibinfo {title} {{Star Formation in Galaxy Mergers with Realistic Models of Stellar Feedback \& the Interstellar Medium}},\ }\href {https://doi.org/10.1093/mnras/stt017} {\bibfield  {journal} {\bibinfo  {journal} {Mon. Not. Roy. Astron. Soc.}\ }\textbf {\bibinfo {volume} {430}},\ \bibinfo {pages} {1901} (\bibinfo {year} {2013})},\ \Eprint {https://arxiv.org/abs/1206.0011} {arXiv:1206.0011 [astro-ph.CO]} \BibitemShut {NoStop}%
\bibitem [{\citenamefont {Cox}\ \emph {et~al.}(2008)\citenamefont {Cox}, \citenamefont {Jonsson}, \citenamefont {Somerville}, \citenamefont {Primack},\ and\ \citenamefont {Dekel}}]{Cox:2007mn}%
  \BibitemOpen
  \bibfield  {author} {\bibinfo {author} {\bibfnamefont {T.~J.}\ \bibnamefont {Cox}}, \bibinfo {author} {\bibfnamefont {P.~B.}\ \bibnamefont {Jonsson}}, \bibinfo {author} {\bibfnamefont {R.~S.}\ \bibnamefont {Somerville}}, \bibinfo {author} {\bibfnamefont {J.~R.}\ \bibnamefont {Primack}},\ and\ \bibinfo {author} {\bibfnamefont {A.}~\bibnamefont {Dekel}},\ }\bibfield  {title} {\bibinfo {title} {{The effect of galaxy mass ratio on merger-driven starbursts}},\ }\href {https://doi.org/10.1111/j.1365-2966.2007.12730.x} {\bibfield  {journal} {\bibinfo  {journal} {Mon. Not. Roy. Astron. Soc.}\ }\textbf {\bibinfo {volume} {384}},\ \bibinfo {pages} {386} (\bibinfo {year} {2008})},\ \Eprint {https://arxiv.org/abs/0709.3511} {arXiv:0709.3511 [astro-ph]} \BibitemShut {NoStop}%
\bibitem [{\citenamefont {Johansson}\ \emph {et~al.}(2009)\citenamefont {Johansson}, \citenamefont {Naab},\ and\ \citenamefont {Burkert}}]{Johansson:2008ib}%
  \BibitemOpen
  \bibfield  {author} {\bibinfo {author} {\bibfnamefont {P.~H.}\ \bibnamefont {Johansson}}, \bibinfo {author} {\bibfnamefont {T.}~\bibnamefont {Naab}},\ and\ \bibinfo {author} {\bibfnamefont {A.}~\bibnamefont {Burkert}},\ }\bibfield  {title} {\bibinfo {title} {{Equal- and unequal-mass mergers of disk and elliptical galaxies with black holes: The M\_BH-sigma and M\_BH-M\_* relations}},\ }\href {https://doi.org/10.1088/0004-637X/690/1/802} {\bibfield  {journal} {\bibinfo  {journal} {Astrophys. J.}\ }\textbf {\bibinfo {volume} {690}},\ \bibinfo {pages} {802} (\bibinfo {year} {2009})},\ \Eprint {https://arxiv.org/abs/0802.0210} {arXiv:0802.0210 [astro-ph]} \BibitemShut {NoStop}%
\bibitem [{\citenamefont {Capelo}\ \emph {et~al.}(2015)\citenamefont {Capelo}, \citenamefont {Volonteri}, \citenamefont {Dotti}, \citenamefont {Bellovary}, \citenamefont {Mayer},\ and\ \citenamefont {Governato}}]{Capelo:2014gqa}%
  \BibitemOpen
  \bibfield  {author} {\bibinfo {author} {\bibfnamefont {P.~R.}\ \bibnamefont {Capelo}}, \bibinfo {author} {\bibfnamefont {M.}~\bibnamefont {Volonteri}}, \bibinfo {author} {\bibfnamefont {M.}~\bibnamefont {Dotti}}, \bibinfo {author} {\bibfnamefont {J.~M.}\ \bibnamefont {Bellovary}}, \bibinfo {author} {\bibfnamefont {L.}~\bibnamefont {Mayer}},\ and\ \bibinfo {author} {\bibfnamefont {F.}~\bibnamefont {Governato}},\ }\bibfield  {title} {\bibinfo {title} {{Growth and activity of black holes in galaxy mergers with varying mass ratios}},\ }\href {https://doi.org/10.1093/mnras/stu2500} {\bibfield  {journal} {\bibinfo  {journal} {Mon. Not. Roy. Astron. Soc.}\ }\textbf {\bibinfo {volume} {447}},\ \bibinfo {pages} {2123} (\bibinfo {year} {2015})},\ \Eprint {https://arxiv.org/abs/1409.0004} {arXiv:1409.0004 [astro-ph.GA]} \BibitemShut {NoStop}%
\bibitem [{\citenamefont {Capelo}\ and\ \citenamefont {Dotti}(2017)}]{Capelo:2016vif}%
  \BibitemOpen
  \bibfield  {author} {\bibinfo {author} {\bibfnamefont {P.~R.}\ \bibnamefont {Capelo}}\ and\ \bibinfo {author} {\bibfnamefont {M.}~\bibnamefont {Dotti}},\ }\bibfield  {title} {\bibinfo {title} {{Shocks and angular momentum flips: a different path to feeding the nuclear regions of merging galaxies}},\ }\href {https://doi.org/10.1093/mnras/stw2872} {\bibfield  {journal} {\bibinfo  {journal} {Mon. Not. Roy. Astron. Soc.}\ }\textbf {\bibinfo {volume} {465}},\ \bibinfo {pages} {2643} (\bibinfo {year} {2017})},\ \Eprint {https://arxiv.org/abs/1610.08507} {arXiv:1610.08507 [astro-ph.GA]} \BibitemShut {NoStop}%
\bibitem [{\citenamefont {Dunhill}\ \emph {et~al.}(2014)\citenamefont {Dunhill}, \citenamefont {Alexander}, \citenamefont {Nixon},\ and\ \citenamefont {King}}]{Dunhill:2014oka}%
  \BibitemOpen
  \bibfield  {author} {\bibinfo {author} {\bibfnamefont {A.}~\bibnamefont {Dunhill}}, \bibinfo {author} {\bibfnamefont {R.}~\bibnamefont {Alexander}}, \bibinfo {author} {\bibfnamefont {C.}~\bibnamefont {Nixon}},\ and\ \bibinfo {author} {\bibfnamefont {A.}~\bibnamefont {King}},\ }\bibfield  {title} {\bibinfo {title} {{Misaligned accretion on to supermassive black hole binaries}},\ }\href {https://doi.org/10.1093/mnras/stu1914} {\bibfield  {journal} {\bibinfo  {journal} {Mon. Not. Roy. Astron. Soc.}\ }\textbf {\bibinfo {volume} {445}},\ \bibinfo {pages} {2285} (\bibinfo {year} {2014})},\ \Eprint {https://arxiv.org/abs/1409.3842} {arXiv:1409.3842 [astro-ph.HE]} \BibitemShut {NoStop}%
\bibitem [{\citenamefont {Goicovic}\ \emph {et~al.}(2016)\citenamefont {Goicovic}, \citenamefont {Cuadra}, \citenamefont {Sesana}, \citenamefont {Stasyszyn}, \citenamefont {Amaro-Seoane},\ and\ \citenamefont {Tanaka}}]{Goicovic:2015kda}%
  \BibitemOpen
  \bibfield  {author} {\bibinfo {author} {\bibfnamefont {F.~G.}\ \bibnamefont {Goicovic}}, \bibinfo {author} {\bibfnamefont {J.}~\bibnamefont {Cuadra}}, \bibinfo {author} {\bibfnamefont {A.}~\bibnamefont {Sesana}}, \bibinfo {author} {\bibfnamefont {F.}~\bibnamefont {Stasyszyn}}, \bibinfo {author} {\bibfnamefont {P.}~\bibnamefont {Amaro-Seoane}},\ and\ \bibinfo {author} {\bibfnamefont {T.~L.}\ \bibnamefont {Tanaka}},\ }\bibfield  {title} {\bibinfo {title} {{Infalling clouds on to supermassive black hole binaries \textendash{} I. Formation of discs, accretion and gas dynamics}},\ }\href {https://doi.org/10.1093/mnras/stv2470} {\bibfield  {journal} {\bibinfo  {journal} {Mon. Not. Roy. Astron. Soc.}\ }\textbf {\bibinfo {volume} {455}},\ \bibinfo {pages} {1989} (\bibinfo {year} {2016})},\ \Eprint {https://arxiv.org/abs/1507.05596} {arXiv:1507.05596 [astro-ph.HE]} \BibitemShut {NoStop}%
\bibitem [{\citenamefont {Goicovic}\ \emph {et~al.}(2017)\citenamefont {Goicovic}, \citenamefont {Sesana}, \citenamefont {Cuadra},\ and\ \citenamefont {Stasyszyn}}]{Goicovic:2016dul}%
  \BibitemOpen
  \bibfield  {author} {\bibinfo {author} {\bibfnamefont {F.~G.}\ \bibnamefont {Goicovic}}, \bibinfo {author} {\bibfnamefont {A.}~\bibnamefont {Sesana}}, \bibinfo {author} {\bibfnamefont {J.}~\bibnamefont {Cuadra}},\ and\ \bibinfo {author} {\bibfnamefont {F.}~\bibnamefont {Stasyszyn}},\ }\bibfield  {title} {\bibinfo {title} {{Infalling clouds on to supermassive black hole binaries \textendash{} II. Binary evolution and the final parsec problem}},\ }\href {https://doi.org/10.1093/mnras/stx1996} {\bibfield  {journal} {\bibinfo  {journal} {Mon. Not. Roy. Astron. Soc.}\ }\textbf {\bibinfo {volume} {472}},\ \bibinfo {pages} {514} (\bibinfo {year} {2017})},\ \Eprint {https://arxiv.org/abs/1602.01966} {arXiv:1602.01966 [astro-ph.HE]} \BibitemShut {NoStop}%
\bibitem [{\citenamefont {Goicovic}\ \emph {et~al.}(2018)\citenamefont {Goicovic}, \citenamefont {Maureira-Fredes}, \citenamefont {Sesana}, \citenamefont {Amaro-Seoane},\ and\ \citenamefont {Cuadra}}]{Goicovic:2018xxi}%
  \BibitemOpen
  \bibfield  {author} {\bibinfo {author} {\bibfnamefont {F.~G.}\ \bibnamefont {Goicovic}}, \bibinfo {author} {\bibfnamefont {C.}~\bibnamefont {Maureira-Fredes}}, \bibinfo {author} {\bibfnamefont {A.}~\bibnamefont {Sesana}}, \bibinfo {author} {\bibfnamefont {P.}~\bibnamefont {Amaro-Seoane}},\ and\ \bibinfo {author} {\bibfnamefont {J.}~\bibnamefont {Cuadra}},\ }\bibfield  {title} {\bibinfo {title} {{Accretion of clumpy cold gas onto massive black hole binaries: a possible fast route to binary coalescence}},\ }\href {https://doi.org/10.1093/mnras/sty1709} {\bibfield  {journal} {\bibinfo  {journal} {Mon. Not. Roy. Astron. Soc.}\ }\textbf {\bibinfo {volume} {479}},\ \bibinfo {pages} {3438} (\bibinfo {year} {2018})},\ \Eprint {https://arxiv.org/abs/1801.04937} {arXiv:1801.04937 [astro-ph.HE]} \BibitemShut {NoStop}%
\bibitem [{\citenamefont {Wang}\ \emph {et~al.}(2025)\citenamefont {Wang}, \citenamefont {Guo}, \citenamefont {Most}, \citenamefont {Hopkins},\ and\ \citenamefont {Lalakos}}]{Wang:2025mit}%
  \BibitemOpen
  \bibfield  {author} {\bibinfo {author} {\bibfnamefont {H.-Y.}\ \bibnamefont {Wang}}, \bibinfo {author} {\bibfnamefont {M.}~\bibnamefont {Guo}}, \bibinfo {author} {\bibfnamefont {E.~R.}\ \bibnamefont {Most}}, \bibinfo {author} {\bibfnamefont {P.~F.}\ \bibnamefont {Hopkins}},\ and\ \bibinfo {author} {\bibfnamefont {A.}~\bibnamefont {Lalakos}},\ }\bibfield  {title} {\bibinfo {title} {{Galactic-scale Feeding Reveals Warped Hypermagnetized Multiphase Circumbinary Accretion Around Supermassive Black Hole Binaries}},\ }\href@noop {} {\bibfield  {journal} {\bibinfo  {journal} {arxiv}\ } (\bibinfo {year} {2025})},\ \Eprint {https://arxiv.org/abs/2504.03874} {arXiv:2504.03874 [astro-ph.HE]} \BibitemShut {NoStop}%
\bibitem [{\citenamefont {Quinlan}(1996)}]{Quinlan:1996vp}%
  \BibitemOpen
  \bibfield  {author} {\bibinfo {author} {\bibfnamefont {G.~D.}\ \bibnamefont {Quinlan}},\ }\bibfield  {title} {\bibinfo {title} {{The dynamical evolution of massive black hole binaries - I. hardening in a fixed stellar background}},\ }\href {https://doi.org/10.1016/S1384-1076(96)00003-6} {\bibfield  {journal} {\bibinfo  {journal} {New Astron.}\ }\textbf {\bibinfo {volume} {1}},\ \bibinfo {pages} {35} (\bibinfo {year} {1996})},\ \Eprint {https://arxiv.org/abs/astro-ph/9601092} {arXiv:astro-ph/9601092} \BibitemShut {NoStop}%
\bibitem [{\citenamefont {Milosavljevic}\ and\ \citenamefont {Merritt}(2001)}]{Milosavljevic:2001vi}%
  \BibitemOpen
  \bibfield  {author} {\bibinfo {author} {\bibfnamefont {M.}~\bibnamefont {Milosavljevic}}\ and\ \bibinfo {author} {\bibfnamefont {D.}~\bibnamefont {Merritt}},\ }\bibfield  {title} {\bibinfo {title} {{Formation of galactic nuclei}},\ }\href {https://doi.org/10.1086/323830} {\bibfield  {journal} {\bibinfo  {journal} {Astrophys. J.}\ }\textbf {\bibinfo {volume} {563}},\ \bibinfo {pages} {34} (\bibinfo {year} {2001})},\ \Eprint {https://arxiv.org/abs/astro-ph/0103350} {arXiv:astro-ph/0103350} \BibitemShut {NoStop}%
\bibitem [{\citenamefont {{Milosavljevi{\'c}}}\ and\ \citenamefont {{Merritt}}(2003)}]{2003AIPC..686..201M}%
  \BibitemOpen
  \bibfield  {author} {\bibinfo {author} {\bibfnamefont {M.}~\bibnamefont {{Milosavljevi{\'c}}}}\ and\ \bibinfo {author} {\bibfnamefont {D.}~\bibnamefont {{Merritt}}},\ }\bibfield  {title} {\bibinfo {title} {{The Final Parsec Problem}},\ }in\ \href {https://doi.org/10.1063/1.1629432} {\emph {\bibinfo {booktitle} {The Astrophysics of Gravitational Wave Sources}}},\ \bibinfo {series} {American Institute of Physics Conference Series}, Vol.\ \bibinfo {volume} {686},\ \bibinfo {editor} {edited by\ \bibinfo {editor} {\bibfnamefont {J.~M.}\ \bibnamefont {{Centrella}}}}\ (\bibinfo {year} {2003})\ pp.\ \bibinfo {pages} {201--210},\ \Eprint {https://arxiv.org/abs/astro-ph/0212270} {arXiv:astro-ph/0212270 [astro-ph]} \BibitemShut {NoStop}%
\bibitem [{\citenamefont {Berczik}\ \emph {et~al.}(2006)\citenamefont {Berczik}, \citenamefont {Merritt}, \citenamefont {Spurzem},\ and\ \citenamefont {Bischof}}]{Berczik:2006tz}%
  \BibitemOpen
  \bibfield  {author} {\bibinfo {author} {\bibfnamefont {P.}~\bibnamefont {Berczik}}, \bibinfo {author} {\bibfnamefont {D.}~\bibnamefont {Merritt}}, \bibinfo {author} {\bibfnamefont {R.}~\bibnamefont {Spurzem}},\ and\ \bibinfo {author} {\bibfnamefont {H.-P.}\ \bibnamefont {Bischof}},\ }\bibfield  {title} {\bibinfo {title} {{Efficient merger of binary supermassive black holes in non-axisymmetric galaxies}},\ }\href {https://doi.org/10.1086/504426} {\bibfield  {journal} {\bibinfo  {journal} {Astrophys. J. Lett.}\ }\textbf {\bibinfo {volume} {642}},\ \bibinfo {pages} {L21} (\bibinfo {year} {2006})},\ \Eprint {https://arxiv.org/abs/astro-ph/0601698} {arXiv:astro-ph/0601698} \BibitemShut {NoStop}%
\bibitem [{\citenamefont {Sesana}\ \emph {et~al.}(2006)\citenamefont {Sesana}, \citenamefont {Haardt},\ and\ \citenamefont {Madau}}]{Sesana:2006xw}%
  \BibitemOpen
  \bibfield  {author} {\bibinfo {author} {\bibfnamefont {A.}~\bibnamefont {Sesana}}, \bibinfo {author} {\bibfnamefont {F.}~\bibnamefont {Haardt}},\ and\ \bibinfo {author} {\bibfnamefont {P.}~\bibnamefont {Madau}},\ }\bibfield  {title} {\bibinfo {title} {{Interaction of massive black hole binaries with their stellar environment. 1. Ejection of hypervelocity stars}},\ }\href {https://doi.org/10.1086/507596} {\bibfield  {journal} {\bibinfo  {journal} {Astrophys. J.}\ }\textbf {\bibinfo {volume} {651}},\ \bibinfo {pages} {392} (\bibinfo {year} {2006})},\ \Eprint {https://arxiv.org/abs/astro-ph/0604299} {arXiv:astro-ph/0604299} \BibitemShut {NoStop}%
\bibitem [{\citenamefont {Rantala}\ \emph {et~al.}(2017)\citenamefont {Rantala}, \citenamefont {Pihajoki}, \citenamefont {Johansson}, \citenamefont {Naab}, \citenamefont {Lah\'en},\ and\ \citenamefont {Sawala}}]{Rantala:2016rng}%
  \BibitemOpen
  \bibfield  {author} {\bibinfo {author} {\bibfnamefont {A.}~\bibnamefont {Rantala}}, \bibinfo {author} {\bibfnamefont {P.}~\bibnamefont {Pihajoki}}, \bibinfo {author} {\bibfnamefont {P.~H.}\ \bibnamefont {Johansson}}, \bibinfo {author} {\bibfnamefont {T.}~\bibnamefont {Naab}}, \bibinfo {author} {\bibfnamefont {N.}~\bibnamefont {Lah\'en}},\ and\ \bibinfo {author} {\bibfnamefont {T.}~\bibnamefont {Sawala}},\ }\bibfield  {title} {\bibinfo {title} {{Post-Newtonian dynamical modeling of supermassive black holes in galactic-scale simulations}},\ }\href {https://doi.org/10.3847/1538-4357/aa6d65} {\bibfield  {journal} {\bibinfo  {journal} {Astrophys. J.}\ }\textbf {\bibinfo {volume} {840}},\ \bibinfo {pages} {53} (\bibinfo {year} {2017})},\ \Eprint {https://arxiv.org/abs/1611.07028} {arXiv:1611.07028 [astro-ph.GA]} \BibitemShut {NoStop}%
\bibitem [{\citenamefont {{Chen}}\ \emph {et~al.}(2022{\natexlab{a}})\citenamefont {{Chen}}, \citenamefont {{Ni}}, \citenamefont {{Tremmel}}, \citenamefont {{Di Matteo}}, \citenamefont {{Bird}}, \citenamefont {{DeGraf}},\ and\ \citenamefont {{Feng}}}]{Chen:2022MNRAS.510..531C}%
  \BibitemOpen
  \bibfield  {author} {\bibinfo {author} {\bibfnamefont {N.}~\bibnamefont {{Chen}}}, \bibinfo {author} {\bibfnamefont {Y.}~\bibnamefont {{Ni}}}, \bibinfo {author} {\bibfnamefont {M.}~\bibnamefont {{Tremmel}}}, \bibinfo {author} {\bibfnamefont {T.}~\bibnamefont {{Di Matteo}}}, \bibinfo {author} {\bibfnamefont {S.}~\bibnamefont {{Bird}}}, \bibinfo {author} {\bibfnamefont {C.}~\bibnamefont {{DeGraf}}},\ and\ \bibinfo {author} {\bibfnamefont {Y.}~\bibnamefont {{Feng}}},\ }\bibfield  {title} {\bibinfo {title} {{Dynamical friction modelling of massive black holes in cosmological simulations and effects on merger rate predictions}},\ }\href {https://doi.org/10.1093/mnras/stab3411} {\bibfield  {journal} {\bibinfo  {journal} {\mnras}\ }\textbf {\bibinfo {volume} {510}},\ \bibinfo {pages} {531} (\bibinfo {year} {2022}{\natexlab{a}})},\ \Eprint {https://arxiv.org/abs/2104.00021} {arXiv:2104.00021 [astro-ph.GA]} \BibitemShut {NoStop}%
\bibitem [{\citenamefont {{Chen}}\ \emph {et~al.}(2022{\natexlab{b}})\citenamefont {{Chen}}, \citenamefont {{Ni}}, \citenamefont {{Holgado}}, \citenamefont {{Di Matteo}}, \citenamefont {{Tremmel}}, \citenamefont {{DeGraf}}, \citenamefont {{Bird}}, \citenamefont {{Croft}},\ and\ \citenamefont {{Feng}}}]{Chen:2022MNRAS.514.2220C}%
  \BibitemOpen
  \bibfield  {author} {\bibinfo {author} {\bibfnamefont {N.}~\bibnamefont {{Chen}}}, \bibinfo {author} {\bibfnamefont {Y.}~\bibnamefont {{Ni}}}, \bibinfo {author} {\bibfnamefont {A.~M.}\ \bibnamefont {{Holgado}}}, \bibinfo {author} {\bibfnamefont {T.}~\bibnamefont {{Di Matteo}}}, \bibinfo {author} {\bibfnamefont {M.}~\bibnamefont {{Tremmel}}}, \bibinfo {author} {\bibfnamefont {C.}~\bibnamefont {{DeGraf}}}, \bibinfo {author} {\bibfnamefont {S.}~\bibnamefont {{Bird}}}, \bibinfo {author} {\bibfnamefont {R.}~\bibnamefont {{Croft}}},\ and\ \bibinfo {author} {\bibfnamefont {Y.}~\bibnamefont {{Feng}}},\ }\bibfield  {title} {\bibinfo {title} {{Massive black hole mergers with orbital information: predictions from the ASTRID simulation}},\ }\href {https://doi.org/10.1093/mnras/stac1432} {\bibfield  {journal} {\bibinfo  {journal} {\mnras}\ }\textbf {\bibinfo {volume} {514}},\ \bibinfo {pages} {2220} (\bibinfo {year} {2022}{\natexlab{b}})},\ \Eprint {https://arxiv.org/abs/2112.08555} {arXiv:2112.08555 [astro-ph.GA]} \BibitemShut {NoStop}%
\bibitem [{\citenamefont {Peters}(1964)}]{Peters:1964zz}%
  \BibitemOpen
  \bibfield  {author} {\bibinfo {author} {\bibfnamefont {P.~C.}\ \bibnamefont {Peters}},\ }\bibfield  {title} {\bibinfo {title} {{Gravitational Radiation and the Motion of Two Point Masses}},\ }\href {https://doi.org/10.1103/PhysRev.136.B1224} {\bibfield  {journal} {\bibinfo  {journal} {Phys. Rev.}\ }\textbf {\bibinfo {volume} {136}},\ \bibinfo {pages} {B1224} (\bibinfo {year} {1964})}\BibitemShut {NoStop}%
\bibitem [{\citenamefont {{Begelman}}\ \emph {et~al.}(1980)\citenamefont {{Begelman}}, \citenamefont {{Blandford}},\ and\ \citenamefont {{Rees}}}]{Begelman1980}%
  \BibitemOpen
  \bibfield  {author} {\bibinfo {author} {\bibfnamefont {M.~C.}\ \bibnamefont {{Begelman}}}, \bibinfo {author} {\bibfnamefont {R.~D.}\ \bibnamefont {{Blandford}}},\ and\ \bibinfo {author} {\bibfnamefont {M.~J.}\ \bibnamefont {{Rees}}},\ }\bibfield  {title} {\bibinfo {title} {{Massive black hole binaries in active galactic nuclei}},\ }\href {https://doi.org/10.1038/287307a0} {\bibfield  {journal} {\bibinfo  {journal} {\nat}\ }\textbf {\bibinfo {volume} {287}},\ \bibinfo {pages} {307} (\bibinfo {year} {1980})}\BibitemShut {NoStop}%
\bibitem [{\citenamefont {Artymowicz}\ and\ \citenamefont {Lubow}(1994)}]{Artymowicz:1994bw}%
  \BibitemOpen
  \bibfield  {author} {\bibinfo {author} {\bibfnamefont {P.}~\bibnamefont {Artymowicz}}\ and\ \bibinfo {author} {\bibfnamefont {S.~H.}\ \bibnamefont {Lubow}},\ }\bibfield  {title} {\bibinfo {title} {{Dynamics of binary-disk interaction. 1: Resonances and disk gap sizes}},\ }\href {https://doi.org/10.1086/173679} {\bibfield  {journal} {\bibinfo  {journal} {Astrophys. J.}\ }\textbf {\bibinfo {volume} {421}},\ \bibinfo {pages} {651} (\bibinfo {year} {1994})}\BibitemShut {NoStop}%
\bibitem [{\citenamefont {{Artymowicz}}\ and\ \citenamefont {{Lubow}}(1996)}]{Artymowicz1996}%
  \BibitemOpen
  \bibfield  {author} {\bibinfo {author} {\bibfnamefont {P.}~\bibnamefont {{Artymowicz}}}\ and\ \bibinfo {author} {\bibfnamefont {S.~H.}\ \bibnamefont {{Lubow}}},\ }\bibfield  {title} {\bibinfo {title} {{Mass Flow through Gaps in Circumbinary Disks}},\ }\href {https://doi.org/10.1086/310200} {\bibfield  {journal} {\bibinfo  {journal} {Astrophys. J. Lett.}\ }\textbf {\bibinfo {volume} {467}},\ \bibinfo {pages} {L77} (\bibinfo {year} {1996})}\BibitemShut {NoStop}%
\bibitem [{\citenamefont {Most}\ \emph {et~al.}(2022)\citenamefont {Most}, \citenamefont {Noronha},\ and\ \citenamefont {Philippov}}]{Most:2021uck}%
  \BibitemOpen
  \bibfield  {author} {\bibinfo {author} {\bibfnamefont {E.~R.}\ \bibnamefont {Most}}, \bibinfo {author} {\bibfnamefont {J.}~\bibnamefont {Noronha}},\ and\ \bibinfo {author} {\bibfnamefont {A.~A.}\ \bibnamefont {Philippov}},\ }\bibfield  {title} {\bibinfo {title} {{Modelling general-relativistic plasmas with collisionless moments and dissipative two-fluid magnetohydrodynamics}},\ }\href {https://doi.org/10.1093/mnras/stac1435} {\bibfield  {journal} {\bibinfo  {journal} {Mon. Not. Roy. Astron. Soc.}\ }\textbf {\bibinfo {volume} {514}},\ \bibinfo {pages} {4989} (\bibinfo {year} {2022})},\ \Eprint {https://arxiv.org/abs/2111.05752} {arXiv:2111.05752 [astro-ph.HE]} \BibitemShut {NoStop}%
\bibitem [{\citenamefont {Bai}\ and\ \citenamefont {Stone}(2013)}]{Bai:2013pi}%
  \BibitemOpen
  \bibfield  {author} {\bibinfo {author} {\bibfnamefont {X.-N.}\ \bibnamefont {Bai}}\ and\ \bibinfo {author} {\bibfnamefont {J.~M.}\ \bibnamefont {Stone}},\ }\bibfield  {title} {\bibinfo {title} {{Non-turbulent Accretion in Protoplanetary Disks. I: Suppression of the Magnetorotational Instability and Launching of the Magnetocentrifugal Wind}},\ }\href {https://doi.org/10.1088/0004-637X/769/1/76} {\bibfield  {journal} {\bibinfo  {journal} {Astrophys. J.}\ }\textbf {\bibinfo {volume} {769}},\ \bibinfo {pages} {76} (\bibinfo {year} {2013})},\ \Eprint {https://arxiv.org/abs/1301.0318} {arXiv:1301.0318 [astro-ph.EP]} \BibitemShut {NoStop}%
\bibitem [{\citenamefont {Jiang}\ \emph {et~al.}(2014)\citenamefont {Jiang}, \citenamefont {Stone},\ and\ \citenamefont {Davis}}]{Jiang:2014tpa}%
  \BibitemOpen
  \bibfield  {author} {\bibinfo {author} {\bibfnamefont {Y.-F.}\ \bibnamefont {Jiang}}, \bibinfo {author} {\bibfnamefont {J.~M.}\ \bibnamefont {Stone}},\ and\ \bibinfo {author} {\bibfnamefont {S.~W.}\ \bibnamefont {Davis}},\ }\bibfield  {title} {\bibinfo {title} {{A Global Three Dimensional Radiation Magneto-hydrodynamic Simulation of Super-Eddington Accretion Disks}},\ }\href {https://doi.org/10.1088/0004-637X/796/2/106} {\bibfield  {journal} {\bibinfo  {journal} {Astrophys. J.}\ }\textbf {\bibinfo {volume} {796}},\ \bibinfo {pages} {106} (\bibinfo {year} {2014})},\ \Eprint {https://arxiv.org/abs/1410.0678} {arXiv:1410.0678 [astro-ph.HE]} \BibitemShut {NoStop}%
\bibitem [{\citenamefont {Mu\~noz}\ \emph {et~al.}(2019)\citenamefont {Mu\~noz}, \citenamefont {Miranda},\ and\ \citenamefont {Lai}}]{Munoz:2018tnj}%
  \BibitemOpen
  \bibfield  {author} {\bibinfo {author} {\bibfnamefont {D.~J.}\ \bibnamefont {Mu\~noz}}, \bibinfo {author} {\bibfnamefont {R.}~\bibnamefont {Miranda}},\ and\ \bibinfo {author} {\bibfnamefont {D.}~\bibnamefont {Lai}},\ }\bibfield  {title} {\bibinfo {title} {{Hydrodynamics of circumbinary accretion: Angular momentum transfer and binary orbital evolution}},\ }\href {https://doi.org/10.3847/1538-4357/aaf867} {\bibfield  {journal} {\bibinfo  {journal} {Astrophys. J.}\ }\textbf {\bibinfo {volume} {871}},\ \bibinfo {pages} {84} (\bibinfo {year} {2019})},\ \Eprint {https://arxiv.org/abs/1810.04676} {arXiv:1810.04676 [astro-ph.HE]} \BibitemShut {NoStop}%
\bibitem [{\citenamefont {{Miranda}}\ and\ \citenamefont {{Lai}}(2015)}]{2015MNRAS.452.2396M}%
  \BibitemOpen
  \bibfield  {author} {\bibinfo {author} {\bibfnamefont {R.}~\bibnamefont {{Miranda}}}\ and\ \bibinfo {author} {\bibfnamefont {D.}~\bibnamefont {{Lai}}},\ }\bibfield  {title} {\bibinfo {title} {{Tidal truncation of inclined circumstellar and circumbinary discs in young stellar binaries}},\ }\href {https://doi.org/10.1093/mnras/stv1450} {\bibfield  {journal} {\bibinfo  {journal} {Mon. Not. Roy. Astron. Soc.}\ }\textbf {\bibinfo {volume} {452}},\ \bibinfo {pages} {2396} (\bibinfo {year} {2015})},\ \Eprint {https://arxiv.org/abs/1504.02917} {arXiv:1504.02917 [astro-ph.EP]} \BibitemShut {NoStop}%
\bibitem [{\citenamefont {{Miranda}}\ \emph {et~al.}(2017)\citenamefont {{Miranda}}, \citenamefont {{Mu{\~n}oz}},\ and\ \citenamefont {{Lai}}}]{Miranda2017}%
  \BibitemOpen
  \bibfield  {author} {\bibinfo {author} {\bibfnamefont {R.}~\bibnamefont {{Miranda}}}, \bibinfo {author} {\bibfnamefont {D.~J.}\ \bibnamefont {{Mu{\~n}oz}}},\ and\ \bibinfo {author} {\bibfnamefont {D.}~\bibnamefont {{Lai}}},\ }\bibfield  {title} {\bibinfo {title} {{Viscous hydrodynamics simulations of circumbinary accretion discs: variability, quasi-steady state and angular momentum transfer}},\ }\href {https://doi.org/10.1093/mnras/stw3189} {\bibfield  {journal} {\bibinfo  {journal} {Mon. Not. Roy. Astron. Soc.}\ }\textbf {\bibinfo {volume} {466}},\ \bibinfo {pages} {1170} (\bibinfo {year} {2017})},\ \Eprint {https://arxiv.org/abs/1610.07263} {arXiv:1610.07263 [astro-ph.SR]} \BibitemShut {NoStop}%
\bibitem [{\citenamefont {{Moody}}\ \emph {et~al.}(2019)\citenamefont {{Moody}}, \citenamefont {{Shi}},\ and\ \citenamefont {{Stone}}}]{Moody2019}%
  \BibitemOpen
  \bibfield  {author} {\bibinfo {author} {\bibfnamefont {M.~S.~L.}\ \bibnamefont {{Moody}}}, \bibinfo {author} {\bibfnamefont {J.-M.}\ \bibnamefont {{Shi}}},\ and\ \bibinfo {author} {\bibfnamefont {J.~M.}\ \bibnamefont {{Stone}}},\ }\bibfield  {title} {\bibinfo {title} {{Hydrodynamic Torques in Circumbinary Accretion Disks}},\ }\href {https://doi.org/10.3847/1538-4357/ab09ee} {\bibfield  {journal} {\bibinfo  {journal} {Astrophys. J.}\ }\textbf {\bibinfo {volume} {875}},\ \bibinfo {eid} {66} (\bibinfo {year} {2019})},\ \Eprint {https://arxiv.org/abs/1903.00008} {arXiv:1903.00008 [astro-ph.HE]} \BibitemShut {NoStop}%
\bibitem [{\citenamefont {{Teyssandier}}\ and\ \citenamefont {{Ogilvie}}(2016)}]{Teyssandier:2016MNRAS.458.3221T}%
  \BibitemOpen
  \bibfield  {author} {\bibinfo {author} {\bibfnamefont {J.}~\bibnamefont {{Teyssandier}}}\ and\ \bibinfo {author} {\bibfnamefont {G.~I.}\ \bibnamefont {{Ogilvie}}},\ }\bibfield  {title} {\bibinfo {title} {{Growth of eccentric modes in disc-planet interactions}},\ }\href {https://doi.org/10.1093/mnras/stw521} {\bibfield  {journal} {\bibinfo  {journal} {Monthly Notices of the Royal Astronomical Society}\ }\textbf {\bibinfo {volume} {458}},\ \bibinfo {pages} {3221} (\bibinfo {year} {2016})},\ \Eprint {https://arxiv.org/abs/1603.00653} {arXiv:1603.00653 [astro-ph.EP]} \BibitemShut {NoStop}%
\bibitem [{\citenamefont {Mu\~noz}\ and\ \citenamefont {Lithwick}(2020)}]{Munoz:2020azx}%
  \BibitemOpen
  \bibfield  {author} {\bibinfo {author} {\bibfnamefont {D.~J.}\ \bibnamefont {Mu\~noz}}\ and\ \bibinfo {author} {\bibfnamefont {Y.}~\bibnamefont {Lithwick}},\ }\bibfield  {title} {\bibinfo {title} {{Long-Lived Eccentric Modes in Circumbinary Disks}},\ }\href {https://doi.org/10.3847/1538-4357/abc74c} {\bibfield  {journal} {\bibinfo  {journal} {Astrophys. J.}\ }\textbf {\bibinfo {volume} {905}},\ \bibinfo {pages} {106} (\bibinfo {year} {2020})},\ \Eprint {https://arxiv.org/abs/2008.08085} {arXiv:2008.08085 [astro-ph.HE]} \BibitemShut {NoStop}%
\bibitem [{\citenamefont {Most}\ and\ \citenamefont {Wang}(2024)}]{paper1}%
  \BibitemOpen
  \bibfield  {author} {\bibinfo {author} {\bibfnamefont {E.~R.}\ \bibnamefont {Most}}\ and\ \bibinfo {author} {\bibfnamefont {H.-Y.}\ \bibnamefont {Wang}},\ }\bibfield  {title} {\bibinfo {title} {{Magnetically Arrested Circumbinary Accretion Flows}},\ }\href {https://doi.org/10.3847/2041-8213/ad7713} {\bibfield  {journal} {\bibinfo  {journal} {Astrophys. J. Lett.}\ }\textbf {\bibinfo {volume} {973}},\ \bibinfo {pages} {L19} (\bibinfo {year} {2024})},\ \Eprint {https://arxiv.org/abs/2408.00757} {arXiv:2408.00757 [astro-ph.HE]} \BibitemShut {NoStop}%
\bibitem [{\citenamefont {Shi}\ \emph {et~al.}(2012)\citenamefont {Shi}, \citenamefont {Krolik}, \citenamefont {Lubow},\ and\ \citenamefont {Hawley}}]{Shi:2011us}%
  \BibitemOpen
  \bibfield  {author} {\bibinfo {author} {\bibfnamefont {J.-M.}\ \bibnamefont {Shi}}, \bibinfo {author} {\bibfnamefont {J.~H.}\ \bibnamefont {Krolik}}, \bibinfo {author} {\bibfnamefont {S.~H.}\ \bibnamefont {Lubow}},\ and\ \bibinfo {author} {\bibfnamefont {J.~F.}\ \bibnamefont {Hawley}},\ }\bibfield  {title} {\bibinfo {title} {{Three Dimensional MHD Simulation of Circumbinary Accretion Disks: Disk Structures and Angular Momentum Transport}},\ }\href {https://doi.org/10.1088/0004-637X/749/2/118} {\bibfield  {journal} {\bibinfo  {journal} {Astrophys. J.}\ }\textbf {\bibinfo {volume} {749}},\ \bibinfo {pages} {118} (\bibinfo {year} {2012})},\ \Eprint {https://arxiv.org/abs/1110.4866} {arXiv:1110.4866 [astro-ph.HE]} \BibitemShut {NoStop}%
\bibitem [{\citenamefont {Noble}\ \emph {et~al.}(2012)\citenamefont {Noble}, \citenamefont {Mundim}, \citenamefont {Nakano}, \citenamefont {Krolik}, \citenamefont {Campanelli}, \citenamefont {Zlochower},\ and\ \citenamefont {Yunes}}]{Noble:2012xz}%
  \BibitemOpen
  \bibfield  {author} {\bibinfo {author} {\bibfnamefont {S.~C.}\ \bibnamefont {Noble}}, \bibinfo {author} {\bibfnamefont {B.~C.}\ \bibnamefont {Mundim}}, \bibinfo {author} {\bibfnamefont {H.}~\bibnamefont {Nakano}}, \bibinfo {author} {\bibfnamefont {J.~H.}\ \bibnamefont {Krolik}}, \bibinfo {author} {\bibfnamefont {M.}~\bibnamefont {Campanelli}}, \bibinfo {author} {\bibfnamefont {Y.}~\bibnamefont {Zlochower}},\ and\ \bibinfo {author} {\bibfnamefont {N.}~\bibnamefont {Yunes}},\ }\bibfield  {title} {\bibinfo {title} {{Circumbinary MHD Accretion into Inspiraling Binary Black Holes}},\ }\href {https://doi.org/10.1088/0004-637X/755/1/51} {\bibfield  {journal} {\bibinfo  {journal} {Astrophys. J.}\ }\textbf {\bibinfo {volume} {755}},\ \bibinfo {pages} {51} (\bibinfo {year} {2012})},\ \Eprint {https://arxiv.org/abs/1204.1073} {arXiv:1204.1073 [astro-ph.HE]} \BibitemShut {NoStop}%
\bibitem [{\citenamefont {Noble}\ \emph {et~al.}(2021)\citenamefont {Noble}, \citenamefont {Krolik}, \citenamefont {Campanelli}, \citenamefont {Zlochower}, \citenamefont {Mundim}, \citenamefont {Nakano},\ and\ \citenamefont {Zilh\~ao}}]{Noble:2021vfg}%
  \BibitemOpen
  \bibfield  {author} {\bibinfo {author} {\bibfnamefont {S.~C.}\ \bibnamefont {Noble}}, \bibinfo {author} {\bibfnamefont {J.~H.}\ \bibnamefont {Krolik}}, \bibinfo {author} {\bibfnamefont {M.}~\bibnamefont {Campanelli}}, \bibinfo {author} {\bibfnamefont {Y.}~\bibnamefont {Zlochower}}, \bibinfo {author} {\bibfnamefont {B.~C.}\ \bibnamefont {Mundim}}, \bibinfo {author} {\bibfnamefont {H.}~\bibnamefont {Nakano}},\ and\ \bibinfo {author} {\bibfnamefont {M.}~\bibnamefont {Zilh\~ao}},\ }\bibfield  {title} {\bibinfo {title} {{Mass-ratio and Magnetic Flux Dependence of Modulated Accretion from Circumbinary Disks}},\ }\href {https://doi.org/10.3847/1538-4357/ac2229} {\bibfield  {journal} {\bibinfo  {journal} {Astrophys. J.}\ }\textbf {\bibinfo {volume} {922}},\ \bibinfo {pages} {175} (\bibinfo {year} {2021})},\ \Eprint {https://arxiv.org/abs/2103.12100} {arXiv:2103.12100 [astro-ph.HE]} \BibitemShut {NoStop}%
\bibitem [{\citenamefont {{Avara}}\ \emph {et~al.}(2024)\citenamefont {{Avara}}, \citenamefont {{Krolik}}, \citenamefont {{Campanelli}}, \citenamefont {{Noble}}, \citenamefont {{Bowen}},\ and\ \citenamefont {{Ryu}}}]{Avara:2023ztw}%
  \BibitemOpen
  \bibfield  {author} {\bibinfo {author} {\bibfnamefont {M.~J.}\ \bibnamefont {{Avara}}}, \bibinfo {author} {\bibfnamefont {J.~H.}\ \bibnamefont {{Krolik}}}, \bibinfo {author} {\bibfnamefont {M.}~\bibnamefont {{Campanelli}}}, \bibinfo {author} {\bibfnamefont {S.~C.}\ \bibnamefont {{Noble}}}, \bibinfo {author} {\bibfnamefont {D.}~\bibnamefont {{Bowen}}},\ and\ \bibinfo {author} {\bibfnamefont {T.}~\bibnamefont {{Ryu}}},\ }\bibfield  {title} {\bibinfo {title} {{Accretion onto a Supermassive Black Hole Binary before Merger}},\ }\href {https://doi.org/10.3847/1538-4357/ad5bda} {\bibfield  {journal} {\bibinfo  {journal} {\apj}\ }\textbf {\bibinfo {volume} {974}},\ \bibinfo {eid} {242} (\bibinfo {year} {2024})},\ \Eprint {https://arxiv.org/abs/2305.18538} {arXiv:2305.18538 [astro-ph.HE]} \BibitemShut {NoStop}%
\bibitem [{\citenamefont {{Lin}}\ and\ \citenamefont {{Papaloizou}}(1979)}]{lin1979}%
  \BibitemOpen
  \bibfield  {author} {\bibinfo {author} {\bibfnamefont {D.~N.~C.}\ \bibnamefont {{Lin}}}\ and\ \bibinfo {author} {\bibfnamefont {J.}~\bibnamefont {{Papaloizou}}},\ }\bibfield  {title} {\bibinfo {title} {{Tidal torques on accretion discs in binary systems with extreme mass ratios.}},\ }\href {https://doi.org/10.1093/mnras/186.4.799} {\bibfield  {journal} {\bibinfo  {journal} {Mon. Not. Roy. Astron. Soc.}\ }\textbf {\bibinfo {volume} {186}},\ \bibinfo {pages} {799} (\bibinfo {year} {1979})}\BibitemShut {NoStop}%
\bibitem [{\citenamefont {Goldreich}\ and\ \citenamefont {Tremaine}(1979)}]{Goldreich:1979zz}%
  \BibitemOpen
  \bibfield  {author} {\bibinfo {author} {\bibfnamefont {P.~M.}\ \bibnamefont {Goldreich}}\ and\ \bibinfo {author} {\bibfnamefont {S.}~\bibnamefont {Tremaine}},\ }\bibfield  {title} {\bibinfo {title} {{The excitation of density waves at the Lindblad and corotation resonances by an external potential}},\ }\href {https://doi.org/10.1086/157448} {\bibfield  {journal} {\bibinfo  {journal} {Astrophys. J.}\ }\textbf {\bibinfo {volume} {233}},\ \bibinfo {pages} {857} (\bibinfo {year} {1979})}\BibitemShut {NoStop}%
\bibitem [{\citenamefont {Goldreich}\ and\ \citenamefont {Tremaine}(1980)}]{Goldreich:1980wa}%
  \BibitemOpen
  \bibfield  {author} {\bibinfo {author} {\bibfnamefont {P.}~\bibnamefont {Goldreich}}\ and\ \bibinfo {author} {\bibfnamefont {S.}~\bibnamefont {Tremaine}},\ }\bibfield  {title} {\bibinfo {title} {{Disk - satellite interactions}},\ }\href {https://doi.org/10.1086/158356} {\bibfield  {journal} {\bibinfo  {journal} {Astrophys. J.}\ }\textbf {\bibinfo {volume} {241}},\ \bibinfo {pages} {425} (\bibinfo {year} {1980})}\BibitemShut {NoStop}%
\bibitem [{\citenamefont {{Papaloizou}}\ and\ \citenamefont {{Lin}}(1984)}]{Lin:1984ApJ...285..818P}%
  \BibitemOpen
  \bibfield  {author} {\bibinfo {author} {\bibfnamefont {J.}~\bibnamefont {{Papaloizou}}}\ and\ \bibinfo {author} {\bibfnamefont {D.~N.~C.}\ \bibnamefont {{Lin}}},\ }\bibfield  {title} {\bibinfo {title} {{On the tidal interaction between protoplanets and the primordial solar nebula. I - Linear calculation of the role of angular momentum exchange}},\ }\href {https://doi.org/10.1086/162561} {\bibfield  {journal} {\bibinfo  {journal} {\apj}\ }\textbf {\bibinfo {volume} {285}},\ \bibinfo {pages} {818} (\bibinfo {year} {1984})}\BibitemShut {NoStop}%
\bibitem [{\citenamefont {{Duffell}}\ \emph {et~al.}(2020)\citenamefont {{Duffell}}, \citenamefont {{D'Orazio}}, \citenamefont {{Derdzinski}}, \citenamefont {{Haiman}}, \citenamefont {{MacFadyen}}, \citenamefont {{Rosen}},\ and\ \citenamefont {{Zrake}}}]{Duffell2020}%
  \BibitemOpen
  \bibfield  {author} {\bibinfo {author} {\bibfnamefont {P.~C.}\ \bibnamefont {{Duffell}}}, \bibinfo {author} {\bibfnamefont {D.}~\bibnamefont {{D'Orazio}}}, \bibinfo {author} {\bibfnamefont {A.}~\bibnamefont {{Derdzinski}}}, \bibinfo {author} {\bibfnamefont {Z.}~\bibnamefont {{Haiman}}}, \bibinfo {author} {\bibfnamefont {A.}~\bibnamefont {{MacFadyen}}}, \bibinfo {author} {\bibfnamefont {A.~L.}\ \bibnamefont {{Rosen}}},\ and\ \bibinfo {author} {\bibfnamefont {J.}~\bibnamefont {{Zrake}}},\ }\bibfield  {title} {\bibinfo {title} {{Circumbinary Disks: Accretion and Torque as a Function of Mass Ratio and Disk Viscosity}},\ }\href {https://doi.org/10.3847/1538-4357/abab95} {\bibfield  {journal} {\bibinfo  {journal} {Astrophys. J.}\ }\textbf {\bibinfo {volume} {901}},\ \bibinfo {eid} {25} (\bibinfo {year} {2020})},\ \Eprint {https://arxiv.org/abs/1911.05506} {arXiv:1911.05506 [astro-ph.SR]} \BibitemShut {NoStop}%
\bibitem [{\citenamefont {Derdzinski}\ \emph {et~al.}(2021)\citenamefont {Derdzinski}, \citenamefont {D'Orazio}, \citenamefont {Duffell}, \citenamefont {Haiman},\ and\ \citenamefont {MacFadyen}}]{Derdzinski:2020wlw}%
  \BibitemOpen
  \bibfield  {author} {\bibinfo {author} {\bibfnamefont {A.}~\bibnamefont {Derdzinski}}, \bibinfo {author} {\bibfnamefont {D.}~\bibnamefont {D'Orazio}}, \bibinfo {author} {\bibfnamefont {P.}~\bibnamefont {Duffell}}, \bibinfo {author} {\bibfnamefont {Z.}~\bibnamefont {Haiman}},\ and\ \bibinfo {author} {\bibfnamefont {A.}~\bibnamefont {MacFadyen}},\ }\bibfield  {title} {\bibinfo {title} {{Evolution of gas disc\textendash{}embedded intermediate mass ratio inspirals in the $LISA$ band}},\ }\href {https://doi.org/10.1093/mnras/staa3976} {\bibfield  {journal} {\bibinfo  {journal} {Mon. Not. Roy. Astron. Soc.}\ }\textbf {\bibinfo {volume} {501}},\ \bibinfo {pages} {3540} (\bibinfo {year} {2021})},\ \Eprint {https://arxiv.org/abs/2005.11333} {arXiv:2005.11333 [astro-ph.HE]} \BibitemShut {NoStop}%
\bibitem [{\citenamefont {{Siwek}}\ \emph {et~al.}(2023{\natexlab{a}})\citenamefont {{Siwek}}, \citenamefont {{Weinberger}},\ and\ \citenamefont {{Hernquist}}}]{Siwek2023b}%
  \BibitemOpen
  \bibfield  {author} {\bibinfo {author} {\bibfnamefont {M.}~\bibnamefont {{Siwek}}}, \bibinfo {author} {\bibfnamefont {R.}~\bibnamefont {{Weinberger}}},\ and\ \bibinfo {author} {\bibfnamefont {L.}~\bibnamefont {{Hernquist}}},\ }\bibfield  {title} {\bibinfo {title} {{Orbital evolution of binaries in circumbinary discs}},\ }\href {https://doi.org/10.1093/mnras/stad1131} {\bibfield  {journal} {\bibinfo  {journal} {Mon. Not. Roy. Astron. Soc.}\ }\textbf {\bibinfo {volume} {522}},\ \bibinfo {pages} {2707} (\bibinfo {year} {2023}{\natexlab{a}})},\ \Eprint {https://arxiv.org/abs/2302.01785} {arXiv:2302.01785 [astro-ph.HE]} \BibitemShut {NoStop}%
\bibitem [{\citenamefont {{Siwek}}\ \emph {et~al.}(2023{\natexlab{b}})\citenamefont {{Siwek}}, \citenamefont {{Weinberger}}, \citenamefont {{Mu{\~n}oz}},\ and\ \citenamefont {{Hernquist}}}]{Siwek2023a}%
  \BibitemOpen
  \bibfield  {author} {\bibinfo {author} {\bibfnamefont {M.}~\bibnamefont {{Siwek}}}, \bibinfo {author} {\bibfnamefont {R.}~\bibnamefont {{Weinberger}}}, \bibinfo {author} {\bibfnamefont {D.~J.}\ \bibnamefont {{Mu{\~n}oz}}},\ and\ \bibinfo {author} {\bibfnamefont {L.}~\bibnamefont {{Hernquist}}},\ }\bibfield  {title} {\bibinfo {title} {{Preferential accretion and circumbinary disc precession in eccentric binary systems}},\ }\href {https://doi.org/10.1093/mnras/stac3263} {\bibfield  {journal} {\bibinfo  {journal} {Mon. Not. Roy. Astron. Soc.}\ }\textbf {\bibinfo {volume} {518}},\ \bibinfo {pages} {5059} (\bibinfo {year} {2023}{\natexlab{b}})},\ \Eprint {https://arxiv.org/abs/2203.02514} {arXiv:2203.02514 [astro-ph.HE]} \BibitemShut {NoStop}%
\bibitem [{\citenamefont {Zrake}\ \emph {et~al.}(2021)\citenamefont {Zrake}, \citenamefont {Tiede}, \citenamefont {MacFadyen},\ and\ \citenamefont {Haiman}}]{Zrake:2020zkw}%
  \BibitemOpen
  \bibfield  {author} {\bibinfo {author} {\bibfnamefont {J.}~\bibnamefont {Zrake}}, \bibinfo {author} {\bibfnamefont {C.}~\bibnamefont {Tiede}}, \bibinfo {author} {\bibfnamefont {A.}~\bibnamefont {MacFadyen}},\ and\ \bibinfo {author} {\bibfnamefont {Z.}~\bibnamefont {Haiman}},\ }\bibfield  {title} {\bibinfo {title} {{Equilibrium Eccentricity of Accreting Binaries}},\ }\href {https://doi.org/10.3847/2041-8213/abdd1c} {\bibfield  {journal} {\bibinfo  {journal} {Astrophys. J. Lett.}\ }\textbf {\bibinfo {volume} {909}},\ \bibinfo {pages} {L13} (\bibinfo {year} {2021})},\ \Eprint {https://arxiv.org/abs/2010.09707} {arXiv:2010.09707 [astro-ph.HE]} \BibitemShut {NoStop}%
\bibitem [{\citenamefont {{D'Orazio}}\ and\ \citenamefont {{Duffell}}(2021)}]{DOrazio2021}%
  \BibitemOpen
  \bibfield  {author} {\bibinfo {author} {\bibfnamefont {D.~J.}\ \bibnamefont {{D'Orazio}}}\ and\ \bibinfo {author} {\bibfnamefont {P.~C.}\ \bibnamefont {{Duffell}}},\ }\bibfield  {title} {\bibinfo {title} {{Orbital Evolution of Equal-mass Eccentric Binaries due to a Gas Disk: Eccentric Inspirals and Circular Outspirals}},\ }\href {https://doi.org/10.3847/2041-8213/ac0621} {\bibfield  {journal} {\bibinfo  {journal} {Astrophys. J. Lett.}\ }\textbf {\bibinfo {volume} {914}},\ \bibinfo {eid} {L21} (\bibinfo {year} {2021})},\ \Eprint {https://arxiv.org/abs/2103.09251} {arXiv:2103.09251 [astro-ph.HE]} \BibitemShut {NoStop}%
\bibitem [{\citenamefont {{Tiede}}\ \emph {et~al.}(2020)\citenamefont {{Tiede}}, \citenamefont {{Zrake}}, \citenamefont {{MacFadyen}},\ and\ \citenamefont {{Haiman}}}]{Tiede2020}%
  \BibitemOpen
  \bibfield  {author} {\bibinfo {author} {\bibfnamefont {C.}~\bibnamefont {{Tiede}}}, \bibinfo {author} {\bibfnamefont {J.}~\bibnamefont {{Zrake}}}, \bibinfo {author} {\bibfnamefont {A.}~\bibnamefont {{MacFadyen}}},\ and\ \bibinfo {author} {\bibfnamefont {Z.}~\bibnamefont {{Haiman}}},\ }\bibfield  {title} {\bibinfo {title} {{Gas-driven Inspiral of Binaries in Thin Accretion Disks}},\ }\href {https://doi.org/10.3847/1538-4357/aba432} {\bibfield  {journal} {\bibinfo  {journal} {Astrophys. J.}\ }\textbf {\bibinfo {volume} {900}},\ \bibinfo {eid} {43} (\bibinfo {year} {2020})},\ \Eprint {https://arxiv.org/abs/2005.09555} {arXiv:2005.09555 [astro-ph.GA]} \BibitemShut {NoStop}%
\bibitem [{\citenamefont {{Dittmann}}\ and\ \citenamefont {{Ryan}}(2024)}]{Dittmann2023b}%
  \BibitemOpen
  \bibfield  {author} {\bibinfo {author} {\bibfnamefont {A.~J.}\ \bibnamefont {{Dittmann}}}\ and\ \bibinfo {author} {\bibfnamefont {G.}~\bibnamefont {{Ryan}}},\ }\bibfield  {title} {\bibinfo {title} {{The Evolution of Accreting Binaries: From Brown Dwarfs to Supermassive Black Holes}},\ }\href {https://doi.org/10.3847/1538-4357/ad2f1e} {\bibfield  {journal} {\bibinfo  {journal} {Astrophys. J.}\ }\textbf {\bibinfo {volume} {967}},\ \bibinfo {eid} {12} (\bibinfo {year} {2024})},\ \Eprint {https://arxiv.org/abs/2310.07758} {arXiv:2310.07758 [astro-ph.GA]} \BibitemShut {NoStop}%
\bibitem [{\citenamefont {Tiede}\ \emph {et~al.}(2025)\citenamefont {Tiede}, \citenamefont {Zrake}, \citenamefont {MacFadyen},\ and\ \citenamefont {Haiman}}]{Tiede:2024mwn}%
  \BibitemOpen
  \bibfield  {author} {\bibinfo {author} {\bibfnamefont {C.}~\bibnamefont {Tiede}}, \bibinfo {author} {\bibfnamefont {J.}~\bibnamefont {Zrake}}, \bibinfo {author} {\bibfnamefont {A.}~\bibnamefont {MacFadyen}},\ and\ \bibinfo {author} {\bibfnamefont {Z.}~\bibnamefont {Haiman}},\ }\bibfield  {title} {\bibinfo {title} {{Suppressed Accretion onto Massive Black Hole Binaries Surrounded by Thin Disks}},\ }\href {https://doi.org/10.3847/1538-4357/adc727} {\bibfield  {journal} {\bibinfo  {journal} {Astrophys. J.}\ }\textbf {\bibinfo {volume} {984}},\ \bibinfo {pages} {144} (\bibinfo {year} {2025})},\ \Eprint {https://arxiv.org/abs/2410.03830} {arXiv:2410.03830 [astro-ph.GA]} \BibitemShut {NoStop}%
\bibitem [{\citenamefont {{Dittmann}}\ \emph {et~al.}(2023)\citenamefont {{Dittmann}}, \citenamefont {{Ryan}},\ and\ \citenamefont {{Miller}}}]{Dittmann2023}%
  \BibitemOpen
  \bibfield  {author} {\bibinfo {author} {\bibfnamefont {A.~J.}\ \bibnamefont {{Dittmann}}}, \bibinfo {author} {\bibfnamefont {G.}~\bibnamefont {{Ryan}}},\ and\ \bibinfo {author} {\bibfnamefont {M.~C.}\ \bibnamefont {{Miller}}},\ }\bibfield  {title} {\bibinfo {title} {{The Decoupling of Binaries from Their Circumbinary Disks}},\ }\href {https://doi.org/10.3847/2041-8213/acd183} {\bibfield  {journal} {\bibinfo  {journal} {Astrophys. J. Lett.}\ }\textbf {\bibinfo {volume} {949}},\ \bibinfo {eid} {L30} (\bibinfo {year} {2023})},\ \Eprint {https://arxiv.org/abs/2303.16204} {arXiv:2303.16204 [astro-ph.HE]} \BibitemShut {NoStop}%
\bibitem [{\citenamefont {{Penzlin}}\ \emph {et~al.}(2024)\citenamefont {{Penzlin}}, \citenamefont {{Booth}}, \citenamefont {{Nelson}}, \citenamefont {{Sch{\"a}fer}},\ and\ \citenamefont {{Kley}}}]{Penzlin:2024MNRAS.532.3166P}%
  \BibitemOpen
  \bibfield  {author} {\bibinfo {author} {\bibfnamefont {A.~B.~T.}\ \bibnamefont {{Penzlin}}}, \bibinfo {author} {\bibfnamefont {R.~A.}\ \bibnamefont {{Booth}}}, \bibinfo {author} {\bibfnamefont {R.~P.}\ \bibnamefont {{Nelson}}}, \bibinfo {author} {\bibfnamefont {C.~M.}\ \bibnamefont {{Sch{\"a}fer}}},\ and\ \bibinfo {author} {\bibfnamefont {W.}~\bibnamefont {{Kley}}},\ }\bibfield  {title} {\bibinfo {title} {{Viscous circumbinary protoplanetary discs - I. Structure of the inner cavity}},\ }\href {https://doi.org/10.1093/mnras/stae1689} {\bibfield  {journal} {\bibinfo  {journal} {\mnras}\ }\textbf {\bibinfo {volume} {532}},\ \bibinfo {pages} {3166} (\bibinfo {year} {2024})},\ \Eprint {https://arxiv.org/abs/2407.07243} {arXiv:2407.07243 [astro-ph.EP]} \BibitemShut {NoStop}%
\bibitem [{\citenamefont {{Penzlin}}\ \emph {et~al.}(2025)\citenamefont {{Penzlin}}, \citenamefont {{Booth}}, \citenamefont {{Nelson}}, \citenamefont {{Sch{\"a}fer}},\ and\ \citenamefont {{Kley}}}]{Penzlin:2025MNRAS.537.2422P}%
  \BibitemOpen
  \bibfield  {author} {\bibinfo {author} {\bibfnamefont {A.~B.~T.}\ \bibnamefont {{Penzlin}}}, \bibinfo {author} {\bibfnamefont {R.~A.}\ \bibnamefont {{Booth}}}, \bibinfo {author} {\bibfnamefont {R.~P.}\ \bibnamefont {{Nelson}}}, \bibinfo {author} {\bibfnamefont {C.~M.}\ \bibnamefont {{Sch{\"a}fer}}},\ and\ \bibinfo {author} {\bibfnamefont {W.}~\bibnamefont {{Kley}}},\ }\bibfield  {title} {\bibinfo {title} {{Viscous circumbinary protoplanetary discs - II. Disc effects on the binary orbit}},\ }\href {https://doi.org/10.1093/mnras/staf177} {\bibfield  {journal} {\bibinfo  {journal} {\mnras}\ }\textbf {\bibinfo {volume} {537}},\ \bibinfo {pages} {2422} (\bibinfo {year} {2025})},\ \Eprint {https://arxiv.org/abs/2501.17055} {arXiv:2501.17055 [astro-ph.EP]} \BibitemShut {NoStop}%
\bibitem [{\citenamefont {{Sudarshan}}\ \emph {et~al.}(2022)\citenamefont {{Sudarshan}}, \citenamefont {{Penzlin}}, \citenamefont {{Ziampras}}, \citenamefont {{Kley}},\ and\ \citenamefont {{Nelson}}}]{Sudarshan2022}%
  \BibitemOpen
  \bibfield  {author} {\bibinfo {author} {\bibfnamefont {P.}~\bibnamefont {{Sudarshan}}}, \bibinfo {author} {\bibfnamefont {A.~B.~T.}\ \bibnamefont {{Penzlin}}}, \bibinfo {author} {\bibfnamefont {A.}~\bibnamefont {{Ziampras}}}, \bibinfo {author} {\bibfnamefont {W.}~\bibnamefont {{Kley}}},\ and\ \bibinfo {author} {\bibfnamefont {R.~P.}\ \bibnamefont {{Nelson}}},\ }\bibfield  {title} {\bibinfo {title} {{How cooling influences circumbinary discs}},\ }\href@noop {} {\bibfield  {journal} {\bibinfo  {journal} {arXiv e-prints}\ ,\ \bibinfo {eid} {arXiv:2206.07749}} (\bibinfo {year} {2022})},\ \Eprint {https://arxiv.org/abs/2206.07749} {arXiv:2206.07749 [astro-ph.EP]} \BibitemShut {NoStop}%
\bibitem [{\citenamefont {{Wang}}\ \emph {et~al.}(2023{\natexlab{a}})\citenamefont {{Wang}}, \citenamefont {{Bai}},\ and\ \citenamefont {{Lai}}}]{Wang2022}%
  \BibitemOpen
  \bibfield  {author} {\bibinfo {author} {\bibfnamefont {H.-Y.}\ \bibnamefont {{Wang}}}, \bibinfo {author} {\bibfnamefont {X.-N.}\ \bibnamefont {{Bai}}},\ and\ \bibinfo {author} {\bibfnamefont {D.}~\bibnamefont {{Lai}}},\ }\bibfield  {title} {\bibinfo {title} {{On the Role of Dynamical Cooling in the Dynamics of Circumbinary Disks}},\ }\href {https://doi.org/10.3847/1538-4357/acac77} {\bibfield  {journal} {\bibinfo  {journal} {Astrophys. J.}\ }\textbf {\bibinfo {volume} {943}},\ \bibinfo {eid} {175} (\bibinfo {year} {2023}{\natexlab{a}})},\ \Eprint {https://arxiv.org/abs/2212.04199} {arXiv:2212.04199 [astro-ph.HE]} \BibitemShut {NoStop}%
\bibitem [{\citenamefont {{Wang}}\ \emph {et~al.}(2023{\natexlab{b}})\citenamefont {{Wang}}, \citenamefont {{Bai}}, \citenamefont {{Lai}},\ and\ \citenamefont {{Lin}}}]{Wang2023}%
  \BibitemOpen
  \bibfield  {author} {\bibinfo {author} {\bibfnamefont {H.-Y.}\ \bibnamefont {{Wang}}}, \bibinfo {author} {\bibfnamefont {X.-N.}\ \bibnamefont {{Bai}}}, \bibinfo {author} {\bibfnamefont {D.}~\bibnamefont {{Lai}}},\ and\ \bibinfo {author} {\bibfnamefont {D.~N.~C.}\ \bibnamefont {{Lin}}},\ }\bibfield  {title} {\bibinfo {title} {{Hydrodynamical simulations of circumbinary accretion: balance between heating and cooling}},\ }\href {https://doi.org/10.1093/mnras/stad2884} {\bibfield  {journal} {\bibinfo  {journal} {Mon. Not. Roy. Astron. Soc.}\ }\textbf {\bibinfo {volume} {526}},\ \bibinfo {pages} {3570} (\bibinfo {year} {2023}{\natexlab{b}})},\ \Eprint {https://arxiv.org/abs/2212.07416} {arXiv:2212.07416 [astro-ph.HE]} \BibitemShut {NoStop}%
\bibitem [{\citenamefont {{Pierens}}\ and\ \citenamefont {{Nelson}}(2023)}]{Pieren2023}%
  \BibitemOpen
  \bibfield  {author} {\bibinfo {author} {\bibfnamefont {A.}~\bibnamefont {{Pierens}}}\ and\ \bibinfo {author} {\bibfnamefont {R.~P.}\ \bibnamefont {{Nelson}}},\ }\bibfield  {title} {\bibinfo {title} {{Three-dimensional evolution of radiative circumbinary discs: The size and shape of the inner cavity}},\ }\href {https://doi.org/10.1051/0004-6361/202244828} {\bibfield  {journal} {\bibinfo  {journal} {A\&A}\ }\textbf {\bibinfo {volume} {670}},\ \bibinfo {eid} {A112} (\bibinfo {year} {2023})},\ \Eprint {https://arxiv.org/abs/2211.03816} {arXiv:2211.03816 [astro-ph.EP]} \BibitemShut {NoStop}%
\bibitem [{\citenamefont {{Franchini}}\ \emph {et~al.}(2021)\citenamefont {{Franchini}}, \citenamefont {{Sesana}},\ and\ \citenamefont {{Dotti}}}]{Franchini2021}%
  \BibitemOpen
  \bibfield  {author} {\bibinfo {author} {\bibfnamefont {A.}~\bibnamefont {{Franchini}}}, \bibinfo {author} {\bibfnamefont {A.}~\bibnamefont {{Sesana}}},\ and\ \bibinfo {author} {\bibfnamefont {M.}~\bibnamefont {{Dotti}}},\ }\bibfield  {title} {\bibinfo {title} {{Circumbinary disc self-gravity governing supermassive black hole binary mergers}},\ }\href {https://doi.org/10.1093/mnras/stab2234} {\bibfield  {journal} {\bibinfo  {journal} {Mon. Not. Roy. Astron. Soc.}\ }\textbf {\bibinfo {volume} {507}},\ \bibinfo {pages} {1458} (\bibinfo {year} {2021})},\ \Eprint {https://arxiv.org/abs/2106.13253} {arXiv:2106.13253 [astro-ph.HE]} \BibitemShut {NoStop}%
\bibitem [{\citenamefont {{Bourne}}\ \emph {et~al.}(2023)\citenamefont {{Bourne}}, \citenamefont {{Fiacconi}}, \citenamefont {{Sijacki}}, \citenamefont {{Piotrowska}},\ and\ \citenamefont {{Koudmani}}}]{Bourne2023}%
  \BibitemOpen
  \bibfield  {author} {\bibinfo {author} {\bibfnamefont {M.~A.}\ \bibnamefont {{Bourne}}}, \bibinfo {author} {\bibfnamefont {D.}~\bibnamefont {{Fiacconi}}}, \bibinfo {author} {\bibfnamefont {D.}~\bibnamefont {{Sijacki}}}, \bibinfo {author} {\bibfnamefont {J.~M.}\ \bibnamefont {{Piotrowska}}},\ and\ \bibinfo {author} {\bibfnamefont {S.}~\bibnamefont {{Koudmani}}},\ }\bibfield  {title} {\bibinfo {title} {{Dynamics and spin alignment in massive, gravito-turbulent circumbinary discs around supermassive black hole binaries}},\ }\href {https://doi.org/10.48550/arXiv.2311.17144} {\bibfield  {journal} {\bibinfo  {journal} {arXiv e-prints}\ ,\ \bibinfo {eid} {arXiv:2311.17144}} (\bibinfo {year} {2023})},\ \Eprint {https://arxiv.org/abs/2311.17144} {arXiv:2311.17144 [astro-ph.HE]} \BibitemShut {NoStop}%
\bibitem [{\citenamefont {{Farris}}\ \emph {et~al.}(2012)\citenamefont {{Farris}}, \citenamefont {{Gold}}, \citenamefont {{Paschalidis}}, \citenamefont {{Etienne}},\ and\ \citenamefont {{Shapiro}}}]{2012PhRvL.109v1102F}%
  \BibitemOpen
  \bibfield  {author} {\bibinfo {author} {\bibfnamefont {B.~D.}\ \bibnamefont {{Farris}}}, \bibinfo {author} {\bibfnamefont {R.}~\bibnamefont {{Gold}}}, \bibinfo {author} {\bibfnamefont {V.}~\bibnamefont {{Paschalidis}}}, \bibinfo {author} {\bibfnamefont {Z.~B.}\ \bibnamefont {{Etienne}}},\ and\ \bibinfo {author} {\bibfnamefont {S.~L.}\ \bibnamefont {{Shapiro}}},\ }\bibfield  {title} {\bibinfo {title} {{Binary Black-Hole Mergers in Magnetized Disks: Simulations in Full General Relativity}},\ }\href {https://doi.org/10.1103/PhysRevLett.109.221102} {\bibfield  {journal} {\bibinfo  {journal} {\prl}\ }\textbf {\bibinfo {volume} {109}},\ \bibinfo {eid} {221102} (\bibinfo {year} {2012})},\ \Eprint {https://arxiv.org/abs/1207.3354} {arXiv:1207.3354 [astro-ph.HE]} \BibitemShut {NoStop}%
\bibitem [{\citenamefont {Gold}\ \emph {et~al.}(2014{\natexlab{a}})\citenamefont {Gold}, \citenamefont {Paschalidis}, \citenamefont {Etienne}, \citenamefont {Shapiro},\ and\ \citenamefont {Pfeiffer}}]{Gold:2013zma}%
  \BibitemOpen
  \bibfield  {author} {\bibinfo {author} {\bibfnamefont {R.}~\bibnamefont {Gold}}, \bibinfo {author} {\bibfnamefont {V.}~\bibnamefont {Paschalidis}}, \bibinfo {author} {\bibfnamefont {Z.~B.}\ \bibnamefont {Etienne}}, \bibinfo {author} {\bibfnamefont {S.~L.}\ \bibnamefont {Shapiro}},\ and\ \bibinfo {author} {\bibfnamefont {H.~P.}\ \bibnamefont {Pfeiffer}},\ }\bibfield  {title} {\bibinfo {title} {{Accretion disks around binary black holes of unequal mass: General relativistic magnetohydrodynamic simulations near decoupling}},\ }\href {https://doi.org/10.1103/PhysRevD.89.064060} {\bibfield  {journal} {\bibinfo  {journal} {Phys. Rev. D}\ }\textbf {\bibinfo {volume} {89}},\ \bibinfo {pages} {064060} (\bibinfo {year} {2014}{\natexlab{a}})},\ \Eprint {https://arxiv.org/abs/1312.0600} {arXiv:1312.0600 [astro-ph.HE]} \BibitemShut {NoStop}%
\bibitem [{\citenamefont {Gold}\ \emph {et~al.}(2014{\natexlab{b}})\citenamefont {Gold}, \citenamefont {Paschalidis}, \citenamefont {Ruiz}, \citenamefont {Shapiro}, \citenamefont {Etienne},\ and\ \citenamefont {Pfeiffer}}]{Gold:2014dta}%
  \BibitemOpen
  \bibfield  {author} {\bibinfo {author} {\bibfnamefont {R.}~\bibnamefont {Gold}}, \bibinfo {author} {\bibfnamefont {V.}~\bibnamefont {Paschalidis}}, \bibinfo {author} {\bibfnamefont {M.}~\bibnamefont {Ruiz}}, \bibinfo {author} {\bibfnamefont {S.~L.}\ \bibnamefont {Shapiro}}, \bibinfo {author} {\bibfnamefont {Z.~B.}\ \bibnamefont {Etienne}},\ and\ \bibinfo {author} {\bibfnamefont {H.~P.}\ \bibnamefont {Pfeiffer}},\ }\bibfield  {title} {\bibinfo {title} {{Accretion disks around binary black holes of unequal mass: General relativistic MHD simulations of postdecoupling and merger}},\ }\href {https://doi.org/10.1103/PhysRevD.90.104030} {\bibfield  {journal} {\bibinfo  {journal} {Phys. Rev. D}\ }\textbf {\bibinfo {volume} {90}},\ \bibinfo {pages} {104030} (\bibinfo {year} {2014}{\natexlab{b}})},\ \Eprint {https://arxiv.org/abs/1410.1543} {arXiv:1410.1543 [astro-ph.GA]} \BibitemShut {NoStop}%
\bibitem [{\citenamefont {Paschalidis}\ \emph {et~al.}(2021)\citenamefont {Paschalidis}, \citenamefont {Bright}, \citenamefont {Ruiz},\ and\ \citenamefont {Gold}}]{Paschalidis:2021ntt}%
  \BibitemOpen
  \bibfield  {author} {\bibinfo {author} {\bibfnamefont {V.}~\bibnamefont {Paschalidis}}, \bibinfo {author} {\bibfnamefont {J.}~\bibnamefont {Bright}}, \bibinfo {author} {\bibfnamefont {M.}~\bibnamefont {Ruiz}},\ and\ \bibinfo {author} {\bibfnamefont {R.}~\bibnamefont {Gold}},\ }\bibfield  {title} {\bibinfo {title} {{Minidisk dynamics in accreting, spinning black hole binaries: Simulations in full general relativity}},\ }\href {https://doi.org/10.3847/2041-8213/abee21} {\bibfield  {journal} {\bibinfo  {journal} {Astrophys. J. Lett.}\ }\textbf {\bibinfo {volume} {910}},\ \bibinfo {pages} {L26} (\bibinfo {year} {2021})},\ \Eprint {https://arxiv.org/abs/2102.06712} {arXiv:2102.06712 [astro-ph.HE]} \BibitemShut {NoStop}%
\bibitem [{\citenamefont {Ennoggi}\ \emph {et~al.}(2025)\citenamefont {Ennoggi} \emph {et~al.}}]{Ennoggi:2025nht}%
  \BibitemOpen
  \bibfield  {author} {\bibinfo {author} {\bibfnamefont {L.}~\bibnamefont {Ennoggi}} \emph {et~al.},\ }\bibfield  {title} {\bibinfo {title} {{Relativistic Gas Accretion onto Supermassive Black Hole Binaries from Inspiral through Merger}},\ }\href@noop {} {\bibfield  {journal} {\bibinfo  {journal} {arXiv preprint}\ } (\bibinfo {year} {2025})},\ \Eprint {https://arxiv.org/abs/2502.06389} {arXiv:2502.06389 [astro-ph.HE]} \BibitemShut {NoStop}%
\bibitem [{\citenamefont {Ressler}\ \emph {et~al.}(2018)\citenamefont {Ressler}, \citenamefont {Quataert},\ and\ \citenamefont {Stone}}]{Ressler:2018yhi}%
  \BibitemOpen
  \bibfield  {author} {\bibinfo {author} {\bibfnamefont {S.~M.}\ \bibnamefont {Ressler}}, \bibinfo {author} {\bibfnamefont {E.}~\bibnamefont {Quataert}},\ and\ \bibinfo {author} {\bibfnamefont {J.~M.}\ \bibnamefont {Stone}},\ }\bibfield  {title} {\bibinfo {title} {{Hydrodynamic simulations of the inner accretion flow of Sagittarius A* fuelled by stellar winds}},\ }\href {https://doi.org/10.1093/mnras/sty1146} {\bibfield  {journal} {\bibinfo  {journal} {Mon. Not. Roy. Astron. Soc.}\ }\textbf {\bibinfo {volume} {478}},\ \bibinfo {pages} {3544} (\bibinfo {year} {2018})},\ \Eprint {https://arxiv.org/abs/1805.00474} {arXiv:1805.00474 [astro-ph.HE]} \BibitemShut {NoStop}%
\bibitem [{\citenamefont {{Guo}}\ \emph {et~al.}(2023)\citenamefont {{Guo}}, \citenamefont {{Stone}}, \citenamefont {{Kim}},\ and\ \citenamefont {{Quataert}}}]{2023ApJ...946...26G}%
  \BibitemOpen
  \bibfield  {author} {\bibinfo {author} {\bibfnamefont {M.}~\bibnamefont {{Guo}}}, \bibinfo {author} {\bibfnamefont {J.~M.}\ \bibnamefont {{Stone}}}, \bibinfo {author} {\bibfnamefont {C.-G.}\ \bibnamefont {{Kim}}},\ and\ \bibinfo {author} {\bibfnamefont {E.}~\bibnamefont {{Quataert}}},\ }\bibfield  {title} {\bibinfo {title} {{Toward Horizon-scale Accretion onto Supermassive Black Holes in Elliptical Galaxies}},\ }\href {https://doi.org/10.3847/1538-4357/acb81e} {\bibfield  {journal} {\bibinfo  {journal} {Astrophys. J.}\ }\textbf {\bibinfo {volume} {946}},\ \bibinfo {eid} {26} (\bibinfo {year} {2023})},\ \Eprint {https://arxiv.org/abs/2211.05131} {arXiv:2211.05131 [astro-ph.HE]} \BibitemShut {NoStop}%
\bibitem [{\citenamefont {{Guo}}\ \emph {et~al.}(2024)\citenamefont {{Guo}}, \citenamefont {{Stone}}, \citenamefont {{Quataert}},\ and\ \citenamefont {{Kim}}}]{Guo2024}%
  \BibitemOpen
  \bibfield  {author} {\bibinfo {author} {\bibfnamefont {M.}~\bibnamefont {{Guo}}}, \bibinfo {author} {\bibfnamefont {J.~M.}\ \bibnamefont {{Stone}}}, \bibinfo {author} {\bibfnamefont {E.}~\bibnamefont {{Quataert}}},\ and\ \bibinfo {author} {\bibfnamefont {C.-G.}\ \bibnamefont {{Kim}}},\ }\bibfield  {title} {\bibinfo {title} {{Magnetized Accretion onto and Feedback from Supermassive Black Holes in Elliptical Galaxies}},\ }\href {https://doi.org/10.48550/arXiv.2405.11711} {\bibfield  {journal} {\bibinfo  {journal} {arXiv e-prints}\ ,\ \bibinfo {eid} {arXiv:2405.11711}} (\bibinfo {year} {2024})},\ \Eprint {https://arxiv.org/abs/2405.11711} {arXiv:2405.11711 [astro-ph.HE]} \BibitemShut {NoStop}%
\bibitem [{\citenamefont {{Hopkins}}\ \emph {et~al.}(2024{\natexlab{a}})\citenamefont {{Hopkins}}, \citenamefont {{Grudic}}, \citenamefont {{Su}}, \citenamefont {{Wellons}}, \citenamefont {{Angles-Alcazar}}, \citenamefont {{Steinwandel}}, \citenamefont {{Guszejnov}}, \citenamefont {{Murray}}, \citenamefont {{Faucher-Giguere}}, \citenamefont {{Quataert}},\ and\ \citenamefont {{Keres}}}]{Hopkins:2023ipv}%
  \BibitemOpen
  \bibfield  {author} {\bibinfo {author} {\bibfnamefont {P.~F.}\ \bibnamefont {{Hopkins}}}, \bibinfo {author} {\bibfnamefont {M.~Y.}\ \bibnamefont {{Grudic}}}, \bibinfo {author} {\bibfnamefont {K.-Y.}\ \bibnamefont {{Su}}}, \bibinfo {author} {\bibfnamefont {S.}~\bibnamefont {{Wellons}}}, \bibinfo {author} {\bibfnamefont {D.}~\bibnamefont {{Angles-Alcazar}}}, \bibinfo {author} {\bibfnamefont {U.~P.}\ \bibnamefont {{Steinwandel}}}, \bibinfo {author} {\bibfnamefont {D.}~\bibnamefont {{Guszejnov}}}, \bibinfo {author} {\bibfnamefont {N.}~\bibnamefont {{Murray}}}, \bibinfo {author} {\bibfnamefont {C.-A.}\ \bibnamefont {{Faucher-Giguere}}}, \bibinfo {author} {\bibfnamefont {E.}~\bibnamefont {{Quataert}}},\ and\ \bibinfo {author} {\bibfnamefont {D.}~\bibnamefont {{Keres}}},\ }\bibfield  {title} {\bibinfo {title} {{FORGE'd in FIRE: Resolving the End of Star Formation and Structure of AGN Accretion Disks from Cosmological Initial Conditions}},\ }\href {https://doi.org/10.21105/astro.2309.13115} {\bibfield  {journal} {\bibinfo  {journal} {The Open Journal of Astrophysics}\ }\textbf {\bibinfo {volume} {7}},\ \bibinfo {eid} {18} (\bibinfo {year} {2024}{\natexlab{a}})},\ \Eprint {https://arxiv.org/abs/2309.13115} {arXiv:2309.13115 [astro-ph.GA]} \BibitemShut {NoStop}%
\bibitem [{\citenamefont {{Hopkins}}\ \emph {et~al.}(2024{\natexlab{b}})\citenamefont {{Hopkins}}, \citenamefont {{Squire}}, \citenamefont {{Su}}, \citenamefont {{Steinwandel}}, \citenamefont {{Kremer}}, \citenamefont {{Shi}}, \citenamefont {{Grudic}}, \citenamefont {{Wellons}}, \citenamefont {{Faucher-Giguere}}, \citenamefont {{Angles-Alcazar}}, \citenamefont {{Murray}},\ and\ \citenamefont {{Quataert}}}]{Hopkins2023}%
  \BibitemOpen
  \bibfield  {author} {\bibinfo {author} {\bibfnamefont {P.~F.}\ \bibnamefont {{Hopkins}}}, \bibinfo {author} {\bibfnamefont {J.}~\bibnamefont {{Squire}}}, \bibinfo {author} {\bibfnamefont {K.-Y.}\ \bibnamefont {{Su}}}, \bibinfo {author} {\bibfnamefont {U.~P.}\ \bibnamefont {{Steinwandel}}}, \bibinfo {author} {\bibfnamefont {K.}~\bibnamefont {{Kremer}}}, \bibinfo {author} {\bibfnamefont {Y.}~\bibnamefont {{Shi}}}, \bibinfo {author} {\bibfnamefont {M.~Y.}\ \bibnamefont {{Grudic}}}, \bibinfo {author} {\bibfnamefont {S.}~\bibnamefont {{Wellons}}}, \bibinfo {author} {\bibfnamefont {C.-A.}\ \bibnamefont {{Faucher-Giguere}}}, \bibinfo {author} {\bibfnamefont {D.}~\bibnamefont {{Angles-Alcazar}}}, \bibinfo {author} {\bibfnamefont {N.}~\bibnamefont {{Murray}}},\ and\ \bibinfo {author} {\bibfnamefont {E.}~\bibnamefont {{Quataert}}},\ }\bibfield  {title} {\bibinfo {title} {{FORGE'd in FIRE II: The Formation of Magnetically-Dominated Quasar Accretion Disks from Cosmological Initial Conditions}},\ }\href {https://doi.org/10.21105/astro.2310.04506} {\bibfield  {journal} {\bibinfo  {journal} {The Open Journal of Astrophysics}\ }\textbf {\bibinfo {volume} {7}},\ \bibinfo {eid} {19} (\bibinfo {year} {2024}{\natexlab{b}})},\ \Eprint {https://arxiv.org/abs/2310.04506} {arXiv:2310.04506 [astro-ph.HE]} \BibitemShut {NoStop}%
\bibitem [{\citenamefont {Cho}\ \emph {et~al.}(2023)\citenamefont {Cho}, \citenamefont {Prather}, \citenamefont {Narayan}, \citenamefont {Natarajan}, \citenamefont {Su}, \citenamefont {Ricarte},\ and\ \citenamefont {Chatterjee}}]{Cho:2023wqr}%
  \BibitemOpen
  \bibfield  {author} {\bibinfo {author} {\bibfnamefont {H.}~\bibnamefont {Cho}}, \bibinfo {author} {\bibfnamefont {B.~S.}\ \bibnamefont {Prather}}, \bibinfo {author} {\bibfnamefont {R.}~\bibnamefont {Narayan}}, \bibinfo {author} {\bibfnamefont {P.}~\bibnamefont {Natarajan}}, \bibinfo {author} {\bibfnamefont {K.-Y.}\ \bibnamefont {Su}}, \bibinfo {author} {\bibfnamefont {A.}~\bibnamefont {Ricarte}},\ and\ \bibinfo {author} {\bibfnamefont {K.}~\bibnamefont {Chatterjee}},\ }\bibfield  {title} {\bibinfo {title} {{Bridging Scales in Black Hole Accretion and Feedback: Magnetized Bondi Accretion in 3D GRMHD}},\ }\href {https://doi.org/10.3847/2041-8213/ad1048} {\bibfield  {journal} {\bibinfo  {journal} {Astrophys. J. Lett.}\ }\textbf {\bibinfo {volume} {959}},\ \bibinfo {pages} {L22} (\bibinfo {year} {2023})},\ \Eprint {https://arxiv.org/abs/2310.19135} {arXiv:2310.19135 [astro-ph.HE]} \BibitemShut {NoStop}%
\bibitem [{\citenamefont {Cho}\ \emph {et~al.}(2024)\citenamefont {Cho}, \citenamefont {Prather}, \citenamefont {Su}, \citenamefont {Narayan},\ and\ \citenamefont {Natarajan}}]{Cho:2024wsp}%
  \BibitemOpen
  \bibfield  {author} {\bibinfo {author} {\bibfnamefont {H.}~\bibnamefont {Cho}}, \bibinfo {author} {\bibfnamefont {B.~S.}\ \bibnamefont {Prather}}, \bibinfo {author} {\bibfnamefont {K.-Y.}\ \bibnamefont {Su}}, \bibinfo {author} {\bibfnamefont {R.}~\bibnamefont {Narayan}},\ and\ \bibinfo {author} {\bibfnamefont {P.}~\bibnamefont {Natarajan}},\ }\bibfield  {title} {\bibinfo {title} {{Multizone Modeling of Black Hole Accretion and Feedback in 3D GRMHD: Bridging Vast Spatial and Temporal Scales}},\ }\href {https://doi.org/10.3847/1538-4357/ad9561} {\bibfield  {journal} {\bibinfo  {journal} {Astrophys. J.}\ }\textbf {\bibinfo {volume} {977}},\ \bibinfo {pages} {200} (\bibinfo {year} {2024})},\ \Eprint {https://arxiv.org/abs/2405.13887} {arXiv:2405.13887 [astro-ph.HE]} \BibitemShut {NoStop}%
\bibitem [{\citenamefont {Shi}\ \emph {et~al.}(2024)\citenamefont {Shi}, \citenamefont {Kremer},\ and\ \citenamefont {Hopkins}}]{Shi:2024wgh}%
  \BibitemOpen
  \bibfield  {author} {\bibinfo {author} {\bibfnamefont {Y.}~\bibnamefont {Shi}}, \bibinfo {author} {\bibfnamefont {K.}~\bibnamefont {Kremer}},\ and\ \bibinfo {author} {\bibfnamefont {P.~F.}\ \bibnamefont {Hopkins}},\ }\bibfield  {title} {\bibinfo {title} {{Feedback-regulated seed black hole growth in star-forming molecular clouds and galactic nuclei}},\ }\href {https://doi.org/10.1051/0004-6361/202450964} {\bibfield  {journal} {\bibinfo  {journal} {Astron. Astrophys.}\ }\textbf {\bibinfo {volume} {691}},\ \bibinfo {pages} {A24} (\bibinfo {year} {2024})},\ \Eprint {https://arxiv.org/abs/2405.12164} {arXiv:2405.12164 [astro-ph.GA]} \BibitemShut {NoStop}%
\bibitem [{\citenamefont {{Shi}}\ \emph {et~al.}(2024)\citenamefont {{Shi}}, \citenamefont {{Kremer}},\ and\ \citenamefont {{Hopkins}}}]{Shi2024}%
  \BibitemOpen
  \bibfield  {author} {\bibinfo {author} {\bibfnamefont {Y.}~\bibnamefont {{Shi}}}, \bibinfo {author} {\bibfnamefont {K.}~\bibnamefont {{Kremer}}},\ and\ \bibinfo {author} {\bibfnamefont {P.~F.}\ \bibnamefont {{Hopkins}}},\ }\bibfield  {title} {\bibinfo {title} {{From Seeds to Supermassive Black Holes: Capture, Growth, Migration, and Pairing in Dense Protobulge Environments}},\ }\href {https://doi.org/10.3847/2041-8213/ad5a95} {\bibfield  {journal} {\bibinfo  {journal} {Astrophys. J. Lett.}\ }\textbf {\bibinfo {volume} {969}},\ \bibinfo {eid} {L31} (\bibinfo {year} {2024})},\ \Eprint {https://arxiv.org/abs/2405.17338} {arXiv:2405.17338 [astro-ph.GA]} \BibitemShut {NoStop}%
\bibitem [{\citenamefont {Akiyama}\ \emph {et~al.}(2021)\citenamefont {Akiyama} \emph {et~al.}}]{EventHorizonTelescope:2021srq}%
  \BibitemOpen
  \bibfield  {author} {\bibinfo {author} {\bibfnamefont {K.}~\bibnamefont {Akiyama}} \emph {et~al.} (\bibinfo {collaboration} {Event Horizon Telescope}),\ }\bibfield  {title} {\bibinfo {title} {{First M87 Event Horizon Telescope Results. VIII. Magnetic Field Structure near The Event Horizon}},\ }\href {https://doi.org/10.3847/2041-8213/abe4de} {\bibfield  {journal} {\bibinfo  {journal} {Astrophys. J. Lett.}\ }\textbf {\bibinfo {volume} {910}},\ \bibinfo {pages} {L13} (\bibinfo {year} {2021})},\ \Eprint {https://arxiv.org/abs/2105.01173} {arXiv:2105.01173 [astro-ph.HE]} \BibitemShut {NoStop}%
\bibitem [{\citenamefont {Yuan}\ \emph {et~al.}(2022)\citenamefont {Yuan}, \citenamefont {Wang},\ and\ \citenamefont {Yang}}]{Yuan:2022mkw}%
  \BibitemOpen
  \bibfield  {author} {\bibinfo {author} {\bibfnamefont {F.}~\bibnamefont {Yuan}}, \bibinfo {author} {\bibfnamefont {H.}~\bibnamefont {Wang}},\ and\ \bibinfo {author} {\bibfnamefont {H.}~\bibnamefont {Yang}},\ }\bibfield  {title} {\bibinfo {title} {{The Accretion Flow in M87 is Really MAD}},\ }\href {https://doi.org/10.3847/1538-4357/ac4714} {\bibfield  {journal} {\bibinfo  {journal} {Astrophys. J.}\ }\textbf {\bibinfo {volume} {924}},\ \bibinfo {pages} {124} (\bibinfo {year} {2022})},\ \Eprint {https://arxiv.org/abs/2201.00512} {arXiv:2201.00512 [astro-ph.HE]} \BibitemShut {NoStop}%
\bibitem [{\citenamefont {Davis}\ and\ \citenamefont {Tchekhovskoy}(2020)}]{Davis:2020wea}%
  \BibitemOpen
  \bibfield  {author} {\bibinfo {author} {\bibfnamefont {S.~W.}\ \bibnamefont {Davis}}\ and\ \bibinfo {author} {\bibfnamefont {A.}~\bibnamefont {Tchekhovskoy}},\ }\bibfield  {title} {\bibinfo {title} {{Magnetohydrodynamics Simulations of Active Galactic Nucleus Disks and Jets}},\ }\href {https://doi.org/10.1146/annurev-astro-081817-051905} {\bibfield  {journal} {\bibinfo  {journal} {Ann. Rev. Astron. Astrophys.}\ }\textbf {\bibinfo {volume} {58}},\ \bibinfo {pages} {407} (\bibinfo {year} {2020})},\ \Eprint {https://arxiv.org/abs/2101.08839} {arXiv:2101.08839 [astro-ph.HE]} \BibitemShut {NoStop}%
\bibitem [{\citenamefont {{Narayan}}\ \emph {et~al.}(2003)\citenamefont {{Narayan}}, \citenamefont {{Igumenshchev}},\ and\ \citenamefont {{Abramowicz}}}]{Narayan2003}%
  \BibitemOpen
  \bibfield  {author} {\bibinfo {author} {\bibfnamefont {R.}~\bibnamefont {{Narayan}}}, \bibinfo {author} {\bibfnamefont {I.~V.}\ \bibnamefont {{Igumenshchev}}},\ and\ \bibinfo {author} {\bibfnamefont {M.~A.}\ \bibnamefont {{Abramowicz}}},\ }\bibfield  {title} {\bibinfo {title} {{Magnetically Arrested Disk: an Energetically Efficient Accretion Flow}},\ }\href {https://doi.org/10.1093/pasj/55.6.L69} {\bibfield  {journal} {\bibinfo  {journal} {PASJ}\ }\textbf {\bibinfo {volume} {55}},\ \bibinfo {pages} {L69} (\bibinfo {year} {2003})},\ \Eprint {https://arxiv.org/abs/astro-ph/0305029} {arXiv:astro-ph/0305029 [astro-ph]} \BibitemShut {NoStop}%
\bibitem [{\citenamefont {Igumenshchev}(2008)}]{Igumenshchev:2007bh}%
  \BibitemOpen
  \bibfield  {author} {\bibinfo {author} {\bibfnamefont {I.~V.}\ \bibnamefont {Igumenshchev}},\ }\bibfield  {title} {\bibinfo {title} {{Magnetically Arrested Disks and Origin of Poynting Jets: Numerical Study}},\ }\href {https://doi.org/10.1086/529025} {\bibfield  {journal} {\bibinfo  {journal} {Astrophys. J.}\ }\textbf {\bibinfo {volume} {677}},\ \bibinfo {pages} {317} (\bibinfo {year} {2008})},\ \Eprint {https://arxiv.org/abs/0711.4391} {arXiv:0711.4391 [astro-ph]} \BibitemShut {NoStop}%
\bibitem [{\citenamefont {Tchekhovskoy}\ \emph {et~al.}(2011)\citenamefont {Tchekhovskoy}, \citenamefont {Narayan},\ and\ \citenamefont {McKinney}}]{Tchekhovskoy:2011zx}%
  \BibitemOpen
  \bibfield  {author} {\bibinfo {author} {\bibfnamefont {A.}~\bibnamefont {Tchekhovskoy}}, \bibinfo {author} {\bibfnamefont {R.}~\bibnamefont {Narayan}},\ and\ \bibinfo {author} {\bibfnamefont {J.~C.}\ \bibnamefont {McKinney}},\ }\bibfield  {title} {\bibinfo {title} {{Efficient Generation of Jets from Magnetically Arrested Accretion on a Rapidly Spinning Black Hole}},\ }\href {https://doi.org/10.1111/j.1745-3933.2011.01147.x} {\bibfield  {journal} {\bibinfo  {journal} {Mon. Not. Roy. Astron. Soc.}\ }\textbf {\bibinfo {volume} {418}},\ \bibinfo {pages} {L79} (\bibinfo {year} {2011})},\ \Eprint {https://arxiv.org/abs/1108.0412} {arXiv:1108.0412 [astro-ph.HE]} \BibitemShut {NoStop}%
\bibitem [{\citenamefont {Parfrey}\ and\ \citenamefont {Tchekhovskoy}(2017)}]{Parfrey:2017nby}%
  \BibitemOpen
  \bibfield  {author} {\bibinfo {author} {\bibfnamefont {K.}~\bibnamefont {Parfrey}}\ and\ \bibinfo {author} {\bibfnamefont {A.}~\bibnamefont {Tchekhovskoy}},\ }\bibfield  {title} {\bibinfo {title} {{General-Relativistic Simulations of Four States of Accretion onto Millisecond Pulsars}},\ }\href {https://doi.org/10.3847/2041-8213/aa9c85} {\bibfield  {journal} {\bibinfo  {journal} {Astrophys. J. Lett.}\ }\textbf {\bibinfo {volume} {851}},\ \bibinfo {pages} {L34} (\bibinfo {year} {2017})},\ \Eprint {https://arxiv.org/abs/1708.06362} {arXiv:1708.06362 [astro-ph.HE]} \BibitemShut {NoStop}%
\bibitem [{\citenamefont {{Zhu}}\ \emph {et~al.}(2024)\citenamefont {{Zhu}}, \citenamefont {{Stone}},\ and\ \citenamefont {{Calvet}}}]{Zhu2024MNRAS.528.2883Z}%
  \BibitemOpen
  \bibfield  {author} {\bibinfo {author} {\bibfnamefont {Z.}~\bibnamefont {{Zhu}}}, \bibinfo {author} {\bibfnamefont {J.~M.}\ \bibnamefont {{Stone}}},\ and\ \bibinfo {author} {\bibfnamefont {N.}~\bibnamefont {{Calvet}}},\ }\bibfield  {title} {\bibinfo {title} {{A global 3D simulation of magnetospheric accretion - I. Magnetically disrupted discs and surface accretion}},\ }\href {https://doi.org/10.1093/mnras/stad3712} {\bibfield  {journal} {\bibinfo  {journal} {Mon. Not. Roy. Astron. Soc.}\ }\textbf {\bibinfo {volume} {528}},\ \bibinfo {pages} {2883} (\bibinfo {year} {2024})},\ \Eprint {https://arxiv.org/abs/2309.15318} {arXiv:2309.15318 [astro-ph.SR]} \BibitemShut {NoStop}%
\bibitem [{\citenamefont {Zhu}(2025)}]{Zhu:2025uhl}%
  \BibitemOpen
  \bibfield  {author} {\bibinfo {author} {\bibfnamefont {Z.}~\bibnamefont {Zhu}},\ }\bibfield  {title} {\bibinfo {title} {{Global 3D simulations of magnetospheric accretion \textendash{} II. Hotspots, equilibrium torque, episodic wind, and mid-plane outflow}},\ }\href {https://doi.org/10.1093/mnras/staf250} {\bibfield  {journal} {\bibinfo  {journal} {Mon. Not. Roy. Astron. Soc.}\ }\textbf {\bibinfo {volume} {537}},\ \bibinfo {pages} {3701} (\bibinfo {year} {2025})},\ \Eprint {https://arxiv.org/abs/2501.08112} {arXiv:2501.08112 [astro-ph.SR]} \BibitemShut {NoStop}%
\bibitem [{\citenamefont {Liska}\ \emph {et~al.}(2022)\citenamefont {Liska}, \citenamefont {Musoke}, \citenamefont {Tchekhovskoy}, \citenamefont {Porth},\ and\ \citenamefont {Beloborodov}}]{Liska:2022jdy}%
  \BibitemOpen
  \bibfield  {author} {\bibinfo {author} {\bibfnamefont {M.~T.~P.}\ \bibnamefont {Liska}}, \bibinfo {author} {\bibfnamefont {G.}~\bibnamefont {Musoke}}, \bibinfo {author} {\bibfnamefont {A.}~\bibnamefont {Tchekhovskoy}}, \bibinfo {author} {\bibfnamefont {O.}~\bibnamefont {Porth}},\ and\ \bibinfo {author} {\bibfnamefont {A.~M.}\ \bibnamefont {Beloborodov}},\ }\bibfield  {title} {\bibinfo {title} {{Formation of Magnetically Truncated Accretion Disks in 3D Radiation-transport Two-temperature GRMHD Simulations}},\ }\href {https://doi.org/10.3847/2041-8213/ac84db} {\bibfield  {journal} {\bibinfo  {journal} {Astrophys. J. Lett.}\ }\textbf {\bibinfo {volume} {935}},\ \bibinfo {pages} {L1} (\bibinfo {year} {2022})},\ \Eprint {https://arxiv.org/abs/2201.03526} {arXiv:2201.03526 [astro-ph.HE]} \BibitemShut {NoStop}%
\bibitem [{\citenamefont {Gammie}(2001)}]{Gammie:2001bw}%
  \BibitemOpen
  \bibfield  {author} {\bibinfo {author} {\bibfnamefont {C.~F.}\ \bibnamefont {Gammie}},\ }\bibfield  {title} {\bibinfo {title} {{Nonlinear Outcome of Gravitational Instability in Cooling, Gaseous Disks}},\ }\href {https://doi.org/10.1086/320631} {\bibfield  {journal} {\bibinfo  {journal} {Astrophys. J.}\ }\textbf {\bibinfo {volume} {553}},\ \bibinfo {pages} {174} (\bibinfo {year} {2001})},\ \Eprint {https://arxiv.org/abs/astro-ph/0101501} {arXiv:astro-ph/0101501} \BibitemShut {NoStop}%
\bibitem [{\citenamefont {{Bisnovatyi-Kogan}}\ and\ \citenamefont {{Ruzmaikin}}(1974)}]{Bisnovatyi-Kogan:1974Ap&SS..28...45B}%
  \BibitemOpen
  \bibfield  {author} {\bibinfo {author} {\bibfnamefont {G.~S.}\ \bibnamefont {{Bisnovatyi-Kogan}}}\ and\ \bibinfo {author} {\bibfnamefont {A.~A.}\ \bibnamefont {{Ruzmaikin}}},\ }\bibfield  {title} {\bibinfo {title} {{The Accretion of Matter by a Collapsing Star in the Presence of a Magnetic Field}},\ }\href {https://doi.org/10.1007/BF00642237} {\bibfield  {journal} {\bibinfo  {journal} {\apss}\ }\textbf {\bibinfo {volume} {28}},\ \bibinfo {pages} {45} (\bibinfo {year} {1974})}\BibitemShut {NoStop}%
\bibitem [{\citenamefont {{Bisnovatyi-Kogan}}\ and\ \citenamefont {{Ruzmaikin}}(1976)}]{Bisnovatyi-Kogan:1976Ap&SS..42..401B}%
  \BibitemOpen
  \bibfield  {author} {\bibinfo {author} {\bibfnamefont {G.~S.}\ \bibnamefont {{Bisnovatyi-Kogan}}}\ and\ \bibinfo {author} {\bibfnamefont {A.~A.}\ \bibnamefont {{Ruzmaikin}}},\ }\bibfield  {title} {\bibinfo {title} {{The Accretion of Matter by a Collapsing Star in the Presence of a Magnetic Field. II: Self-consistent Stationary Picture}},\ }\href {https://doi.org/10.1007/BF01225967} {\bibfield  {journal} {\bibinfo  {journal} {\apss}\ }\textbf {\bibinfo {volume} {42}},\ \bibinfo {pages} {401} (\bibinfo {year} {1976})}\BibitemShut {NoStop}%
\bibitem [{\citenamefont {Igumenshchev}\ \emph {et~al.}(2003)\citenamefont {Igumenshchev}, \citenamefont {Narayan},\ and\ \citenamefont {Abramowicz}}]{Igumenshchev:2003rt}%
  \BibitemOpen
  \bibfield  {author} {\bibinfo {author} {\bibfnamefont {I.~V.}\ \bibnamefont {Igumenshchev}}, \bibinfo {author} {\bibfnamefont {R.}~\bibnamefont {Narayan}},\ and\ \bibinfo {author} {\bibfnamefont {M.~A.}\ \bibnamefont {Abramowicz}},\ }\bibfield  {title} {\bibinfo {title} {{Three-dimensional mhd simulations of radiatively inefficient accretion flows}},\ }\href {https://doi.org/10.1086/375769} {\bibfield  {journal} {\bibinfo  {journal} {Astrophys. J.}\ }\textbf {\bibinfo {volume} {592}},\ \bibinfo {pages} {1042} (\bibinfo {year} {2003})},\ \Eprint {https://arxiv.org/abs/astro-ph/0301402} {arXiv:astro-ph/0301402} \BibitemShut {NoStop}%
\bibitem [{\citenamefont {Yuan}\ and\ \citenamefont {Narayan}(2014)}]{Yuan:2014gma}%
  \BibitemOpen
  \bibfield  {author} {\bibinfo {author} {\bibfnamefont {F.}~\bibnamefont {Yuan}}\ and\ \bibinfo {author} {\bibfnamefont {R.}~\bibnamefont {Narayan}},\ }\bibfield  {title} {\bibinfo {title} {{Hot Accretion Flows Around Black Holes}},\ }\href {https://doi.org/10.1146/annurev-astro-082812-141003} {\bibfield  {journal} {\bibinfo  {journal} {Ann. Rev. Astron. Astrophys.}\ }\textbf {\bibinfo {volume} {52}},\ \bibinfo {pages} {529} (\bibinfo {year} {2014})},\ \Eprint {https://arxiv.org/abs/1401.0586} {arXiv:1401.0586 [astro-ph.HE]} \BibitemShut {NoStop}%
\bibitem [{\citenamefont {{Most}}\ and\ \citenamefont {{Wang}}(2024)}]{Most:2024onq}%
  \BibitemOpen
  \bibfield  {author} {\bibinfo {author} {\bibfnamefont {E.~R.}\ \bibnamefont {{Most}}}\ and\ \bibinfo {author} {\bibfnamefont {H.-Y.}\ \bibnamefont {{Wang}}},\ }\bibfield  {title} {\bibinfo {title} {{Decoupling of a supermassive black hole binary from its magnetically arrested circumbinary accretion disk}},\ }\href {https://doi.org/10.48550/arXiv.2410.23264} {\bibfield  {journal} {\bibinfo  {journal} {arXiv e-prints}\ ,\ \bibinfo {eid} {arXiv:2410.23264}} (\bibinfo {year} {2024})},\ \Eprint {https://arxiv.org/abs/2410.23264} {arXiv:2410.23264 [astro-ph.HE]} \BibitemShut {NoStop}%
\bibitem [{\citenamefont {Ripperda}\ \emph {et~al.}(2022)\citenamefont {Ripperda}, \citenamefont {Liska}, \citenamefont {Chatterjee}, \citenamefont {Musoke}, \citenamefont {Philippov}, \citenamefont {Markoff}, \citenamefont {Tchekhovskoy},\ and\ \citenamefont {Younsi}}]{Ripperda:2021zpn}%
  \BibitemOpen
  \bibfield  {author} {\bibinfo {author} {\bibfnamefont {B.}~\bibnamefont {Ripperda}}, \bibinfo {author} {\bibfnamefont {M.}~\bibnamefont {Liska}}, \bibinfo {author} {\bibfnamefont {K.}~\bibnamefont {Chatterjee}}, \bibinfo {author} {\bibfnamefont {G.}~\bibnamefont {Musoke}}, \bibinfo {author} {\bibfnamefont {A.~A.}\ \bibnamefont {Philippov}}, \bibinfo {author} {\bibfnamefont {S.~B.}\ \bibnamefont {Markoff}}, \bibinfo {author} {\bibfnamefont {A.}~\bibnamefont {Tchekhovskoy}},\ and\ \bibinfo {author} {\bibfnamefont {Z.}~\bibnamefont {Younsi}},\ }\bibfield  {title} {\bibinfo {title} {{Black Hole Flares: Ejection of Accreted Magnetic Flux through 3D Plasmoid-mediated Reconnection}},\ }\href {https://doi.org/10.3847/2041-8213/ac46a1} {\bibfield  {journal} {\bibinfo  {journal} {Astrophys. J. Lett.}\ }\textbf {\bibinfo {volume} {924}},\ \bibinfo {pages} {L32} (\bibinfo {year} {2022})},\ \Eprint {https://arxiv.org/abs/2109.15115} {arXiv:2109.15115 [astro-ph.HE]} \BibitemShut {NoStop}%
\bibitem [{\citenamefont {Begelman}\ \emph {et~al.}(2022)\citenamefont {Begelman}, \citenamefont {Scepi},\ and\ \citenamefont {Dexter}}]{Begelman:2021ufo}%
  \BibitemOpen
  \bibfield  {author} {\bibinfo {author} {\bibfnamefont {M.~C.}\ \bibnamefont {Begelman}}, \bibinfo {author} {\bibfnamefont {N.}~\bibnamefont {Scepi}},\ and\ \bibinfo {author} {\bibfnamefont {J.}~\bibnamefont {Dexter}},\ }\bibfield  {title} {\bibinfo {title} {{What really makes an accretion disc MAD}},\ }\href {https://doi.org/10.1093/mnras/stab3790} {\bibfield  {journal} {\bibinfo  {journal} {Mon. Not. Roy. Astron. Soc.}\ }\textbf {\bibinfo {volume} {511}},\ \bibinfo {pages} {2040} (\bibinfo {year} {2022})},\ \Eprint {https://arxiv.org/abs/2111.02439} {arXiv:2111.02439 [astro-ph.HE]} \BibitemShut {NoStop}%
\bibitem [{\citenamefont {Porth}\ \emph {et~al.}(2021)\citenamefont {Porth}, \citenamefont {Mizuno}, \citenamefont {Younsi},\ and\ \citenamefont {Fromm}}]{Porth:2020txf}%
  \BibitemOpen
  \bibfield  {author} {\bibinfo {author} {\bibfnamefont {O.}~\bibnamefont {Porth}}, \bibinfo {author} {\bibfnamefont {Y.}~\bibnamefont {Mizuno}}, \bibinfo {author} {\bibfnamefont {Z.}~\bibnamefont {Younsi}},\ and\ \bibinfo {author} {\bibfnamefont {C.~M.}\ \bibnamefont {Fromm}},\ }\bibfield  {title} {\bibinfo {title} {{Flares in the Galactic Centre \textendash{} I. Orbiting flux tubes in magnetically arrested black hole accretion discs}},\ }\href {https://doi.org/10.1093/mnras/stab163} {\bibfield  {journal} {\bibinfo  {journal} {Mon. Not. Roy. Astron. Soc.}\ }\textbf {\bibinfo {volume} {502}},\ \bibinfo {pages} {2023} (\bibinfo {year} {2021})},\ \Eprint {https://arxiv.org/abs/2006.03658} {arXiv:2006.03658 [astro-ph.HE]} \BibitemShut {NoStop}%
\bibitem [{\citenamefont {McKinney}\ \emph {et~al.}(2012)\citenamefont {McKinney}, \citenamefont {Tchekhovskoy},\ and\ \citenamefont {Blandford}}]{McKinney:2012vh}%
  \BibitemOpen
  \bibfield  {author} {\bibinfo {author} {\bibfnamefont {J.~C.}\ \bibnamefont {McKinney}}, \bibinfo {author} {\bibfnamefont {A.}~\bibnamefont {Tchekhovskoy}},\ and\ \bibinfo {author} {\bibfnamefont {R.~D.}\ \bibnamefont {Blandford}},\ }\bibfield  {title} {\bibinfo {title} {{General Relativistic Magnetohydrodynamic Simulations of Magnetically Choked Accretion Flows around Black Holes}},\ }\href {https://doi.org/10.1111/j.1365-2966.2012.21074.x} {\bibfield  {journal} {\bibinfo  {journal} {Mon. Not. Roy. Astron. Soc.}\ }\textbf {\bibinfo {volume} {423}},\ \bibinfo {pages} {3083} (\bibinfo {year} {2012})},\ \Eprint {https://arxiv.org/abs/1201.4163} {arXiv:1201.4163 [astro-ph.HE]} \BibitemShut {NoStop}%
\bibitem [{\citenamefont {{Chatterjee}}\ and\ \citenamefont {{Narayan}}(2022)}]{Chatterjee2022}%
  \BibitemOpen
  \bibfield  {author} {\bibinfo {author} {\bibfnamefont {K.}~\bibnamefont {{Chatterjee}}}\ and\ \bibinfo {author} {\bibfnamefont {R.}~\bibnamefont {{Narayan}}},\ }\bibfield  {title} {\bibinfo {title} {{Flux Eruption Events Drive Angular Momentum Transport in Magnetically Arrested Accretion Flows}},\ }\href {https://doi.org/10.3847/1538-4357/ac9d97} {\bibfield  {journal} {\bibinfo  {journal} {Astrophys. J.}\ }\textbf {\bibinfo {volume} {941}},\ \bibinfo {eid} {30} (\bibinfo {year} {2022})},\ \Eprint {https://arxiv.org/abs/2210.08045} {arXiv:2210.08045 [astro-ph.HE]} \BibitemShut {NoStop}%
\bibitem [{\citenamefont {Davelaar}\ \emph {et~al.}(2019)\citenamefont {Davelaar}, \citenamefont {Olivares}, \citenamefont {Porth}, \citenamefont {Bronzwaer}, \citenamefont {Janssen}, \citenamefont {Roelofs}, \citenamefont {Mizuno}, \citenamefont {Fromm}, \citenamefont {Falcke},\ and\ \citenamefont {Rezzolla}}]{Davelaar:2019jxr}%
  \BibitemOpen
  \bibfield  {author} {\bibinfo {author} {\bibfnamefont {J.}~\bibnamefont {Davelaar}}, \bibinfo {author} {\bibfnamefont {H.}~\bibnamefont {Olivares}}, \bibinfo {author} {\bibfnamefont {O.}~\bibnamefont {Porth}}, \bibinfo {author} {\bibfnamefont {T.}~\bibnamefont {Bronzwaer}}, \bibinfo {author} {\bibfnamefont {M.}~\bibnamefont {Janssen}}, \bibinfo {author} {\bibfnamefont {F.}~\bibnamefont {Roelofs}}, \bibinfo {author} {\bibfnamefont {Y.}~\bibnamefont {Mizuno}}, \bibinfo {author} {\bibfnamefont {C.~M.}\ \bibnamefont {Fromm}}, \bibinfo {author} {\bibfnamefont {H.}~\bibnamefont {Falcke}},\ and\ \bibinfo {author} {\bibfnamefont {L.}~\bibnamefont {Rezzolla}},\ }\bibfield  {title} {\bibinfo {title} {{Modeling non-thermal emission from the jet-launching region of M 87 with adaptive mesh refinement}},\ }\href {https://doi.org/10.1051/0004-6361/201936150} {\bibfield  {journal} {\bibinfo  {journal} {Astron. Astrophys.}\ }\textbf {\bibinfo {volume} {632}},\ \bibinfo {pages} {A2} (\bibinfo {year} {2019})},\ \Eprint {https://arxiv.org/abs/1906.10065} {arXiv:1906.10065 [astro-ph.HE]} \BibitemShut {NoStop}%
\bibitem [{\citenamefont {Dexter}\ \emph {et~al.}(2020)\citenamefont {Dexter} \emph {et~al.}}]{Dexter:2020cuv}%
  \BibitemOpen
  \bibfield  {author} {\bibinfo {author} {\bibfnamefont {J.}~\bibnamefont {Dexter}} \emph {et~al.},\ }\bibfield  {title} {\bibinfo {title} {{Sgr A* near-infrared flares from reconnection events in a magnetically arrested disc}},\ }\href {https://doi.org/10.1093/mnras/staa2288} {\bibfield  {journal} {\bibinfo  {journal} {Mon. Not. Roy. Astron. Soc.}\ }\textbf {\bibinfo {volume} {497}},\ \bibinfo {pages} {4999} (\bibinfo {year} {2020})},\ \Eprint {https://arxiv.org/abs/2006.03657} {arXiv:2006.03657 [astro-ph.HE]} \BibitemShut {NoStop}%
\bibitem [{\citenamefont {Hakobyan}\ \emph {et~al.}(2023)\citenamefont {Hakobyan}, \citenamefont {Ripperda},\ and\ \citenamefont {Philippov}}]{Hakobyan:2022alv}%
  \BibitemOpen
  \bibfield  {author} {\bibinfo {author} {\bibfnamefont {H.}~\bibnamefont {Hakobyan}}, \bibinfo {author} {\bibfnamefont {B.}~\bibnamefont {Ripperda}},\ and\ \bibinfo {author} {\bibfnamefont {A.}~\bibnamefont {Philippov}},\ }\bibfield  {title} {\bibinfo {title} {{Radiative Reconnection-powered TeV Flares from the Black Hole Magnetosphere in M87}},\ }\href {https://doi.org/10.3847/2041-8213/acb264} {\bibfield  {journal} {\bibinfo  {journal} {Astrophys. J. Lett.}\ }\textbf {\bibinfo {volume} {943}},\ \bibinfo {pages} {L29} (\bibinfo {year} {2023})},\ \Eprint {https://arxiv.org/abs/2209.02105} {arXiv:2209.02105 [astro-ph.HE]} \BibitemShut {NoStop}%
\bibitem [{\citenamefont {Zhdankin}\ \emph {et~al.}(2023)\citenamefont {Zhdankin}, \citenamefont {Ripperda},\ and\ \citenamefont {Philippov}}]{Zhdankin:2023wch}%
  \BibitemOpen
  \bibfield  {author} {\bibinfo {author} {\bibfnamefont {V.}~\bibnamefont {Zhdankin}}, \bibinfo {author} {\bibfnamefont {B.}~\bibnamefont {Ripperda}},\ and\ \bibinfo {author} {\bibfnamefont {A.~A.}\ \bibnamefont {Philippov}},\ }\bibfield  {title} {\bibinfo {title} {{Particle acceleration by magnetic Rayleigh-Taylor instability: Mechanism for flares in black hole accretion flows}},\ }\href {https://doi.org/10.1103/PhysRevResearch.5.043023} {\bibfield  {journal} {\bibinfo  {journal} {Phys. Rev. Res.}\ }\textbf {\bibinfo {volume} {5}},\ \bibinfo {pages} {043023} (\bibinfo {year} {2023})},\ \Eprint {https://arxiv.org/abs/2302.05276} {arXiv:2302.05276 [astro-ph.HE]} \BibitemShut {NoStop}%
\bibitem [{\citenamefont {D'Orazio}\ \emph {et~al.}(2013)\citenamefont {D'Orazio}, \citenamefont {Haiman},\ and\ \citenamefont {MacFadyen}}]{DOrazio:2012vqt}%
  \BibitemOpen
  \bibfield  {author} {\bibinfo {author} {\bibfnamefont {D.~J.}\ \bibnamefont {D'Orazio}}, \bibinfo {author} {\bibfnamefont {Z.}~\bibnamefont {Haiman}},\ and\ \bibinfo {author} {\bibfnamefont {A.}~\bibnamefont {MacFadyen}},\ }\bibfield  {title} {\bibinfo {title} {{Accretion into the Central Cavity of a Circumbinary Disk}},\ }\href {https://doi.org/10.1093/mnras/stt1787} {\bibfield  {journal} {\bibinfo  {journal} {Mon. Not. Roy. Astron. Soc.}\ }\textbf {\bibinfo {volume} {436}},\ \bibinfo {pages} {2997} (\bibinfo {year} {2013})},\ \Eprint {https://arxiv.org/abs/1210.0536} {arXiv:1210.0536 [astro-ph.GA]} \BibitemShut {NoStop}%
\bibitem [{\citenamefont {D'Orazio}\ \emph {et~al.}(2016)\citenamefont {D'Orazio}, \citenamefont {Haiman}, \citenamefont {Duffell}, \citenamefont {MacFadyen},\ and\ \citenamefont {Farris}}]{DOrazio:2015shf}%
  \BibitemOpen
  \bibfield  {author} {\bibinfo {author} {\bibfnamefont {D.~J.}\ \bibnamefont {D'Orazio}}, \bibinfo {author} {\bibfnamefont {Z.}~\bibnamefont {Haiman}}, \bibinfo {author} {\bibfnamefont {P.}~\bibnamefont {Duffell}}, \bibinfo {author} {\bibfnamefont {A.~I.}\ \bibnamefont {MacFadyen}},\ and\ \bibinfo {author} {\bibfnamefont {B.~D.}\ \bibnamefont {Farris}},\ }\bibfield  {title} {\bibinfo {title} {{A transition in circumbinary accretion discs at a binary mass ratio of 1:25}},\ }\href {https://doi.org/10.1093/mnras/stw792} {\bibfield  {journal} {\bibinfo  {journal} {Mon. Not. Roy. Astron. Soc.}\ }\textbf {\bibinfo {volume} {459}},\ \bibinfo {pages} {2379} (\bibinfo {year} {2016})},\ \Eprint {https://arxiv.org/abs/1512.05788} {arXiv:1512.05788 [astro-ph.HE]} \BibitemShut {NoStop}%
\bibitem [{\citenamefont {Balbus}\ and\ \citenamefont {Hawley}(1991)}]{Balbus:1991ay}%
  \BibitemOpen
  \bibfield  {author} {\bibinfo {author} {\bibfnamefont {S.~A.}\ \bibnamefont {Balbus}}\ and\ \bibinfo {author} {\bibfnamefont {J.~F.}\ \bibnamefont {Hawley}},\ }\bibfield  {title} {\bibinfo {title} {{A powerful local shear instability in weakly magnetized disks. 1. Linear analysis. 2. Nonlinear evolution}},\ }\href {https://doi.org/10.1086/170270} {\bibfield  {journal} {\bibinfo  {journal} {Astrophys. J.}\ }\textbf {\bibinfo {volume} {376}},\ \bibinfo {pages} {214} (\bibinfo {year} {1991})}\BibitemShut {NoStop}%
\bibitem [{\citenamefont {Guo}\ \emph {et~al.}(2025{\natexlab{a}})\citenamefont {Guo}, \citenamefont {Stone}, \citenamefont {Quataert},\ and\ \citenamefont {Springel}}]{Guo:2025sjb}%
  \BibitemOpen
  \bibfield  {author} {\bibinfo {author} {\bibfnamefont {M.}~\bibnamefont {Guo}}, \bibinfo {author} {\bibfnamefont {J.~M.}\ \bibnamefont {Stone}}, \bibinfo {author} {\bibfnamefont {E.}~\bibnamefont {Quataert}},\ and\ \bibinfo {author} {\bibfnamefont {V.}~\bibnamefont {Springel}},\ }\bibfield  {title} {\bibinfo {title} {{Cyclic Zoom: Multi-scale GRMHD Modeling of Black Hole Accretion and Feedback}},\ }\href@noop {} {\bibfield  {journal} {\bibinfo  {journal} {arxiv}\ } (\bibinfo {year} {2025}{\natexlab{a}})},\ \Eprint {https://arxiv.org/abs/2504.16802} {arXiv:2504.16802 [astro-ph.HE]} \BibitemShut {NoStop}%
\bibitem [{\citenamefont {{Lalakos}}\ \emph {et~al.}(2025)\citenamefont {{Lalakos}}, \citenamefont {{Tchekhovskoy}}, \citenamefont {{Most}}, \citenamefont {{Ripperda}}, \citenamefont {{Chatterjee}},\ and\ \citenamefont {{Liska}}}]{Lalakos:2025msz}%
  \BibitemOpen
  \bibfield  {author} {\bibinfo {author} {\bibfnamefont {A.}~\bibnamefont {{Lalakos}}}, \bibinfo {author} {\bibfnamefont {A.}~\bibnamefont {{Tchekhovskoy}}}, \bibinfo {author} {\bibfnamefont {E.~R.}\ \bibnamefont {{Most}}}, \bibinfo {author} {\bibfnamefont {B.}~\bibnamefont {{Ripperda}}}, \bibinfo {author} {\bibfnamefont {K.}~\bibnamefont {{Chatterjee}}},\ and\ \bibinfo {author} {\bibfnamefont {M.}~\bibnamefont {{Liska}}},\ }\bibfield  {title} {\bibinfo {title} {{Universal Radial Scaling of Large-Scale Black Hole Accretion for Magnetically Arrested And Rocking Accretion Disks}},\ }\href {https://doi.org/10.48550/arXiv.2505.23888} {\bibfield  {journal} {\bibinfo  {journal} {arXiv e-prints}\ ,\ \bibinfo {eid} {arXiv:2505.23888}} (\bibinfo {year} {2025})},\ \Eprint {https://arxiv.org/abs/2505.23888} {arXiv:2505.23888 [astro-ph.HE]} \BibitemShut {NoStop}%
\bibitem [{\citenamefont {Spruit}\ \emph {et~al.}(1995)\citenamefont {Spruit}, \citenamefont {Stehle},\ and\ \citenamefont {Papaloizou}}]{Spruit:1995fr}%
  \BibitemOpen
  \bibfield  {author} {\bibinfo {author} {\bibfnamefont {H.~C.}\ \bibnamefont {Spruit}}, \bibinfo {author} {\bibfnamefont {R.}~\bibnamefont {Stehle}},\ and\ \bibinfo {author} {\bibfnamefont {J.~C.~B.}\ \bibnamefont {Papaloizou}},\ }\bibfield  {title} {\bibinfo {title} {{Interchange instability in an accretion disc with a poloidal magnetic field}},\ }\href {https://doi.org/10.1093/mnras/275.4.1223} {\bibfield  {journal} {\bibinfo  {journal} {Mon. Not. Roy. Astron. Soc.}\ }\textbf {\bibinfo {volume} {275}},\ \bibinfo {pages} {1223} (\bibinfo {year} {1995})},\ \Eprint {https://arxiv.org/abs/astro-ph/9504043} {arXiv:astro-ph/9504043} \BibitemShut {NoStop}%
\bibitem [{\citenamefont {Kulkarni}\ and\ \citenamefont {Romanova}(2008)}]{Kulkarni:2008vk}%
  \BibitemOpen
  \bibfield  {author} {\bibinfo {author} {\bibfnamefont {A.~K.}\ \bibnamefont {Kulkarni}}\ and\ \bibinfo {author} {\bibfnamefont {M.~M.}\ \bibnamefont {Romanova}},\ }\bibfield  {title} {\bibinfo {title} {{Accretion to Magnetized Stars through the Rayleigh-Taylor Instability: Global Three-Dimensional Simulations}},\ }\href {https://doi.org/10.1111/j.1365-2966.2008.13094.x} {\bibfield  {journal} {\bibinfo  {journal} {Mon. Not. Roy. Astron. Soc.}\ }\textbf {\bibinfo {volume} {386}},\ \bibinfo {pages} {673} (\bibinfo {year} {2008})},\ \Eprint {https://arxiv.org/abs/0802.1759} {arXiv:0802.1759 [astro-ph]} \BibitemShut {NoStop}%
\bibitem [{\citenamefont {Takasao}\ \emph {et~al.}(2022)\citenamefont {Takasao}, \citenamefont {Tomida}, \citenamefont {Iwasaki},\ and\ \citenamefont {Suzuki}}]{Takasao:2022glf}%
  \BibitemOpen
  \bibfield  {author} {\bibinfo {author} {\bibfnamefont {S.}~\bibnamefont {Takasao}}, \bibinfo {author} {\bibfnamefont {K.}~\bibnamefont {Tomida}}, \bibinfo {author} {\bibfnamefont {K.}~\bibnamefont {Iwasaki}},\ and\ \bibinfo {author} {\bibfnamefont {T.~K.}\ \bibnamefont {Suzuki}},\ }\bibfield  {title} {\bibinfo {title} {{Three-dimensional Simulations of Magnetospheric Accretion in a T Tauri Star: Accretion and Wind Structures Just Around the Star}},\ }\href {https://doi.org/10.3847/1538-4357/ac9eb1} {\bibfield  {journal} {\bibinfo  {journal} {Astrophys. J.}\ }\textbf {\bibinfo {volume} {941}},\ \bibinfo {pages} {73} (\bibinfo {year} {2022})},\ \bibinfo {note} {[Erratum: Astrophys.J. 946, 57 (2023)]},\ \Eprint {https://arxiv.org/abs/2211.01072} {arXiv:2211.01072 [astro-ph.SR]} \BibitemShut {NoStop}%
\bibitem [{\citenamefont {Parfrey}\ and\ \citenamefont {Tchekhovskoy}(2024)}]{Parfrey:2023swe}%
  \BibitemOpen
  \bibfield  {author} {\bibinfo {author} {\bibfnamefont {K.}~\bibnamefont {Parfrey}}\ and\ \bibinfo {author} {\bibfnamefont {A.}~\bibnamefont {Tchekhovskoy}},\ }\bibfield  {title} {\bibinfo {title} {{Accreting Neutron Stars in 3D General-relativistic Magnetohydrodynamic Simulations: Jets, Magnetic Polarity, and the Interchange Slingshot}},\ }\href {https://doi.org/10.3847/1538-4357/ad737b} {\bibfield  {journal} {\bibinfo  {journal} {Astrophys. J.}\ }\textbf {\bibinfo {volume} {975}},\ \bibinfo {pages} {57} (\bibinfo {year} {2024})},\ \Eprint {https://arxiv.org/abs/2311.04291} {arXiv:2311.04291 [astro-ph.HE]} \BibitemShut {NoStop}%
\bibitem [{\citenamefont {{GRAVITY Collaboration}}\ \emph {et~al.}(2018)\citenamefont {{GRAVITY Collaboration}} \emph {et~al.}}]{Gravity2018}%
  \BibitemOpen
  \bibfield  {author} {\bibinfo {author} {\bibnamefont {{GRAVITY Collaboration}}} \emph {et~al.},\ }\bibfield  {title} {\bibinfo {title} {{Detection of orbital motions near the last stable circular orbit of the massive black hole SgrA*}},\ }\href {https://doi.org/10.1051/0004-6361/201834294} {\bibfield  {journal} {\bibinfo  {journal} {A\&A}\ }\textbf {\bibinfo {volume} {618}},\ \bibinfo {eid} {L10} (\bibinfo {year} {2018})},\ \Eprint {https://arxiv.org/abs/1810.12641} {arXiv:1810.12641 [astro-ph.GA]} \BibitemShut {NoStop}%
\bibitem [{\citenamefont {Abuter}\ \emph {et~al.}(2021)\citenamefont {Abuter} \emph {et~al.}}]{Gravity2021}%
  \BibitemOpen
  \bibfield  {author} {\bibinfo {author} {\bibfnamefont {R.}~\bibnamefont {Abuter}} \emph {et~al.} (\bibinfo {collaboration} {GRAVITY}),\ }\bibfield  {title} {\bibinfo {title} {{Constraining particle acceleration in Sgr A{\ensuremath{\star}} with simultaneous GRAVITY, Spitzer, NuSTAR, and Chandra observations}},\ }\href {https://doi.org/10.1051/0004-6361/202140981} {\bibfield  {journal} {\bibinfo  {journal} {Astron. Astrophys.}\ }\textbf {\bibinfo {volume} {654}},\ \bibinfo {pages} {A22} (\bibinfo {year} {2021})},\ \Eprint {https://arxiv.org/abs/2107.01096} {arXiv:2107.01096 [astro-ph.HE]} \BibitemShut {NoStop}%
\bibitem [{\citenamefont {Ripperda}\ \emph {et~al.}(2020)\citenamefont {Ripperda}, \citenamefont {Bacchini},\ and\ \citenamefont {Philippov}}]{Ripperda:2020bpz}%
  \BibitemOpen
  \bibfield  {author} {\bibinfo {author} {\bibfnamefont {B.}~\bibnamefont {Ripperda}}, \bibinfo {author} {\bibfnamefont {F.}~\bibnamefont {Bacchini}},\ and\ \bibinfo {author} {\bibfnamefont {A.}~\bibnamefont {Philippov}},\ }\bibfield  {title} {\bibinfo {title} {{Magnetic Reconnection and Hot Spot Formation in Black Hole Accretion Disks}},\ }\href {https://doi.org/10.3847/1538-4357/ababab} {\bibfield  {journal} {\bibinfo  {journal} {Astrophys. J.}\ }\textbf {\bibinfo {volume} {900}},\ \bibinfo {pages} {100} (\bibinfo {year} {2020})},\ \Eprint {https://arxiv.org/abs/2003.04330} {arXiv:2003.04330 [astro-ph.HE]} \BibitemShut {NoStop}%
\bibitem [{\citenamefont {Tchekhovskoy}\ \emph {et~al.}(2010)\citenamefont {Tchekhovskoy}, \citenamefont {Narayan},\ and\ \citenamefont {McKinney}}]{Tchekhovskoy:2009ba}%
  \BibitemOpen
  \bibfield  {author} {\bibinfo {author} {\bibfnamefont {A.}~\bibnamefont {Tchekhovskoy}}, \bibinfo {author} {\bibfnamefont {R.}~\bibnamefont {Narayan}},\ and\ \bibinfo {author} {\bibfnamefont {J.~C.}\ \bibnamefont {McKinney}},\ }\bibfield  {title} {\bibinfo {title} {{Black Hole Spin and the Radio Loud/Quiet Dichotomy of Active Galactic Nuclei}},\ }\href {https://doi.org/10.1088/0004-637X/711/1/50} {\bibfield  {journal} {\bibinfo  {journal} {Astrophys. J.}\ }\textbf {\bibinfo {volume} {711}},\ \bibinfo {pages} {50} (\bibinfo {year} {2010})},\ \Eprint {https://arxiv.org/abs/0911.2228} {arXiv:0911.2228 [astro-ph.HE]} \BibitemShut {NoStop}%
\bibitem [{\citenamefont {Manikantan}\ \emph {et~al.}(2024)\citenamefont {Manikantan}, \citenamefont {Kaaz}, \citenamefont {Jacquemin-Ide}, \citenamefont {Musoke}, \citenamefont {Chatterjee}, \citenamefont {Liska},\ and\ \citenamefont {Tchekhovskoy}}]{Manikantan:2023vcw}%
  \BibitemOpen
  \bibfield  {author} {\bibinfo {author} {\bibfnamefont {V.}~\bibnamefont {Manikantan}}, \bibinfo {author} {\bibfnamefont {N.}~\bibnamefont {Kaaz}}, \bibinfo {author} {\bibfnamefont {J.}~\bibnamefont {Jacquemin-Ide}}, \bibinfo {author} {\bibfnamefont {G.}~\bibnamefont {Musoke}}, \bibinfo {author} {\bibfnamefont {K.}~\bibnamefont {Chatterjee}}, \bibinfo {author} {\bibfnamefont {M.}~\bibnamefont {Liska}},\ and\ \bibinfo {author} {\bibfnamefont {A.}~\bibnamefont {Tchekhovskoy}},\ }\bibfield  {title} {\bibinfo {title} {{Winds and Disk Turbulence Exert Equal Torques on Thick Magnetically Arrested Disks}},\ }\href {https://doi.org/10.3847/1538-4357/ad323d} {\bibfield  {journal} {\bibinfo  {journal} {Astrophys. J.}\ }\textbf {\bibinfo {volume} {965}},\ \bibinfo {pages} {175} (\bibinfo {year} {2024})},\ \Eprint {https://arxiv.org/abs/2310.11490} {arXiv:2310.11490 [astro-ph.HE]} \BibitemShut {NoStop}%
\bibitem [{\citenamefont {Blandford}\ and\ \citenamefont {Payne}(1982)}]{Blandford:1982xxl}%
  \BibitemOpen
  \bibfield  {author} {\bibinfo {author} {\bibfnamefont {R.~D.}\ \bibnamefont {Blandford}}\ and\ \bibinfo {author} {\bibfnamefont {D.~G.}\ \bibnamefont {Payne}},\ }\bibfield  {title} {\bibinfo {title} {{Hydromagnetic flows from accretion discs and the production of radio jets}},\ }\href {https://doi.org/10.1093/mnras/199.4.883} {\bibfield  {journal} {\bibinfo  {journal} {Mon. Not. Roy. Astron. Soc.}\ }\textbf {\bibinfo {volume} {199}},\ \bibinfo {pages} {883} (\bibinfo {year} {1982})}\BibitemShut {NoStop}%
\bibitem [{\citenamefont {Duffell}\ \emph {et~al.}(2020)\citenamefont {Duffell}, \citenamefont {D'Orazio}, \citenamefont {Derdzinski}, \citenamefont {Haiman}, \citenamefont {MacFadyen}, \citenamefont {Rosen},\ and\ \citenamefont {Zrake}}]{Duffell:2019uuk}%
  \BibitemOpen
  \bibfield  {author} {\bibinfo {author} {\bibfnamefont {P.~C.}\ \bibnamefont {Duffell}}, \bibinfo {author} {\bibfnamefont {D.}~\bibnamefont {D'Orazio}}, \bibinfo {author} {\bibfnamefont {A.}~\bibnamefont {Derdzinski}}, \bibinfo {author} {\bibfnamefont {Z.}~\bibnamefont {Haiman}}, \bibinfo {author} {\bibfnamefont {A.}~\bibnamefont {MacFadyen}}, \bibinfo {author} {\bibfnamefont {A.~L.}\ \bibnamefont {Rosen}},\ and\ \bibinfo {author} {\bibfnamefont {J.}~\bibnamefont {Zrake}},\ }\bibfield  {title} {\bibinfo {title} {{Circumbinary Disks: Accretion and Torque as a Function of Mass Ratio and Disk Viscosity}},\ }\href {https://doi.org/10.3847/1538-4357/abab95} {\bibfield  {journal} {\bibinfo  {journal} {Astrophys. J.}\ }\textbf {\bibinfo {volume} {901}},\ \bibinfo {pages} {25} (\bibinfo {year} {2020})},\ \Eprint {https://arxiv.org/abs/1911.05506} {arXiv:1911.05506 [astro-ph.SR]} \BibitemShut {NoStop}%
\bibitem [{\citenamefont {Dittmann}\ and\ \citenamefont {Ryan}(2022)}]{Dittmann:2022obl}%
  \BibitemOpen
  \bibfield  {author} {\bibinfo {author} {\bibfnamefont {A.~J.}\ \bibnamefont {Dittmann}}\ and\ \bibinfo {author} {\bibfnamefont {G.}~\bibnamefont {Ryan}},\ }\bibfield  {title} {\bibinfo {title} {{A survey of disc thickness and viscosity in circumbinary accretion: Binary evolution, variability, and disc morphology}},\ }\href {https://doi.org/10.1093/mnras/stac935} {\bibfield  {journal} {\bibinfo  {journal} {Mon. Not. Roy. Astron. Soc.}\ }\textbf {\bibinfo {volume} {513}},\ \bibinfo {pages} {6158} (\bibinfo {year} {2022})},\ \Eprint {https://arxiv.org/abs/2201.07816} {arXiv:2201.07816 [astro-ph.HE]} \BibitemShut {NoStop}%
\bibitem [{\citenamefont {Dittmann}\ and\ \citenamefont {Ryan}(2024)}]{Dittmann:2023ztg}%
  \BibitemOpen
  \bibfield  {author} {\bibinfo {author} {\bibfnamefont {A.~J.}\ \bibnamefont {Dittmann}}\ and\ \bibinfo {author} {\bibfnamefont {G.}~\bibnamefont {Ryan}},\ }\bibfield  {title} {\bibinfo {title} {{The Evolution of Accreting Binaries: From Brown Dwarfs to Supermassive Black Holes}},\ }\href {https://doi.org/10.3847/1538-4357/ad2f1e} {\bibfield  {journal} {\bibinfo  {journal} {Astrophys. J.}\ }\textbf {\bibinfo {volume} {967}},\ \bibinfo {pages} {12} (\bibinfo {year} {2024})},\ \Eprint {https://arxiv.org/abs/2310.07758} {arXiv:2310.07758 [astro-ph.GA]} \BibitemShut {NoStop}%
\bibitem [{\citenamefont {{Lynden-Bell}}(1996)}]{Lynden-Bell:1996MNRAS.279..389L}%
  \BibitemOpen
  \bibfield  {author} {\bibinfo {author} {\bibfnamefont {D.}~\bibnamefont {{Lynden-Bell}}},\ }\bibfield  {title} {\bibinfo {title} {{Magnetic collimation by accretion discs of quasars and stars}},\ }\href {https://doi.org/10.1093/mnras/279.2.389} {\bibfield  {journal} {\bibinfo  {journal} {\mnras}\ }\textbf {\bibinfo {volume} {279}},\ \bibinfo {pages} {389} (\bibinfo {year} {1996})}\BibitemShut {NoStop}%
\bibitem [{\citenamefont {{Lynden-Bell}}(2003)}]{Lynden-Bell:2003MNRAS.341.1360L}%
  \BibitemOpen
  \bibfield  {author} {\bibinfo {author} {\bibfnamefont {D.}~\bibnamefont {{Lynden-Bell}}},\ }\bibfield  {title} {\bibinfo {title} {{On why discs generate magnetic towers and collimate jets}},\ }\href {https://doi.org/10.1046/j.1365-8711.2003.06506.x} {\bibfield  {journal} {\bibinfo  {journal} {\mnras}\ }\textbf {\bibinfo {volume} {341}},\ \bibinfo {pages} {1360} (\bibinfo {year} {2003})},\ \Eprint {https://arxiv.org/abs/astro-ph/0208388} {arXiv:astro-ph/0208388 [astro-ph]} \BibitemShut {NoStop}%
\bibitem [{\citenamefont {Blandford}\ and\ \citenamefont {Znajek}(1977)}]{Blandford:1977ds}%
  \BibitemOpen
  \bibfield  {author} {\bibinfo {author} {\bibfnamefont {R.~D.}\ \bibnamefont {Blandford}}\ and\ \bibinfo {author} {\bibfnamefont {R.~L.}\ \bibnamefont {Znajek}},\ }\bibfield  {title} {\bibinfo {title} {{Electromagnetic extractions of energy from Kerr black holes}},\ }\href {https://doi.org/10.1093/mnras/179.3.433} {\bibfield  {journal} {\bibinfo  {journal} {Mon. Not. Roy. Astron. Soc.}\ }\textbf {\bibinfo {volume} {179}},\ \bibinfo {pages} {433} (\bibinfo {year} {1977})}\BibitemShut {NoStop}%
\bibitem [{\citenamefont {{Heyvaerts}}\ and\ \citenamefont {{Norman}}(1989)}]{Heyvaerts:1989ApJ...347.1055H}%
  \BibitemOpen
  \bibfield  {author} {\bibinfo {author} {\bibfnamefont {J.}~\bibnamefont {{Heyvaerts}}}\ and\ \bibinfo {author} {\bibfnamefont {C.}~\bibnamefont {{Norman}}},\ }\bibfield  {title} {\bibinfo {title} {{The Collimation of Magnetized Winds}},\ }\href {https://doi.org/10.1086/168195} {\bibfield  {journal} {\bibinfo  {journal} {\apj}\ }\textbf {\bibinfo {volume} {347}},\ \bibinfo {pages} {1055} (\bibinfo {year} {1989})}\BibitemShut {NoStop}%
\bibitem [{\citenamefont {{Bromberg}}\ and\ \citenamefont {{Tchekhovskoy}}(2016)}]{Bromberg:2016MNRAS.456.1739B}%
  \BibitemOpen
  \bibfield  {author} {\bibinfo {author} {\bibfnamefont {O.}~\bibnamefont {{Bromberg}}}\ and\ \bibinfo {author} {\bibfnamefont {A.}~\bibnamefont {{Tchekhovskoy}}},\ }\bibfield  {title} {\bibinfo {title} {{Relativistic MHD simulations of core-collapse GRB jets: 3D instabilities and magnetic dissipation}},\ }\href {https://doi.org/10.1093/mnras/stv2591} {\bibfield  {journal} {\bibinfo  {journal} {\mnras}\ }\textbf {\bibinfo {volume} {456}},\ \bibinfo {pages} {1739} (\bibinfo {year} {2016})},\ \Eprint {https://arxiv.org/abs/1508.02721} {arXiv:1508.02721 [astro-ph.HE]} \BibitemShut {NoStop}%
\bibitem [{\citenamefont {{Blandford}}\ and\ \citenamefont {{Globus}}(2022{\natexlab{a}})}]{Blandford:2022Galax..10...89B}%
  \BibitemOpen
  \bibfield  {author} {\bibinfo {author} {\bibfnamefont {R.}~\bibnamefont {{Blandford}}}\ and\ \bibinfo {author} {\bibfnamefont {N.}~\bibnamefont {{Globus}}},\ }\bibfield  {title} {\bibinfo {title} {{Jets, Disks and Winds from Spinning Black Holes: Nature or Nurture?}},\ }\href {https://doi.org/10.3390/galaxies10040089} {\bibfield  {journal} {\bibinfo  {journal} {Galaxies}\ }\textbf {\bibinfo {volume} {10}},\ \bibinfo {eid} {89} (\bibinfo {year} {2022}{\natexlab{a}})},\ \Eprint {https://arxiv.org/abs/2207.05839} {arXiv:2207.05839 [astro-ph.HE]} \BibitemShut {NoStop}%
\bibitem [{\citenamefont {{Blandford}}\ and\ \citenamefont {{Globus}}(2022{\natexlab{b}})}]{Blandford:2022MNRAS.514.5141B}%
  \BibitemOpen
  \bibfield  {author} {\bibinfo {author} {\bibfnamefont {R.}~\bibnamefont {{Blandford}}}\ and\ \bibinfo {author} {\bibfnamefont {N.}~\bibnamefont {{Globus}}},\ }\bibfield  {title} {\bibinfo {title} {{Ergomagnetosphere, ejection disc, magnetopause in M87 - I. Global flow of mass, angular momentum, energy, and current}},\ }\href {https://doi.org/10.1093/mnras/stac1682} {\bibfield  {journal} {\bibinfo  {journal} {Monthly Notices of the Royal Astronomical Society}\ }\textbf {\bibinfo {volume} {514}},\ \bibinfo {pages} {5141} (\bibinfo {year} {2022}{\natexlab{b}})},\ \Eprint {https://arxiv.org/abs/2204.11995} {arXiv:2204.11995 [astro-ph.HE]} \BibitemShut {NoStop}%
\bibitem [{\citenamefont {Stone}\ \emph {et~al.}(2008)\citenamefont {Stone}, \citenamefont {Gardiner}, \citenamefont {Teuben}, \citenamefont {Hawley},\ and\ \citenamefont {Simon}}]{Stone:2008mh}%
  \BibitemOpen
  \bibfield  {author} {\bibinfo {author} {\bibfnamefont {J.~M.}\ \bibnamefont {Stone}}, \bibinfo {author} {\bibfnamefont {T.~A.}\ \bibnamefont {Gardiner}}, \bibinfo {author} {\bibfnamefont {P.}~\bibnamefont {Teuben}}, \bibinfo {author} {\bibfnamefont {J.~F.}\ \bibnamefont {Hawley}},\ and\ \bibinfo {author} {\bibfnamefont {J.~B.}\ \bibnamefont {Simon}},\ }\bibfield  {title} {\bibinfo {title} {{Athena: A New Code for Astrophysical MHD}},\ }\href {https://doi.org/10.1086/588755} {\bibfield  {journal} {\bibinfo  {journal} {Astrophys. J. Suppl.}\ }\textbf {\bibinfo {volume} {178}},\ \bibinfo {pages} {137} (\bibinfo {year} {2008})},\ \Eprint {https://arxiv.org/abs/0804.0402} {arXiv:0804.0402 [astro-ph]} \BibitemShut {NoStop}%
\bibitem [{\citenamefont {{White}}\ \emph {et~al.}(2023)\citenamefont {{White}}, \citenamefont {{Mullen}}, \citenamefont {{Jiang}}, \citenamefont {{Davis}}, \citenamefont {{Stone}}, \citenamefont {{Morozova}},\ and\ \citenamefont {{Zhang}}}]{2023ApJ...949..103W}%
  \BibitemOpen
  \bibfield  {author} {\bibinfo {author} {\bibfnamefont {C.~J.}\ \bibnamefont {{White}}}, \bibinfo {author} {\bibfnamefont {P.~D.}\ \bibnamefont {{Mullen}}}, \bibinfo {author} {\bibfnamefont {Y.-F.}\ \bibnamefont {{Jiang}}}, \bibinfo {author} {\bibfnamefont {S.~W.}\ \bibnamefont {{Davis}}}, \bibinfo {author} {\bibfnamefont {J.~M.}\ \bibnamefont {{Stone}}}, \bibinfo {author} {\bibfnamefont {V.}~\bibnamefont {{Morozova}}},\ and\ \bibinfo {author} {\bibfnamefont {L.}~\bibnamefont {{Zhang}}},\ }\bibfield  {title} {\bibinfo {title} {{An Extension of the Athena++ Code Framework for Radiation-magnetohydrodynamics in General Relativity Using a Finite-solid-angle Discretization}},\ }\href {https://doi.org/10.3847/1538-4357/acc8cf} {\bibfield  {journal} {\bibinfo  {journal} {Astrophys. J.}\ }\textbf {\bibinfo {volume} {949}},\ \bibinfo {eid} {103} (\bibinfo {year} {2023})},\ \Eprint {https://arxiv.org/abs/2302.04283} {arXiv:2302.04283 [astro-ph.HE]} \BibitemShut {NoStop}%
\bibitem [{\citenamefont {Zhang}\ \emph {et~al.}(2025)\citenamefont {Zhang}, \citenamefont {Stone}, \citenamefont {Mullen}, \citenamefont {Davis}, \citenamefont {Jiang},\ and\ \citenamefont {White}}]{Zhang:2025uug}%
  \BibitemOpen
  \bibfield  {author} {\bibinfo {author} {\bibfnamefont {L.}~\bibnamefont {Zhang}}, \bibinfo {author} {\bibfnamefont {J.~M.}\ \bibnamefont {Stone}}, \bibinfo {author} {\bibfnamefont {P.~D.}\ \bibnamefont {Mullen}}, \bibinfo {author} {\bibfnamefont {S.~W.}\ \bibnamefont {Davis}}, \bibinfo {author} {\bibfnamefont {Y.-F.}\ \bibnamefont {Jiang}},\ and\ \bibinfo {author} {\bibfnamefont {C.~J.}\ \bibnamefont {White}},\ }\bibfield  {title} {\bibinfo {title} {{Radiation GRMHD Models of Accretion onto Stellar-Mass Black Holes: I. Survey of Eddington Ratios}},\ }\href@noop {} {\bibfield  {journal} {\bibinfo  {journal} {arxiv}\ } (\bibinfo {year} {2025})},\ \Eprint {https://arxiv.org/abs/2506.02289} {arXiv:2506.02289 [astro-ph.HE]} \BibitemShut {NoStop}%
\bibitem [{\citenamefont {{Mu{\~n}oz}}\ and\ \citenamefont {{Lai}}(2016)}]{Munoz:2016ApJ...827...43M}%
  \BibitemOpen
  \bibfield  {author} {\bibinfo {author} {\bibfnamefont {D.~J.}\ \bibnamefont {{Mu{\~n}oz}}}\ and\ \bibinfo {author} {\bibfnamefont {D.}~\bibnamefont {{Lai}}},\ }\bibfield  {title} {\bibinfo {title} {{Pulsed Accretion onto Eccentric and Circular Binaries}},\ }\href {https://doi.org/10.3847/0004-637X/827/1/43} {\bibfield  {journal} {\bibinfo  {journal} {\apj}\ }\textbf {\bibinfo {volume} {827}},\ \bibinfo {eid} {43} (\bibinfo {year} {2016})},\ \Eprint {https://arxiv.org/abs/1604.00004} {arXiv:1604.00004 [astro-ph.EP]} \BibitemShut {NoStop}%
\bibitem [{\citenamefont {{Duffell}}\ \emph {et~al.}(2024)\citenamefont {{Duffell}}, \citenamefont {{Dittmann}}, \citenamefont {{D'Orazio}}, \citenamefont {{Franchini}}, \citenamefont {{Kratter}}, \citenamefont {{Penzlin}}, \citenamefont {{Ragusa}}, \citenamefont {{Siwek}}, \citenamefont {{Tiede}}, \citenamefont {{Wang}}, \citenamefont {{Zrake}}, \citenamefont {{Dempsey}}, \citenamefont {{Haiman}}, \citenamefont {{Lupi}}, \citenamefont {{Pirog}},\ and\ \citenamefont {{Ryan}}}]{Duffell:2024fwy}%
  \BibitemOpen
  \bibfield  {author} {\bibinfo {author} {\bibfnamefont {P.~C.}\ \bibnamefont {{Duffell}}}, \bibinfo {author} {\bibfnamefont {A.~J.}\ \bibnamefont {{Dittmann}}}, \bibinfo {author} {\bibfnamefont {D.~J.}\ \bibnamefont {{D'Orazio}}}, \bibinfo {author} {\bibfnamefont {A.}~\bibnamefont {{Franchini}}}, \bibinfo {author} {\bibfnamefont {K.~M.}\ \bibnamefont {{Kratter}}}, \bibinfo {author} {\bibfnamefont {A.~B.~T.}\ \bibnamefont {{Penzlin}}}, \bibinfo {author} {\bibfnamefont {E.}~\bibnamefont {{Ragusa}}}, \bibinfo {author} {\bibfnamefont {M.}~\bibnamefont {{Siwek}}}, \bibinfo {author} {\bibfnamefont {C.}~\bibnamefont {{Tiede}}}, \bibinfo {author} {\bibfnamefont {H.}~\bibnamefont {{Wang}}}, \bibinfo {author} {\bibfnamefont {J.}~\bibnamefont {{Zrake}}}, \bibinfo {author} {\bibfnamefont {A.~M.}\ \bibnamefont {{Dempsey}}}, \bibinfo {author} {\bibfnamefont {Z.}~\bibnamefont {{Haiman}}}, \bibinfo {author} {\bibfnamefont {A.}~\bibnamefont {{Lupi}}}, \bibinfo {author} {\bibfnamefont {M.}~\bibnamefont {{Pirog}}},\ and\ \bibinfo {author} {\bibfnamefont {G.}~\bibnamefont {{Ryan}}},\ }\bibfield  {title} {\bibinfo {title} {{The Santa Barbara Binary‑disk Code Comparison}},\ }\href {https://doi.org/10.3847/1538-4357/ad5a7e} {\bibfield  {journal} {\bibinfo  {journal} {\apj}\ }\textbf {\bibinfo {volume} {970}},\ \bibinfo {eid} {156} (\bibinfo {year} {2024})},\ \Eprint {https://arxiv.org/abs/2402.13039} {arXiv:2402.13039 [astro-ph.SR]} \BibitemShut {NoStop}%
\bibitem [{\citenamefont {{Tchekhovskoy}}\ and\ \citenamefont {{McKinney}}(2012)}]{Tchekhovskoy:2012MNRAS.423L..55T}%
  \BibitemOpen
  \bibfield  {author} {\bibinfo {author} {\bibfnamefont {A.}~\bibnamefont {{Tchekhovskoy}}}\ and\ \bibinfo {author} {\bibfnamefont {J.~C.}\ \bibnamefont {{McKinney}}},\ }\bibfield  {title} {\bibinfo {title} {{Prograde and retrograde black holes: whose jet is more powerful?}},\ }\href {https://doi.org/10.1111/j.1745-3933.2012.01256.x} {\bibfield  {journal} {\bibinfo  {journal} {\mnras}\ }\textbf {\bibinfo {volume} {423}},\ \bibinfo {pages} {L55} (\bibinfo {year} {2012})},\ \Eprint {https://arxiv.org/abs/1201.4385} {arXiv:1201.4385 [astro-ph.HE]} \BibitemShut {NoStop}%
\bibitem [{\citenamefont {Liska}\ \emph {et~al.}(2020)\citenamefont {Liska}, \citenamefont {Tchekhovskoy},\ and\ \citenamefont {Quataert}}]{Liska:2018btr}%
  \BibitemOpen
  \bibfield  {author} {\bibinfo {author} {\bibfnamefont {M.~T.~P.}\ \bibnamefont {Liska}}, \bibinfo {author} {\bibfnamefont {A.}~\bibnamefont {Tchekhovskoy}},\ and\ \bibinfo {author} {\bibfnamefont {E.}~\bibnamefont {Quataert}},\ }\bibfield  {title} {\bibinfo {title} {{Large-Scale Poloidal Magnetic Field Dynamo Leads to Powerful Jets in GRMHD Simulations of Black Hole Accretion with Toroidal Field}},\ }\href {https://doi.org/10.1093/mnras/staa955} {\bibfield  {journal} {\bibinfo  {journal} {Mon. Not. Roy. Astron. Soc.}\ }\textbf {\bibinfo {volume} {494}},\ \bibinfo {pages} {3656} (\bibinfo {year} {2020})},\ \Eprint {https://arxiv.org/abs/1809.04608} {arXiv:1809.04608 [astro-ph.HE]} \BibitemShut {NoStop}%
\bibitem [{\citenamefont {Jacquemin-Ide}\ \emph {et~al.}(2024)\citenamefont {Jacquemin-Ide}, \citenamefont {Rincon}, \citenamefont {Tchekhovskoy},\ and\ \citenamefont {Liska}}]{Jacquemin-Ide:2023qrj}%
  \BibitemOpen
  \bibfield  {author} {\bibinfo {author} {\bibfnamefont {J.}~\bibnamefont {Jacquemin-Ide}}, \bibinfo {author} {\bibfnamefont {F.}~\bibnamefont {Rincon}}, \bibinfo {author} {\bibfnamefont {A.}~\bibnamefont {Tchekhovskoy}},\ and\ \bibinfo {author} {\bibfnamefont {M.}~\bibnamefont {Liska}},\ }\bibfield  {title} {\bibinfo {title} {{Magnetorotational dynamo can generate large-scale vertical magnetic fields in 3D GRMHD simulations of accreting black holes}},\ }\href {https://doi.org/10.1093/mnras/stae1538} {\bibfield  {journal} {\bibinfo  {journal} {Mon. Not. Roy. Astron. Soc.}\ }\textbf {\bibinfo {volume} {532}},\ \bibinfo {pages} {1522} (\bibinfo {year} {2024})},\ \Eprint {https://arxiv.org/abs/2311.00034} {arXiv:2311.00034 [astro-ph.HE]} \BibitemShut {NoStop}%
\bibitem [{\citenamefont {{Dempsey}}\ \emph {et~al.}(2020)\citenamefont {{Dempsey}}, \citenamefont {{Mu{\~n}oz}},\ and\ \citenamefont {{Lithwick}}}]{2020ApJ...892L..29D}%
  \BibitemOpen
  \bibfield  {author} {\bibinfo {author} {\bibfnamefont {A.~M.}\ \bibnamefont {{Dempsey}}}, \bibinfo {author} {\bibfnamefont {D.}~\bibnamefont {{Mu{\~n}oz}}},\ and\ \bibinfo {author} {\bibfnamefont {Y.}~\bibnamefont {{Lithwick}}},\ }\bibfield  {title} {\bibinfo {title} {{Inner Boundary Condition in Quasi-Lagrangian Simulations of Accretion Disks}},\ }\href {https://doi.org/10.3847/2041-8213/ab800e} {\bibfield  {journal} {\bibinfo  {journal} {Astrophys. J. Lett.}\ }\textbf {\bibinfo {volume} {892}},\ \bibinfo {eid} {L29} (\bibinfo {year} {2020})},\ \Eprint {https://arxiv.org/abs/2002.05164} {arXiv:2002.05164 [astro-ph.EP]} \BibitemShut {NoStop}%
\bibitem [{\citenamefont {Dittmann}\ and\ \citenamefont {Ryan}(2021)}]{Dittmann:2021wzj}%
  \BibitemOpen
  \bibfield  {author} {\bibinfo {author} {\bibfnamefont {A.~J.}\ \bibnamefont {Dittmann}}\ and\ \bibinfo {author} {\bibfnamefont {G.}~\bibnamefont {Ryan}},\ }\bibfield  {title} {\bibinfo {title} {{Preventing Anomalous Torques in Circumbinary Accretion Simulations}},\ }\href {https://doi.org/10.3847/1538-4357/ac1bbd} {\bibfield  {journal} {\bibinfo  {journal} {Astrophys. J.}\ }\textbf {\bibinfo {volume} {921}},\ \bibinfo {pages} {71} (\bibinfo {year} {2021})},\ \Eprint {https://arxiv.org/abs/2102.05684} {arXiv:2102.05684 [astro-ph.HE]} \BibitemShut {NoStop}%
\bibitem [{\citenamefont {{Stone}}\ \emph {et~al.}(2024)\citenamefont {{Stone}}, \citenamefont {{Mullen}}, \citenamefont {{Fielding}}, \citenamefont {{Grete}}, \citenamefont {{Guo}}, \citenamefont {{Kempski}}, \citenamefont {{Most}}, \citenamefont {{White}},\ and\ \citenamefont {{Wong}}}]{athenak}%
  \BibitemOpen
  \bibfield  {author} {\bibinfo {author} {\bibfnamefont {J.~M.}\ \bibnamefont {{Stone}}}, \bibinfo {author} {\bibfnamefont {P.~D.}\ \bibnamefont {{Mullen}}}, \bibinfo {author} {\bibfnamefont {D.}~\bibnamefont {{Fielding}}}, \bibinfo {author} {\bibfnamefont {P.}~\bibnamefont {{Grete}}}, \bibinfo {author} {\bibfnamefont {M.}~\bibnamefont {{Guo}}}, \bibinfo {author} {\bibfnamefont {P.}~\bibnamefont {{Kempski}}}, \bibinfo {author} {\bibfnamefont {E.~R.}\ \bibnamefont {{Most}}}, \bibinfo {author} {\bibfnamefont {C.~J.}\ \bibnamefont {{White}}},\ and\ \bibinfo {author} {\bibfnamefont {G.~N.}\ \bibnamefont {{Wong}}},\ }\bibfield  {title} {\bibinfo {title} {{AthenaK: A Performance-Portable Version of the Athena++ AMR Framework}},\ }\href {https://doi.org/10.48550/arXiv.2409.16053} {\bibfield  {journal} {\bibinfo  {journal} {arXiv e-prints}\ ,\ \bibinfo {eid} {arXiv:2409.16053}} (\bibinfo {year} {2024})},\ \Eprint {https://arxiv.org/abs/2409.16053} {arXiv:2409.16053 [astro-ph.IM]} \BibitemShut {NoStop}%
\bibitem [{\citenamefont {{Stone}}\ \emph {et~al.}(2020)\citenamefont {{Stone}}, \citenamefont {{Tomida}}, \citenamefont {{White}},\ and\ \citenamefont {{Felker}}}]{Stone2020}%
  \BibitemOpen
  \bibfield  {author} {\bibinfo {author} {\bibfnamefont {J.~M.}\ \bibnamefont {{Stone}}}, \bibinfo {author} {\bibfnamefont {K.}~\bibnamefont {{Tomida}}}, \bibinfo {author} {\bibfnamefont {C.~J.}\ \bibnamefont {{White}}},\ and\ \bibinfo {author} {\bibfnamefont {K.~G.}\ \bibnamefont {{Felker}}},\ }\bibfield  {title} {\bibinfo {title} {{The Athena++ Adaptive Mesh Refinement Framework: Design and Magnetohydrodynamic Solvers}},\ }\href {https://doi.org/10.3847/1538-4365/ab929b} {\bibfield  {journal} {\bibinfo  {journal} {Astrophys. J.s}\ }\textbf {\bibinfo {volume} {249}},\ \bibinfo {eid} {4} (\bibinfo {year} {2020})},\ \Eprint {https://arxiv.org/abs/2005.06651} {arXiv:2005.06651 [astro-ph.IM]} \BibitemShut {NoStop}%
\bibitem [{\citenamefont {{Trott}}\ \emph {et~al.}(2021)\citenamefont {{Trott}}, \citenamefont {{Berger-Vergiat}}, \citenamefont {{Poliakoff}}, \citenamefont {{Rajamanickam}}, \citenamefont {{Lebrun-Grandie}}, \citenamefont {{Madsen}}, \citenamefont {{Al Awar}}, \citenamefont {{Gligoric}}, \citenamefont {{Shipman}},\ and\ \citenamefont {{Womeldorff}}}]{Trott2021}%
  \BibitemOpen
  \bibfield  {author} {\bibinfo {author} {\bibfnamefont {C.}~\bibnamefont {{Trott}}}, \bibinfo {author} {\bibfnamefont {L.}~\bibnamefont {{Berger-Vergiat}}}, \bibinfo {author} {\bibfnamefont {D.}~\bibnamefont {{Poliakoff}}}, \bibinfo {author} {\bibfnamefont {S.}~\bibnamefont {{Rajamanickam}}}, \bibinfo {author} {\bibfnamefont {D.}~\bibnamefont {{Lebrun-Grandie}}}, \bibinfo {author} {\bibfnamefont {J.}~\bibnamefont {{Madsen}}}, \bibinfo {author} {\bibfnamefont {N.}~\bibnamefont {{Al Awar}}}, \bibinfo {author} {\bibfnamefont {M.}~\bibnamefont {{Gligoric}}}, \bibinfo {author} {\bibfnamefont {G.}~\bibnamefont {{Shipman}}},\ and\ \bibinfo {author} {\bibfnamefont {G.}~\bibnamefont {{Womeldorff}}},\ }\bibfield  {title} {\bibinfo {title} {{The Kokkos EcoSystem: Comprehensive Performance Portability for High Performance Computing}},\ }\href {https://doi.org/10.1109/MCSE.2021.3098509} {\bibfield  {journal} {\bibinfo  {journal} {Computing in Science and Engineering}\ }\textbf {\bibinfo {volume} {23}},\ \bibinfo {pages} {10} (\bibinfo {year} {2021})}\BibitemShut {NoStop}%
\bibitem [{\citenamefont {{Colella}}\ and\ \citenamefont {{Woodward}}(1984)}]{Colella1984}%
  \BibitemOpen
  \bibfield  {author} {\bibinfo {author} {\bibfnamefont {P.}~\bibnamefont {{Colella}}}\ and\ \bibinfo {author} {\bibfnamefont {P.~R.}\ \bibnamefont {{Woodward}}},\ }\bibfield  {title} {\bibinfo {title} {{The Piecewise Parabolic Method (PPM) for Gas-Dynamical Simulations}},\ }\href {https://doi.org/10.1016/0021-9991(84)90143-8} {\bibfield  {journal} {\bibinfo  {journal} {Journal of Computational Physics}\ }\textbf {\bibinfo {volume} {54}},\ \bibinfo {pages} {174} (\bibinfo {year} {1984})}\BibitemShut {NoStop}%
\bibitem [{\citenamefont {{Miyoshi}}\ and\ \citenamefont {{Kusano}}(2005)}]{Miyoshi2005}%
  \BibitemOpen
  \bibfield  {author} {\bibinfo {author} {\bibfnamefont {T.}~\bibnamefont {{Miyoshi}}}\ and\ \bibinfo {author} {\bibfnamefont {K.}~\bibnamefont {{Kusano}}},\ }\bibfield  {title} {\bibinfo {title} {{A multi-state HLL approximate Riemann solver for ideal magnetohydrodynamics}},\ }\href {https://doi.org/10.1016/j.jcp.2005.02.017} {\bibfield  {journal} {\bibinfo  {journal} {Journal of Computational Physics}\ }\textbf {\bibinfo {volume} {208}},\ \bibinfo {pages} {315} (\bibinfo {year} {2005})}\BibitemShut {NoStop}%
\bibitem [{\citenamefont {{Gardiner}}\ and\ \citenamefont {{Stone}}(2008)}]{Gardiner2008}%
  \BibitemOpen
  \bibfield  {author} {\bibinfo {author} {\bibfnamefont {T.~A.}\ \bibnamefont {{Gardiner}}}\ and\ \bibinfo {author} {\bibfnamefont {J.~M.}\ \bibnamefont {{Stone}}},\ }\bibfield  {title} {\bibinfo {title} {{An unsplit Godunov method for ideal MHD via constrained transport in three dimensions}},\ }\href {https://doi.org/10.1016/j.jcp.2007.12.017} {\bibfield  {journal} {\bibinfo  {journal} {Journal of Computational Physics}\ }\textbf {\bibinfo {volume} {227}},\ \bibinfo {pages} {4123} (\bibinfo {year} {2008})},\ \Eprint {https://arxiv.org/abs/0712.2634} {arXiv:0712.2634 [astro-ph]} \BibitemShut {NoStop}%
\bibitem [{\citenamefont {Gaburov}\ \emph {et~al.}(2012)\citenamefont {Gaburov}, \citenamefont {Johansen},\ and\ \citenamefont {Levin}}]{Gaburov:2012jd}%
  \BibitemOpen
  \bibfield  {author} {\bibinfo {author} {\bibfnamefont {E.}~\bibnamefont {Gaburov}}, \bibinfo {author} {\bibfnamefont {A.}~\bibnamefont {Johansen}},\ and\ \bibinfo {author} {\bibfnamefont {Y.}~\bibnamefont {Levin}},\ }\bibfield  {title} {\bibinfo {title} {{Magnetically-levitating disks around supermassive black holes}},\ }\href {https://doi.org/10.1088/0004-637X/758/2/103} {\bibfield  {journal} {\bibinfo  {journal} {Astrophys. J.}\ }\textbf {\bibinfo {volume} {758}},\ \bibinfo {pages} {103} (\bibinfo {year} {2012})},\ \Eprint {https://arxiv.org/abs/1201.4873} {arXiv:1201.4873 [astro-ph.GA]} \BibitemShut {NoStop}%
\bibitem [{\citenamefont {Hopkins}\ \emph {et~al.}(2023)\citenamefont {Hopkins}, \citenamefont {Squire}, \citenamefont {Quataert}, \citenamefont {Murray}, \citenamefont {Su}, \citenamefont {Steinwandel}, \citenamefont {Kremer}, \citenamefont {Faucher-Giguere},\ and\ \citenamefont {Wellons}}]{Hopkins:2023lgk}%
  \BibitemOpen
  \bibfield  {author} {\bibinfo {author} {\bibfnamefont {P.~F.}\ \bibnamefont {Hopkins}}, \bibinfo {author} {\bibfnamefont {J.}~\bibnamefont {Squire}}, \bibinfo {author} {\bibfnamefont {E.}~\bibnamefont {Quataert}}, \bibinfo {author} {\bibfnamefont {N.}~\bibnamefont {Murray}}, \bibinfo {author} {\bibfnamefont {K.-Y.}\ \bibnamefont {Su}}, \bibinfo {author} {\bibfnamefont {U.~P.}\ \bibnamefont {Steinwandel}}, \bibinfo {author} {\bibfnamefont {K.}~\bibnamefont {Kremer}}, \bibinfo {author} {\bibfnamefont {C.-A.}\ \bibnamefont {Faucher-Giguere}},\ and\ \bibinfo {author} {\bibfnamefont {S.}~\bibnamefont {Wellons}},\ }\bibfield  {title} {\bibinfo {title} {{An Analytic Model For Magnetically-Dominated Accretion Disks}},\ }\href@noop {} {\bibfield  {journal} {\bibinfo  {journal} {arxiv}\ } (\bibinfo {year} {2023})},\ \Eprint {https://arxiv.org/abs/2310.04507} {arXiv:2310.04507 [astro-ph.HE]} \BibitemShut {NoStop}%
\bibitem [{\citenamefont {Guo}\ \emph {et~al.}(2025{\natexlab{b}})\citenamefont {Guo}, \citenamefont {Quataert}, \citenamefont {Squire}, \citenamefont {Hopkins},\ and\ \citenamefont {Stone}}]{Guo:2025glc}%
  \BibitemOpen
  \bibfield  {author} {\bibinfo {author} {\bibfnamefont {M.}~\bibnamefont {Guo}}, \bibinfo {author} {\bibfnamefont {E.}~\bibnamefont {Quataert}}, \bibinfo {author} {\bibfnamefont {J.}~\bibnamefont {Squire}}, \bibinfo {author} {\bibfnamefont {P.~F.}\ \bibnamefont {Hopkins}},\ and\ \bibinfo {author} {\bibfnamefont {J.~M.}\ \bibnamefont {Stone}},\ }\bibfield  {title} {\bibinfo {title} {{Idealized Global Models of Accretion Disks with Strong Toroidal Magnetic Fields}},\ }\href@noop {} {\bibfield  {journal} {\bibinfo  {journal} {arxiv}\ } (\bibinfo {year} {2025}{\natexlab{b}})},\ \Eprint {https://arxiv.org/abs/2505.12671} {arXiv:2505.12671 [astro-ph.HE]} \BibitemShut {NoStop}%
\bibitem [{\citenamefont {Palenzuela}\ \emph {et~al.}(2010)\citenamefont {Palenzuela}, \citenamefont {Lehner},\ and\ \citenamefont {Liebling}}]{Palenzuela:2010nf}%
  \BibitemOpen
  \bibfield  {author} {\bibinfo {author} {\bibfnamefont {C.}~\bibnamefont {Palenzuela}}, \bibinfo {author} {\bibfnamefont {L.}~\bibnamefont {Lehner}},\ and\ \bibinfo {author} {\bibfnamefont {S.~L.}\ \bibnamefont {Liebling}},\ }\bibfield  {title} {\bibinfo {title} {{Dual Jets from Binary Black Holes}},\ }\href {https://doi.org/10.1126/science.1191766} {\bibfield  {journal} {\bibinfo  {journal} {Science}\ }\textbf {\bibinfo {volume} {329}},\ \bibinfo {pages} {927} (\bibinfo {year} {2010})},\ \Eprint {https://arxiv.org/abs/1005.1067} {arXiv:1005.1067 [astro-ph.HE]} \BibitemShut {NoStop}%
\bibitem [{\citenamefont {Ressler}\ \emph {et~al.}(2025)\citenamefont {Ressler}, \citenamefont {Combi}, \citenamefont {Ripperda},\ and\ \citenamefont {Most}}]{Ressler:2024tan}%
  \BibitemOpen
  \bibfield  {author} {\bibinfo {author} {\bibfnamefont {S.~M.}\ \bibnamefont {Ressler}}, \bibinfo {author} {\bibfnamefont {L.}~\bibnamefont {Combi}}, \bibinfo {author} {\bibfnamefont {B.}~\bibnamefont {Ripperda}},\ and\ \bibinfo {author} {\bibfnamefont {E.~R.}\ \bibnamefont {Most}},\ }\bibfield  {title} {\bibinfo {title} {{Dual Jet Interaction, Magnetically Arrested Flows, and Flares in Accreting Binary Black Holes}},\ }\href {https://doi.org/10.3847/2041-8213/ad9eb5} {\bibfield  {journal} {\bibinfo  {journal} {Astrophys. J. Lett.}\ }\textbf {\bibinfo {volume} {979}},\ \bibinfo {pages} {L24} (\bibinfo {year} {2025})},\ \Eprint {https://arxiv.org/abs/2410.10944} {arXiv:2410.10944 [astro-ph.HE]} \BibitemShut {NoStop}%
\bibitem [{\citenamefont {{Zhu}}\ and\ \citenamefont {{Stone}}(2018)}]{2018ApJ...857...34Z}%
  \BibitemOpen
  \bibfield  {author} {\bibinfo {author} {\bibfnamefont {Z.}~\bibnamefont {{Zhu}}}\ and\ \bibinfo {author} {\bibfnamefont {J.~M.}\ \bibnamefont {{Stone}}},\ }\bibfield  {title} {\bibinfo {title} {{Global Evolution of an Accretion Disk with a Net Vertical Field: Coronal Accretion, Flux Transport, and Disk Winds}},\ }\href {https://doi.org/10.3847/1538-4357/aaafc9} {\bibfield  {journal} {\bibinfo  {journal} {Astrophys. J.}\ }\textbf {\bibinfo {volume} {857}},\ \bibinfo {eid} {34} (\bibinfo {year} {2018})},\ \Eprint {https://arxiv.org/abs/1701.04627} {arXiv:1701.04627 [astro-ph.EP]} \BibitemShut {NoStop}%
\bibitem [{\citenamefont {Yuan}\ \emph {et~al.}(2021)\citenamefont {Yuan}, \citenamefont {Murase}, \citenamefont {Zhang}, \citenamefont {Kimura},\ and\ \citenamefont {M{\'e}sz{\'a}ros}}]{Yuan:2021jjt}%
  \BibitemOpen
  \bibfield  {author} {\bibinfo {author} {\bibfnamefont {C.}~\bibnamefont {Yuan}}, \bibinfo {author} {\bibfnamefont {K.}~\bibnamefont {Murase}}, \bibinfo {author} {\bibfnamefont {B.~T.}\ \bibnamefont {Zhang}}, \bibinfo {author} {\bibfnamefont {S.~S.}\ \bibnamefont {Kimura}},\ and\ \bibinfo {author} {\bibfnamefont {P.}~\bibnamefont {M{\'e}sz{\'a}ros}},\ }\bibfield  {title} {\bibinfo {title} {{Post-Merger Jets from Supermassive Black Hole Coalescences as Electromagnetic Counterparts of Gravitational Wave Emission}},\ }\href {https://doi.org/10.3847/2041-8213/abee24} {\bibfield  {journal} {\bibinfo  {journal} {Astrophys. J. Lett.}\ }\textbf {\bibinfo {volume} {911}},\ \bibinfo {pages} {L15} (\bibinfo {year} {2021})},\ \Eprint {https://arxiv.org/abs/2101.05788} {arXiv:2101.05788 [astro-ph.HE]} \BibitemShut {NoStop}%
\bibitem [{\citenamefont {{Sheikhnezami}}\ and\ \citenamefont {{Fendt}}(2022)}]{2022ApJ...925..161S}%
  \BibitemOpen
  \bibfield  {author} {\bibinfo {author} {\bibfnamefont {S.}~\bibnamefont {{Sheikhnezami}}}\ and\ \bibinfo {author} {\bibfnamefont {C.}~\bibnamefont {{Fendt}}},\ }\bibfield  {title} {\bibinfo {title} {{The Physics of the MHD Disk-Jet Transition in Binary Systems: Jetted Spiral Walls Launched from Disk Spiral Arms}},\ }\href {https://doi.org/10.3847/1538-4357/ac3f31} {\bibfield  {journal} {\bibinfo  {journal} {Astrophys. J.}\ }\textbf {\bibinfo {volume} {925}},\ \bibinfo {eid} {161} (\bibinfo {year} {2022})},\ \Eprint {https://arxiv.org/abs/2112.02111} {arXiv:2112.02111 [astro-ph.HE]} \BibitemShut {NoStop}%
\bibitem [{\citenamefont {Begelman}\ \emph {et~al.}(1980)\citenamefont {Begelman}, \citenamefont {Blandford},\ and\ \citenamefont {Rees}}]{Begelman:1980vb}%
  \BibitemOpen
  \bibfield  {author} {\bibinfo {author} {\bibfnamefont {M.~C.}\ \bibnamefont {Begelman}}, \bibinfo {author} {\bibfnamefont {R.~D.}\ \bibnamefont {Blandford}},\ and\ \bibinfo {author} {\bibfnamefont {M.~J.}\ \bibnamefont {Rees}},\ }\bibfield  {title} {\bibinfo {title} {{Massive black hole binaries in active galactic nuclei}},\ }\href {https://doi.org/10.1038/287307a0} {\bibfield  {journal} {\bibinfo  {journal} {Nature}\ }\textbf {\bibinfo {volume} {287}},\ \bibinfo {pages} {307} (\bibinfo {year} {1980})}\BibitemShut {NoStop}%
\bibitem [{\citenamefont {D'Orazio}\ and\ \citenamefont {Duffell}(2021)}]{DOrazio:2021kob}%
  \BibitemOpen
  \bibfield  {author} {\bibinfo {author} {\bibfnamefont {D.~J.}\ \bibnamefont {D'Orazio}}\ and\ \bibinfo {author} {\bibfnamefont {P.~C.}\ \bibnamefont {Duffell}},\ }\bibfield  {title} {\bibinfo {title} {{Orbital Evolution of Equal-mass Eccentric Binaries due to a Gas Disk: Eccentric Inspirals and Circular Outspirals}},\ }\href {https://doi.org/10.3847/2041-8213/ac0621} {\bibfield  {journal} {\bibinfo  {journal} {Astrophys. J. Lett.}\ }\textbf {\bibinfo {volume} {914}},\ \bibinfo {pages} {L21} (\bibinfo {year} {2021})},\ \Eprint {https://arxiv.org/abs/2103.09251} {arXiv:2103.09251 [astro-ph.HE]} \BibitemShut {NoStop}%
\bibitem [{\citenamefont {Siwek}\ \emph {et~al.}(2023)\citenamefont {Siwek}, \citenamefont {Weinberger},\ and\ \citenamefont {Hernquist}}]{Siwek:2023rlk}%
  \BibitemOpen
  \bibfield  {author} {\bibinfo {author} {\bibfnamefont {M.}~\bibnamefont {Siwek}}, \bibinfo {author} {\bibfnamefont {R.}~\bibnamefont {Weinberger}},\ and\ \bibinfo {author} {\bibfnamefont {L.}~\bibnamefont {Hernquist}},\ }\bibfield  {title} {\bibinfo {title} {{Orbital evolution of binaries in circumbinary discs}},\ }\href {https://doi.org/10.1093/mnras/stad1131} {\bibfield  {journal} {\bibinfo  {journal} {Mon. Not. Roy. Astron. Soc.}\ }\textbf {\bibinfo {volume} {522}},\ \bibinfo {pages} {2707} (\bibinfo {year} {2023})},\ \Eprint {https://arxiv.org/abs/2302.01785} {arXiv:2302.01785 [astro-ph.HE]} \BibitemShut {NoStop}%
\bibitem [{\citenamefont {Chatterjee}\ \emph {et~al.}(2021)\citenamefont {Chatterjee} \emph {et~al.}}]{Chatterjee:2020wef}%
  \BibitemOpen
  \bibfield  {author} {\bibinfo {author} {\bibfnamefont {K.}~\bibnamefont {Chatterjee}} \emph {et~al.},\ }\bibfield  {title} {\bibinfo {title} {{General relativistic MHD simulations of non-thermal flaring in Sagittarius A*}},\ }\href {https://doi.org/10.1093/mnras/stab2466} {\bibfield  {journal} {\bibinfo  {journal} {Mon. Not. Roy. Astron. Soc.}\ }\textbf {\bibinfo {volume} {507}},\ \bibinfo {pages} {5281} (\bibinfo {year} {2021})},\ \Eprint {https://arxiv.org/abs/2011.08904} {arXiv:2011.08904 [astro-ph.HE]} \BibitemShut {NoStop}%
\bibitem [{\citenamefont {Moody}\ \emph {et~al.}(2019)\citenamefont {Moody}, \citenamefont {Shi},\ and\ \citenamefont {Stone}}]{Moody:2019nes}%
  \BibitemOpen
  \bibfield  {author} {\bibinfo {author} {\bibfnamefont {M.~S.~L.}\ \bibnamefont {Moody}}, \bibinfo {author} {\bibfnamefont {J.-M.}\ \bibnamefont {Shi}},\ and\ \bibinfo {author} {\bibfnamefont {J.~M.}\ \bibnamefont {Stone}},\ }\bibfield  {title} {\bibinfo {title} {{Hydrodynamic Torques in Circumbinary Accretion Disks}},\ }\href {https://doi.org/10.3847/1538-4357/ab09ee} {\bibfield  {journal} {\bibinfo  {journal} {Astrophys. J.}\ }\textbf {\bibinfo {volume} {875}},\ \bibinfo {pages} {66} (\bibinfo {year} {2019})},\ \Eprint {https://arxiv.org/abs/1903.00008} {arXiv:1903.00008 [astro-ph.HE]} \BibitemShut {NoStop}%
\bibitem [{\citenamefont {{Dittmann}}\ and\ \citenamefont {{Ryan}}(2022)}]{Dittmann2022}%
  \BibitemOpen
  \bibfield  {author} {\bibinfo {author} {\bibfnamefont {A.~J.}\ \bibnamefont {{Dittmann}}}\ and\ \bibinfo {author} {\bibfnamefont {G.}~\bibnamefont {{Ryan}}},\ }\bibfield  {title} {\bibinfo {title} {{A Survey of Disc Thickness and Viscosity in Circumbinary Accretion: Binary Evolution, Variability, and Disc Morphology}},\ }\href@noop {} {\bibfield  {journal} {\bibinfo  {journal} {arXiv e-prints}\ ,\ \bibinfo {eid} {arXiv:2201.07816}} (\bibinfo {year} {2022})},\ \Eprint {https://arxiv.org/abs/2201.07816} {arXiv:2201.07816 [astro-ph.HE]} \BibitemShut {NoStop}%
\bibitem [{\citenamefont {Tiede}\ and\ \citenamefont {D'Orazio}(2025)}]{Tiede:2025llq}%
  \BibitemOpen
  \bibfield  {author} {\bibinfo {author} {\bibfnamefont {C.}~\bibnamefont {Tiede}}\ and\ \bibinfo {author} {\bibfnamefont {D.~J.}\ \bibnamefont {D'Orazio}},\ }\bibfield  {title} {\bibinfo {title} {{Hot, cold, and multi-component accretion flows around supermassive black hole binaries}},\ }\href@noop {} {\bibfield  {journal} {\bibinfo  {journal} {arxiv}\ } (\bibinfo {year} {2025})},\ \Eprint {https://arxiv.org/abs/2508.11748} {arXiv:2508.11748 [astro-ph.HE]} \BibitemShut {NoStop}%
\bibitem [{\citenamefont {Wen}\ and\ \citenamefont {Paschalidis}(2025)}]{Wen:2025xpa}%
  \BibitemOpen
  \bibfield  {author} {\bibinfo {author} {\bibfnamefont {S.}~\bibnamefont {Wen}}\ and\ \bibinfo {author} {\bibfnamefont {V.}~\bibnamefont {Paschalidis}},\ }\bibfield  {title} {\bibinfo {title} {{A pseudo-Newtonian stationary circumbinary slim disk model}},\ }\href@noop {} {\bibfield  {journal} {\bibinfo  {journal} {arxiv}\ } (\bibinfo {year} {2025})},\ \Eprint {https://arxiv.org/abs/2508.15150} {arXiv:2508.15150 [astro-ph.HE]} \BibitemShut {NoStop}%
\bibitem [{\citenamefont {Hakobyan}\ \emph {et~al.}(2025)\citenamefont {Hakobyan}, \citenamefont {Levinson}, \citenamefont {Sironi}, \citenamefont {Philippov},\ and\ \citenamefont {Ripperda}}]{Hakobyan:2025ywq}%
  \BibitemOpen
  \bibfield  {author} {\bibinfo {author} {\bibfnamefont {H.}~\bibnamefont {Hakobyan}}, \bibinfo {author} {\bibfnamefont {A.}~\bibnamefont {Levinson}}, \bibinfo {author} {\bibfnamefont {L.}~\bibnamefont {Sironi}}, \bibinfo {author} {\bibfnamefont {A.}~\bibnamefont {Philippov}},\ and\ \bibinfo {author} {\bibfnamefont {B.}~\bibnamefont {Ripperda}},\ }\bibfield  {title} {\bibinfo {title} {{Reconnection-driven Flares in M87*: Proton-synchrotron Powered GeV Emission}},\ }\href@noop {} {\bibfield  {journal} {\bibinfo  {journal} {arxiv}\ } (\bibinfo {year} {2025})},\ \Eprint {https://arxiv.org/abs/2507.14002} {arXiv:2507.14002 [astro-ph.HE]} \BibitemShut {NoStop}%
\bibitem [{\citenamefont {Vetter}\ \emph {et~al.}(2024)\citenamefont {Vetter}, \citenamefont {R{\"o}pke}, \citenamefont {Schneider}, \citenamefont {Pakmor}, \citenamefont {Ohlmann}, \citenamefont {Lau},\ and\ \citenamefont {Andrassy}}]{Vetter:2024loo}%
  \BibitemOpen
  \bibfield  {author} {\bibinfo {author} {\bibfnamefont {M.}~\bibnamefont {Vetter}}, \bibinfo {author} {\bibfnamefont {F.~K.}\ \bibnamefont {R{\"o}pke}}, \bibinfo {author} {\bibfnamefont {F.~R.~N.}\ \bibnamefont {Schneider}}, \bibinfo {author} {\bibfnamefont {R.}~\bibnamefont {Pakmor}}, \bibinfo {author} {\bibfnamefont {S.~T.}\ \bibnamefont {Ohlmann}}, \bibinfo {author} {\bibfnamefont {M.~Y.~M.}\ \bibnamefont {Lau}},\ and\ \bibinfo {author} {\bibfnamefont {R.}~\bibnamefont {Andrassy}},\ }\bibfield  {title} {\bibinfo {title} {{From spherical stars to disk-like structures: 3D common-envelope evolution of massive binaries beyond inspiral}},\ }\href {https://doi.org/10.1051/0004-6361/202451579} {\bibfield  {journal} {\bibinfo  {journal} {Astron. Astrophys.}\ }\textbf {\bibinfo {volume} {691}},\ \bibinfo {pages} {A244} (\bibinfo {year} {2024})},\ \Eprint {https://arxiv.org/abs/2410.07841} {arXiv:2410.07841 [astro-ph.SR]} \BibitemShut {NoStop}%
\bibitem [{\citenamefont {Vetter}\ \emph {et~al.}(2025)\citenamefont {Vetter}, \citenamefont {Roepke}, \citenamefont {Schneider}, \citenamefont {Pakmor}, \citenamefont {Ohlmann}, \citenamefont {Mor{\'a}n-Fraile}, \citenamefont {Lau}, \citenamefont {Leidi}, \citenamefont {Gagnier},\ and\ \citenamefont {Andrassy}}]{Vetter:2025fby}%
  \BibitemOpen
  \bibfield  {author} {\bibinfo {author} {\bibfnamefont {M.}~\bibnamefont {Vetter}}, \bibinfo {author} {\bibfnamefont {F.~K.}\ \bibnamefont {Roepke}}, \bibinfo {author} {\bibfnamefont {F.~R.~N.}\ \bibnamefont {Schneider}}, \bibinfo {author} {\bibfnamefont {R.}~\bibnamefont {Pakmor}}, \bibinfo {author} {\bibfnamefont {S.}~\bibnamefont {Ohlmann}}, \bibinfo {author} {\bibfnamefont {J.}~\bibnamefont {Mor{\'a}n-Fraile}}, \bibinfo {author} {\bibfnamefont {M.~Y.~M.}\ \bibnamefont {Lau}}, \bibinfo {author} {\bibfnamefont {G.}~\bibnamefont {Leidi}}, \bibinfo {author} {\bibfnamefont {D.}~\bibnamefont {Gagnier}},\ and\ \bibinfo {author} {\bibfnamefont {R.}~\bibnamefont {Andrassy}},\ }\bibfield  {title} {\bibinfo {title} {{Magnetically driven outflows in the 3D common envelope evolution of massive stars}},\ }\href {https://doi.org/10.1051/0004-6361/202554685} {\bibfield  {journal} {\bibinfo  {journal} {Astron. Astrophys.}\ }\textbf {\bibinfo {volume} {698}},\ \bibinfo {pages} {A133} (\bibinfo {year} {2025})},\ \Eprint {https://arxiv.org/abs/2504.12213} {arXiv:2504.12213 [astro-ph.SR]} \BibitemShut {NoStop}%
\bibitem [{\citenamefont {{Luo}}\ and\ \citenamefont {{Shlosman}}(2024)}]{luo:2024.zoomin.bh.1star}%
  \BibitemOpen
  \bibfield  {author} {\bibinfo {author} {\bibfnamefont {Y.}~\bibnamefont {{Luo}}}\ and\ \bibinfo {author} {\bibfnamefont {I.}~\bibnamefont {{Shlosman}}},\ }\bibfield  {title} {\bibinfo {title} {{Direct Collapse Accretion Disks within Dark Matter Halos: Saturation of the Magnetorotational Instability and the Field Expulsion}},\ }\href {https://doi.org/10.3847/1538-4357/ad7fec} {\bibfield  {journal} {\bibinfo  {journal} {\apj}\ }\textbf {\bibinfo {volume} {976}},\ \bibinfo {eid} {85} (\bibinfo {year} {2024})},\ \Eprint {https://arxiv.org/abs/2409.17247} {arXiv:2409.17247 [astro-ph.GA]} \BibitemShut {NoStop}%
\bibitem [{\citenamefont {{Kato}}\ \emph {et~al.}(2004)\citenamefont {{Kato}}, \citenamefont {{Mineshige}},\ and\ \citenamefont {{Shibata}}}]{Kato2004}%
  \BibitemOpen
  \bibfield  {author} {\bibinfo {author} {\bibfnamefont {Y.}~\bibnamefont {{Kato}}}, \bibinfo {author} {\bibfnamefont {S.}~\bibnamefont {{Mineshige}}},\ and\ \bibinfo {author} {\bibfnamefont {K.}~\bibnamefont {{Shibata}}},\ }\bibfield  {title} {\bibinfo {title} {{Magnetohydrodynamic Accretion Flows: Formation of Magnetic Tower Jet and Subsequent Quasi-Steady State}},\ }\href {https://doi.org/10.1086/381234} {\bibfield  {journal} {\bibinfo  {journal} {Astrophys. J.}\ }\textbf {\bibinfo {volume} {605}},\ \bibinfo {pages} {307} (\bibinfo {year} {2004})},\ \Eprint {https://arxiv.org/abs/astro-ph/0307306} {arXiv:astro-ph/0307306 [astro-ph]} \BibitemShut {NoStop}%
\end{thebibliography}%
\end{document}